\documentclass[iop]{emulateapj-rtx4} 
\shortauthors{Sekanina}
\shorttitle{Formation and Evolution of SOHO Kreutz Sungrazers. II}
\slugcomment{Version \today }
\newcommand{\Rsun}{$\!R_{\mbox{\scriptsize \boldmath $\odot$}}$}

\newcommand{\lapeq}{$\;$\raisebox{0.3ex}{$<$}\hspace{-0.28cm}\raisebox{-0.75ex}{$\sim$}$\;$}

\begin{document}
\title{Formation and Evolution of the Stream of SOHO Kreutz Sungrazers.\\
 II. Results and Implications of Monte Carlo Simulation\\[-1.5cm]}
\author{Zdenek Sekanina}
\affil{La Canada Flintridge, California 91011, U.S.A.; {\sl ZdenSek@gmail.com}}

\begin{abstract} 
I present the results of the first comprehensive effort aimed at modeling a
major component~of~the stream of SOHO sungrazers, 5000 of which have been
detected by the onboard coronagraphs~since~1996.  The stream of Population~I
of the Kreutz system, investigated by a Monte Carlo simulation technique,
is treated as a product of cascading fragmentation due to unstable
rotation of a ``seed,'' a subkilometer-sized object that separated with
others from comet X/1106~C1 at perihelion.  The stream's activity~is
predicted to last for 200~years from $\sim$1950 to $\sim$2150, culminating
in the 2010s, when~a~swarm~of~bright SOHO sungrazers (peak mags not fainter
than 3) of Population~I was observed.  By the end of 2023 about 42~percent
of the stream had already arrived.  Scatter amounts to 7$^\circ$ in the
longitude of the ascending node and at most 0.2~\Rsun\,\,in the perihelion
distance.  On its initial orbit the seed would pass perihelion in 2036,
193~years after C/1843~D1, the principal fragment of X/1106~C1.~Comet
C/1668~E1 is proposed as another major fragment and yet another is predicted
to arrive in the 2050s or 2060s.
\end{abstract}
\keywords{individual comets: X/1106 C1, C/1668 E1, X/1702 D1, C/1843 D1, C/1882 R1, C/1887 B1, C/1963 R1, C/1965 S1, C/2011 W3; methods: data analysis\vspace{0.1cm}}

\section{Introduction} 
To model any class of cosmic objects is risky, because models are tested
by the degree of their compatibility with the results of observations,
which are a function of time.  A model may be consistent with existing
data one day, but get in conflict with new data the next day.  To model
a process of formation and evolution is particularly perilous because
of the delicate nature of the subject.  It is recalled that the
discovery and strength of the stream of Kreutz sungrazers in images
taken by the LASCO coronagraphs on board the Solar and Heliospheric
Observatory (SOHO) was a surprise, even though a fair number of similar
(though brighter) sungrazers was detected by coronagraphs on board two
previously launched spacecraft, operational over a decade only years
before SOHO began its mission.  Yet, it did not take long and the
space observatory became the most prolific discoverer of comets.  And
even though SOHO has been imaging all sorts of comets, the members of
the Kreutz system prevail by a wide margin.

A model for the formation and evolution of the sungrazer stream, the
observed part of which contains at present approximately 4300~objects, most
of them discovered over a period of 28~years, needs to be conceptually
anchored.  Accordingly, aiming at understanding the attributes of the
stream, I provide a comprehensive description of the proposed rationale,
including features of the most massive objects in the Kreutz system,
the source or sources of the stream, the nature of the processes that
govern the stream's formation and evolution, as well as the period of
time and the volume of space over which the activity has been continuing.
To assist in solving these complex intertwined topics, in Part~I of this
investigation I examined the annual arrival rate variations of the
Kreutz sungrazers in the stream, including the tendency toward swarming
in both time and the longitude of the ascending node.

\section{Perihelion Fragmentation}  
The two main populations of the Kreutz system, equivalent to Marsden's
(1967) original Subgroups~I and II, still provide one of the fundamental
dividing lines today, when the number of known members is orders of
magnitude greater.  Below I demonstrate that by paying attention to the two
populations in the SOHO database as well as among the brightest objects,
one gains new insights into the stream's formation and evolution.

Even though Populations I and II are known to differ from each other greatly
in a number of respects, the enormous disparities have generally been
acknowledged and accepted with no comment, as if they were expected.  The
most obvious among these imbalancies is the huge discrepancy in the arrival
rates:\ even though different authors report different numbers, the
Population~I membership among the SOHO sungrazers does always greatly
outnumber Population~II.  Recently, I have derived a ratio of 14:1 from
a SOHO database for the years 1996 through mid-2010 that lists objects
seen exclusively in the C2 coronagraph (Sekanina 2022a).  This enormous
ratio implies that the arrival rate of the SOHO sungrazers for one of
the two populations is necessarily anomalous:\ either it is much too
high for Population~I or much too low for Population~II.

Similarly disparate are the properties of the two populations' brightest
members and presumably the largest surviving masses of the Kreutz
progenitor:\ the Great March Comet of 1843 (C/1843~D1), of Population~I;
and the Great September Comet of 1882 (C/1882~R1), of Population~II.
While they may have been about equally bright intrinsically before
perihelion (when too few observations were made to be sure), the 1882
sungrazer was much brighter long after perihelion, apparently because
of its near-perihelion splitting.  By contrast, the 1843 sungrazer
exhibited no signs of breakup and faded rapidly when receding from
the Sun.

\begin{figure}[t] 
\vspace{0.15cm}
\hspace{-0.2cm}
\centerline{ 
\scalebox{0.88}{
\includegraphics{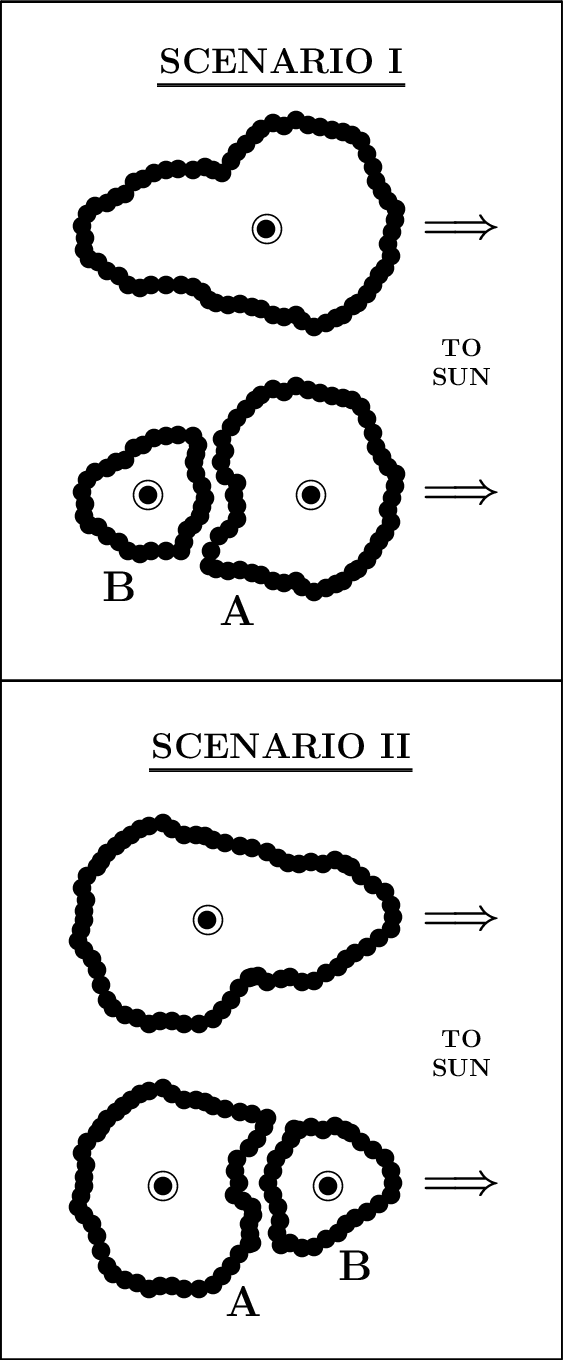}}}
\vspace{0cm}
\caption{A sungrazer's nucleus in close proximity of perihelion shortly
before and after breaking up tidally into two uneven fragments.  In
Scenario~I (top), it is the more sizable fragment~A that begins its
existence on the sunward side of the parent's nucleus (to the right).
It ends up in an orbit of shorter orbital period than the parent's.
On the other hand, the smaller fragment~B, on the far side from the
Sun, enters an orbit of longer orbital period than was the parent's.
In Scenario~II, the positions of the two fragments are swapped and so
are their future orbital periods.  The circled disks are the centers
of mass of the parent comet and the two fragments. (From Sekanina \&
Kracht 2022.)\vspace{-0.3cm}}
\end{figure}

\subsection{Perihelion Breakup of the 1882 Sungrazer} 
As discussed later in this paper, the phenomenon of perihelion breakup,
triggered presumably by the Sun's tidal forces, is likely to be the
initial stage of the process of formation of the stream of SOHO sungrazers.
Using the well-known breakup of the Great September Comet of 1882, I
will, in Section~2.3, point to a new implication of the differences
between Populations~I and II.

First of all, I employ a simple, but criticallly
important rule on the orbital transformation, that Sekanina \& Kracht
(2022) used in their orbit-integration computations.  The rule is
illustrated in Figure~1:\ at the instant of breakup, the parent
sungrazer and its center of mass, moving with a given orbital velocity,
suddenly turn into two or more fragments, each with its own center
of mass, but both/all still moving with the orbital velocity of the
original body.  The difference between the distances from the Sun of the
centers of mass of the parent and each of the fragments makes the latter
enter a new orbit with a different orbital period.  The fragment whose
center of mass is farther from the Sun than the parent's ends up in an
orbit of a longer period and vice versa.  It is straightforward to show
that the orbital period of a fragment, $P_{\rm frg}$,
is related\footnote{The only approximation used in the derivation of this
expression is \mbox{$r_{\rm frg} \!+\! U_{\rm frg} \doteq r_{\rm frg}$},
which for the Kreutz sungrazers{\vspace{-0.06cm}} involves typically
an error on the order of 10$^{-6}$.} to the orbital period of the
parent, $P_{\rm par}$, by
\begin{equation} 
P_{\rm frg} = P_{\rm par} \!\left(\!1 - \frac{2 U_{\rm frg}}{r_{\rm frg}^2}
 P_{\rm par}^{\frac{2}{3}} \!\!\!\:\right)^{\!\!-\frac{3}{2}} \!\!,
\end{equation}
where $r_{\rm frg}$ is the heliocentric distance at fragmentation,
$U_{\rm frg}$ is the difference between the heliocentric distances of the
fragment's and the parent's centers of mass; $U_{\rm frg}$ is positive when
the fragment is farther from the Sun than the parent.  In Equation~(1)
both $r_{\rm frg}$ and $U_{\rm frg}$ must be in AU and $P_{\rm par}$ and
$P_{\rm frg}$ in yr.  The relation can also~be written as
\begin{equation} 
U_{\rm frg} = {\textstyle \frac{1}{2}}\, r_{\rm frg}^2 \!\!\:\left(\! P_{\rm
 par}^{-\frac{2}{3}} \!-\! P_{\rm frg}^{-\frac{2}{3}} \!\!\:\right) \!.
\end{equation}

The minimum absolute value of $U_{\rm frg}$ for a given pair of periods
$P_{\rm par}$ and $P_{\rm frg}$ is obviously reached at perihelion.
Assuming that sungrazers could be tidally disrupted anywhere between
perihelion and a point in the orbit of up to, say, 3.4 {\Rsun} from the
Sun (crudely estimated~from~the Roche limit) and given that the perihelion
distance of some sungrazers is as low as 1.1~{\Rsun}, Equation~(2) shows
that the dimensions of sungrazers' fragments could not be estimated from
effects on the orbital period with accuracy much better than a factor of
ten.  In any case the differences between the orbital periods of the
parent and a fragment (or between different{\vspace{-0.04cm}} fragments)
provide information on the quantity of $U_{\rm frg}/r_{\rm frg}^2$.

\begin{table*}[t] 
\vspace{0.2cm}
\hspace{-0.2cm}
\centerline{
\scalebox{1}{
\includegraphics{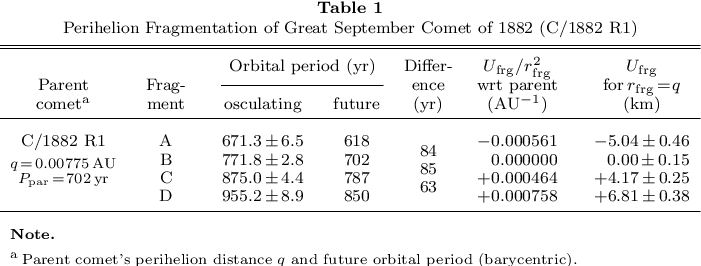}}}
\vspace{0.6cm}
\end{table*}

Kreutz (1891) investigated the orbital motions of four fragments of the
nucleus of the 1882 sungrazer in considerable detail.  He called the
fragments No.~1 through No.~4, although they are nowadays referred to
usually as A through D.  Although the number of observed fragments was
at times greater, Kreutz was not able to collect enough data to compute
satisfactory orbits for the additional ones.

The osculating orbital periods derived by Kreutz for the fragments A
through D are presented in column~3~of Table~1.  Fragment~B is assumed
to be the primary, by far the most massive fragment, which after the
perihelion-breakup event continued to move essentially in the orbit of
the parent sungrazer, C/1882~R1.\footnote{Throughout this paper, Kreutz's
(1891) nonrelativistic set of orbital elements rather than Hufnagel's
(1919) relativistic set is used for fragment~B of the 1882 sungrazer,
because the latter is likely to be erroneous.  Hufnagel's numbers
show that inclusion of the relativistic effect diminished the orbital
period by 1.4~percent.  However, Marsden's (1967) orbit determination
of comet Ikeya-Seki indicated that inclusion of the relativistic
effect {\it increased\/} the orbital period by less than 0.2~percent.
Also, his integration of Kreutz's nonrelativistic orbit back to the
12th century implied the parent comet's perihelion time in April 1138,
which matched the parent sungrazer's proposed time to within an
incredible 0.04~percent of the orbital period!  Integration of Hufnagel's
orbit, on the other hand, resulted in a meaningless date in November 1849.}
Based on the computations by Sekanina \& Chodas (2007), the corresponding
future barycentric orbital periods are given in column~4, while column~5
shows the differences.  The ratio of $U_{\rm frg}/r_{\rm frg}^2$, derived
{\vspace{-0.04cm}}from Equation~(2), is displayed in column~6. The last
column provides the shift in the heliocentric distance of each fragment's
center of mass relative to the parent's for fragmentation taking place
at perihelion.  Yes, a shift of mere 5~km in the radial position of the
center of mass implies a change of 85~years in the fragment's orbital
period!

I used a method devised for other purposes in one of my recent papers
(Sekanina 2021a) to check whether it is at all possible to satisfy a
condition that B be located at the center of mass of the parent while
the centers of mass of A, C, and D be distributed in line with the
distances determined by their post-breakup orbital periods in Table~1.
The method, based on the properties of a spheroid, shows that this
indeed is possible.  For a tidal event at perihelion, the condition is
satisfied when the length of the spheroid is 18.7~km and the masses
of the fragments are in the ratios of \mbox{A\,:\,B\,:\,C\,:\,D = 
0.27\,:\,0.46\,:\,0.16\,:\,0.11}.  The primary would thus retain
almost 50~percent of the pre-breakup mass.  It is likely that the
method underestimates the mass of B.  Given that $r_{\rm frg}$
could be up to twice the perihelion distance, the spheroid nucleus
of the 1882 sungrazer before its perihelion breakup may have been
up to about 75~km in length.  In a previous paper (Sekanina 2002)
I crudely estimated the comet's nucleus to be about 50 km across.
These dimensions would imply that the tidal fragmentation occurred
at a heliocentric distance of 2.7~{\Rsun}.

\subsection{Perihelion Breakup of the 1882 Sungrazer's\\Parent Comet}
Marsden's (1967) proof that the sungrazers C/1882~R1 and C/1965~S1
were a single object on their way to the 12th century perihelion has
motivated me to consider applying this procedure to their parent,
recently identified as the Chinese Comet of 1138 reported in
September of that year (Sekanina \& Kracht 2022).  The Great September
Comet of 1882 should serve as the principal fragment.  I have searched
for additional fragments and found two potential ones.  In Strom's
(2002) list of previously unrecognized Chinese daytime observations
of ``sun-comets'', one event is dated to April or May of 1792,
reported from the Shandong Province.  While it is by no means certain,
or even probable, that this indeed was a Kreutz sungrazer, I count it
in for the sake of argument.

The second candidate is comet X/1702~D1, which as 1702a was considered
by Kreutz (1901).  Even though the poor quality of the observations did
not allow him to compute an orbit, his opinion was that the comet's
motion could better be fitted with the orbit of C/1882~R1 than
C/1843~D1, and that the perihelion time, converted to TT, was close
to February 15.0.

\begin{table*}[t] 
\vspace{0.2cm}
\hspace{-0.2cm}
\centerline{
\scalebox{1}{
\includegraphics{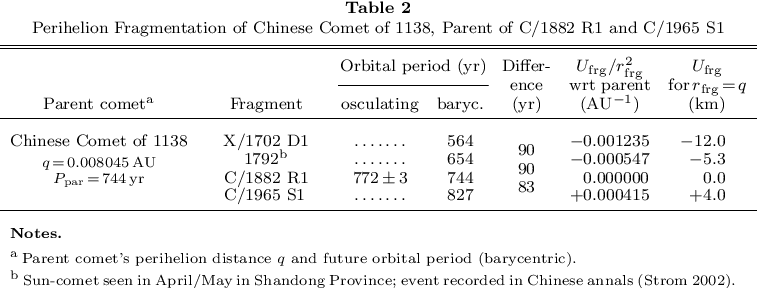}}}
\vspace{0.6cm}
\end{table*}

The potential perihelion-fragmentation products of the Chinese comet
of 1138 are presented in Table~2, which is organized in a manner
resembling that of Table~1.  The reader will note that if the breakup
took place at perihelion, the overall range of $U_{\rm frg}$ was about
12~km (implying a spheroid nearly 19~km in length) for the Great
September Comet of 1882, but 16~km for its parent.  In any case,
the similarity between the numbers of $U_{\rm frg}$ for fragments A
and C on the one hand and for 1792 and C/1965~S1 on the other hand
is stunning.  As it is by no means obvious that comet X/1702~D1 was
actually a fragment of the Chinese comet of 1138, one similarly finds
no reason why another fragment could not in due time follow comet
Ikeya-Seki.

This argument does in turn tempt one to directly ask a highly
contentious question:\ ``{\it So what about the future?\/}''  A
potential follower of comet Ikeya-Seki in the chain would arrive
at a time determined by its own value of $U_{\rm frg}$.  For
example, at \mbox{$r_{\rm frg} = q$} a doubled Ikeya-Seki value,
\mbox{$U_{\rm frg} = 8$ km}, would lead to \mbox{$P_{\rm frg} = 927$
yr} and a perihelion time in 2065.  If, instead, $U_{\rm frg}$ for
the next fragment were close to that for fragment~D of the Great
September Comet of 1882, it should arrive in 2032.  Given the
current low level of SOHO Population~II activity, this possibility
does not look likely.

\subsection{Presumed Perihelion Breakup of the Great Comet\\of 1106
 (X/1106~C1)}  
From the limited results of available observations of the nuclear region
made with modest telescopes of the time, such as the descriptions by
C.\ Piazzi Smyth at Cape (Warner 1980), the Great March Comet of 1843
displayed no obvious signs of breakup.  In addition, no records exist
about a bright Population~I sungrazer at equivalent times, about 80 to
90~years before or following the 1843~spectacle.

Comets C/1880~C1 and C/1887~B1 would not do.  They arrived at
inappropriate times and C/1887~B1~was almost certainly a subfragment of
C/1880~C1, which itself appeared to have been the product of an episode
of non\-tidal fragmentation that took place at fairly large heliocentric
{\vspace{-0.04cm}}distance (\mbox{Sekanina} 2021b).  Besides, either
object was much too small\footnote{For example, Gould (1891) remarked
that positional measurements of comet C/1880~C1, observed over a period
of only two weeks after perihelion, ``{\it were rendered difficult by
the lack of a nucleus or condensation in the head, which appeared like
a cloud, elongated in the direction of the tail and of but slightly
greater brilliancy.\/}''  This description is strongly reminiscent of
the words depicting cometary nuclei in the process of disintegration.}
to represent a major fragment of such a magnificent comet as X/1106~C1
was.

Comet Pereyra (C/1963~R1) would fit sizewise, but it moved in a wrong
orbit, as illustrated~by~\mbox{Marsden's} (1989) difficulties with
its motion.  Also, the associated population is classified as a
separate branch of Population~I, with a distinct range of nodal
longitudes (Sekanina 2021b), and the fragmentation history of comet
Pereyra had apparently not been directly linked to the Great Comet
of 1106, but to a presumed sungrazer of 1041 (Sekanina \& Kracht 2022).

If the Great Comet of 1106 did not split at perihelion, why do we witness
such an astonishingly prominent presence of Population~I in the stream of
SOHO sungrazers?  A possible but rather shocking solution to this dilemma
is{\vspace{-0.06cm}} provided by the expression $U_{\rm frg}/r_{\rm frg}^2$
in Equations~(1) and (2).  If the positions of the fragments' centers of
mass share similar patterns for the sungrazers of either population in
that the values of $U_{\rm frg}$ are comparable, the substantially smaller
perihelion distances of Population~I objects cause that {\it the arrivals
of these fragments should stretch\/} over time periods at least {\it twice
as wide\/} as do the Population~II sungrazers.  It is likely that X/1106~C1
{\it did split\/}, but the fragments have been scattered too far apart to
be easily recognized as such.

This disparity shows up when I apply the conditions
for fragments A and B of C/1882~R1 from Table~1 (referred to as
Scenario~Y$^-$) and for comets 1792 and C/1882~R1 as fragments of comet
1138 from Table~2 (Scenario~Z$^-$) to an equivalent episode of perihelion
fragmentation experienced by X/1106~C1 to see when the 1843 sungrazer's
sibling that preceded it should have arrived.  I next apply the conditions
for fragments B and C of C/1882~R1 from Table~1 (Scenario Y$^+$) and
sungrazers C/1882~R1 vs C/1965~S1 from Table~2 (Scenario Z$^+$) to an
equivalent fragmentation event of X/1106~C1 to see when
the 1843 sungrazer's sibling that follows it may arrive.

The results in Table~3 confirm what was said above, yet they do offer a
surprise.  If the fragmentation constants derived for the Population~II
sungrazers apply to X/1106~C1 as well, the fragment preceding C/1843~D1
should have been C/1668~E1, believed by some in the mid-19th century to
have been the 1843 sungrazer's previous return to the Sun (e.g., Henderson
1843)!  A map of the 1668 comet's path as seen from Goa, India, was examined
by Kreutz (1901).  He concluded that the comet moved in the orbital plane
of C/1843~D1 and for an appropriately selected perihelion time the
elements did not contradict the observations; comparison with C/1882~R1
was less satisfactory.  Marsden (1967) did include the comet of 1668
among his eight Kreutz objects.

The younger generation of comet enthusiasts might~be interested to learn
that \mbox{\it this procedure predicts\/} the appearance of a naked-eye
Population~I sungrazer within the next 50~years or so --- the X/1106~C1
fragment that follows C/1843~D1.  I noted in Section~2.2 that a major member
of Population~II, a sibling of comet Ikeya-Seki, is also expected to arrive
at about the same time.  

\begin{table}[b] 
\vspace{0.6cm}
\hspace{-0.22cm}
\centerline{
\scalebox{1}{
\includegraphics{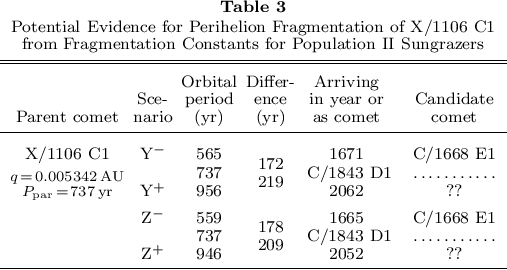}}}
\vspace{0.11cm}
\end{table}

Approximately equivalent timelines for arrivals of new naked-eye members
of Populations~I and II clearly contradict the uneven contributions to
the stream of SOHO sungrazers.  However, the dominance of Population~I
could be interpreted in various ways.  Besides the one to regard it
as supporting evidence for an ``imminent'' arrival of a spectacular
member of Population~I, one could take an opposite view and argue
that the overwhelming preponderance of Population~I among the SOHO
sungrazers represents the debris of the disintegrated naked-eye object
that we have been waiting for in vain.  A problem with this view is that
if a sungrazer kilometers across fell apart, the debris stream should be
orders of magnitude more plentiful.  The problem is further complicated
by the existence of temporally limited enhancements of dwarf comets, as
noted in Part~I.

In a striking example, a burst of SOHO/STEREO sungrazers reaching
magnitude~3 or brighter at maximum light was reported by
\mbox{Sekanina} \& Kracht (2013).  The period of time covered by the
study was \mbox{2004--2013}, but the highly elevated arrival rates
were restricted to a few years centered on the peak of 4.6~objects
per year at the end of 2010.  This happened to be almost exactly one
year before the appearance of comet Lovejoy (C/2011~W3), which had
no direct association with the burst consisting exclusively of
Population~I members.  By 2013 the rate was back to normal,
\mbox{1--2}~comets brighter than magnitude~3 per year.

Because of uncertainties beyond the limits of the applied procedure,
the prospects for the arrival of a brilliant Population~I sungrazer
in a much nearer future than $\sim$40 or so years from now cannot be
ruled out.  Overall, it appears to be more likely than the early
arrival of a naked-eye Population~II sungrazer.

%
%
\subsection{Peculiar Kreutz Sungrazers}  
One may be tempted to recognize two kinds of a Kreutz sungrazer:\ its
nucleus is large enough to survive its return to the Sun either essentially
intact or split into two or more fragments; or it is too small to survive
and perishes just before perihelion.

Comet Lovejoy defied either scenario.  This sungrazer, discovered before
perihelion from the ground but observed by a number of space observatories
throughout the perihelion arc of its sungrazing orbit and starting as early
as 1~day after perihelion again from the ground, did survive perihelion
and began to immediately develop its new dust tail.  However, on a
ground-based image taken 3.4~days after perihelion, a narrow streamer at
least 200,000~km long showed up, suggesting that the comet suffered a modest
outburst.  Twenty-four hours later, the object's morphology changed radically,
its nuclear condensation {\it very suddenly\/} disappearing.  Examination
of systematic ground-based imaging observations of the new straight,
ribbon-like ``spine'' or trail of material, which replaced the ordinary
dust tail, allowed us to determine that the disintegration process
commenced as early as 1.6\,$\pm$\,0.2~days after perihelion (Sekanina \&
Chodas 2012).  We estimated that the nucleus was about 400~meters across
upon approach to the Sun.

In retrospect, comet Lovejoy was not the only Kreutz sungrazer that {\it
survived perihelion but not the entire perihelion return\/}.  The strange
comet C/1887~B1, observed as a headless tail over a period of 10~days
starting 8~days after perihelion, must have been subjected to the same
type of catastrophic event 0.24\,$\pm$\,0.03~day after perihelion (Sekanina
1984), suggesting that the initial dimensions of its nucleus were smaller
than Lovejoy's.  I suspect that C/1880~C1 was close to experiencing the
Lovejoy-type event, avoiding it just narrowly.

Another potential member of this exceptional category of Kreutz sungrazers,
comet du Toit (C/1945~X1), does not as yet contribute to our knowledge of
this type of events, but a comprehensive examination of the relevant Boyden
plates --- now that their digitized version has become available --- could
offer new information.  The comet was last photographed about 13~days
before perihelion and not seen ever since.  Unless some positive evidence
does show up in the future, this object could rank as a dwarf Kreutz comet
(Sekanina \& Kracht 2015a), a view that Seargent (2009) also appears to be
in favor of.  However, the comet was then exceptionally bright before
perihelion, an anomaly that needs to be explained.

\subsection{Sunlight Striking An Eroding Sungrazer\\Near Perihelion}
An important part of interaction with the Sun near perihelion is the
amount of radiation that is incident on an eroding sungrazer's surface.  It
is standard to assume that the solar flux drops with increasing heliocentric
distance $r$ at a rate of \mbox{$f_\odot(r) = R_\odot^2/r^2$}, where
$R_\odot$~is~the~radius of the Sun's photosphere.  This point-source
approximation is fine as long as \mbox{$r \!\gg\! R_\odot$}, but does not
apply near the Sun, when the radiation reaching the sungrazer's surface comes
from only a small part of the Sun's surface, which however is much closer
to the sungrazer than $r$.

\begin{figure}[b]
\vspace{0.6cm}
\hspace{-0.2cm}
\centerline{
\scalebox{0.8}{
\includegraphics{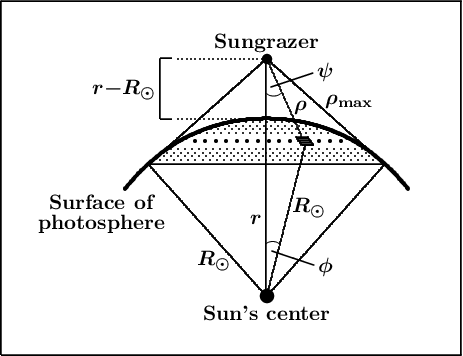}}}
\vspace{0cm}
\caption{Sunlight incident on a sungrazer in a close-up.  Only the
radiation from the dotted area of the photosphere reaches the comet.  The
rhombus shows an infinetesimal area on the Sun, whose distance from the
sungrazer is $\rho$ and the angles from the Sun-comet line are $\phi$ and
$\psi$, reckoned at their centers, respectively.{\vspace{-0.09cm}}}
\end{figure}
\begin{table*}[t] 
\vspace{0.15cm}
\hspace{-0.18cm}
\centerline{
\scalebox{1}{
\includegraphics{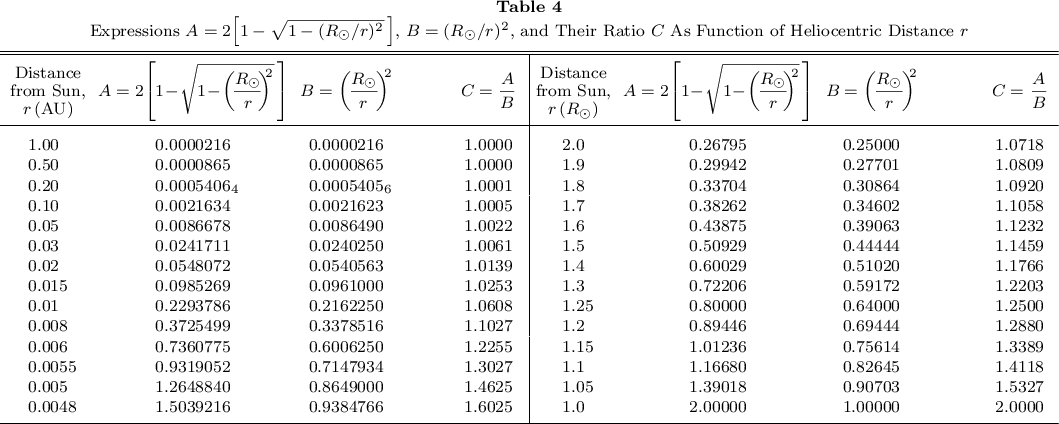}}}
\vspace{0.6cm}
\end{table*}

The correct expression for $f_\odot$ is derived by following a schematic
picture in Figure~2.  The area of the Sun's photosphere, from which the
radiation reaches the sungrazer, is the shaded cap centered on the
sungrazer's~sub\-solar point, whose distance from the comet is \mbox{$r
\!-\!R_\odot$}.~The~radiation from a sum of infinitesimal areas of
width \mbox{$R_\odot \, d\phi$}, like the one projected in the figure as
a rhombus, integrated along the dotted circle of radius \mbox{$R_\odot
\sin \phi$} on the Sun's surface and at a distance $\rho$ from the
comet, amounts to \mbox{$2\pi R_\odot^2 \cos(\phi \!+\! \psi) \sin \phi \,
d\phi/\rho^2$}, where \mbox{$\phi \!+\! \psi$} is the angle that the
normal to the radiation area subtends with the direction to the sungrazer.
Normalizing and integrating over all relevant angles $\phi$, one has
\begin{equation} 
f_\odot(r) = 2 R_\odot^2 \!\!\int_{0}^{\arccos(\!R_\odot\!\!\:/r)}
 \frac{\cos(\phi \!+\! \psi) \sin \phi}{\rho^2} \, d\phi,
\end{equation}
where
\begin{eqnarray} 
\rho \sin \psi & = & R_\odot \sin \phi, \nonumber \\
\rho \cos \psi & = & r \!-\! R_\odot \cos \phi,
\end{eqnarray}
so that
\begin{equation} 
\cos(\phi \!+\! \psi) = \frac{r \cos \phi \!-\! R_\odot}{\left(r^2 \!+\!
 R_\odot^2 \!-\! 2 r R_\odot \cos \phi\right)^{\frac{1}{2}}}.
\end{equation}
Inserting this expression into Equation (3) one gets, after substituting
\mbox{$\cos \phi = x$}, the result:
\begin{eqnarray} 
f_\odot(r) & = & 2 R_\odot^2 \!\!\int_{\!R_\odot\!\!\:/r}^{1}
 \frac{rx \!-\! R_\odot}{\left(r^2 \!+\! R_\odot^2 \!-\! 2rR_\odot
 x\right)^{\frac{3}{2}}} \, dx
 \nonumber \\[0.2cm]
 & = & 2 \!\left[1 - \sqrt{1 \!-\! \left( \!\frac{R_\odot}{r}
 \!\right)^{\!\!2}} \,\, \right] \!\!.
\end{eqnarray}
As expected,
\begin{equation} 
\lim_{r \gg R_\odot} \!\!\!f_\odot(r) = \left( \! \frac{R_\odot}{r} \!
 \right)^{\!\!2} \!\!.
\end{equation}

The significance of the correct procedure is apparent from Table~4, in which
the accurate expression, referred to as $A$, is compared with the approximate
one, $B$, in a wide range of heliocentric distances.  It turns out that the
amount of solar radiation reaching the surface of a sungrazer is always {\it
higher\/} than the inverse square power law suggests.  However, while the
difference is still only a fraction of 1~percent at a distance as small as
0.03~AU from the Sun, at 0.01~AU it already reaches 6~percent.  Table~4 shows
that the magnitude of the effect increases progressively at close proximity
of the Sun but stays smaller than a factor of two in any sungrazing orbit.

To what extent does his effect influence the sungrazers' rate of erosion?
A straightforward answer can readily be provided by illustrating the
progressing sublimation{\vspace{-0.055cm}} of a column of water ice
integrated along a sungrazing parabolic orbit.  Let
$\dot{Z}(r)$ be the rate of{\vspace{-0.04cm}} sublimation (reckoned in
a number of molecules per cm$^2$ per second) from a sphere of water
ice at $r$, as determined from a standard, surface-averaged energy balance
equation using an inverse square power-law approximation (e.g., Delsemme \&
Miller 1971).  If $m$ is the mass of a water molecule and \mbox{$\rho_0 =
0.5$ g cm$^{-3}$} a bulk density of the sphere, a layer of ice lost by the
sphere between aphelion and time $t-t_\pi$, measured from perihelion, has
a thickness $\tau$ equaling
\begin{equation} 
\tau = \!\!\int_{-\infty}^{t-t_\pi} \!\frac{m \dot{Z}(r)}{\rho_0} \,dt =
 \frac{\sqrt{2}mq^{\frac{3}{2}}}{\rho_0k_0}\!\!\int_{-\frac{1}{2}\pi}^{\frac{1}{2}u}
 \frac{\dot{Z}(r)}{\cos^4\frac{1}{2}u} \,d({\textstyle \frac{1}{2}}u),
\end{equation}
where $u$ is the true anomaly at time $t$ and heliocentric distance $r$,
and $k_0$ is the Gaussian{\vspace{-0.035cm}} gravitational constant.
The sublimation rates $\dot{Z}(r)$, derived from the standard inverse
square power law, underestimate the thickness of the eroded layer of
ice when inserted into Equation~(8).  To account approximately for the
effect expressed by Equation~(6) and thus obtain a corrected
thickness of the eroded ice layer, $\widehat{\tau}$, one needs in each
integration step in Equation (8) to substitute $\widehat{r}$ for $r$,
\begin{equation} 
\widehat{r} = \frac{R_\odot}{\sqrt{2}} \!\left[1 \!-\! \sqrt{1 \!-\! \left(
 \! \frac{R_\odot}{r} \!\right)^{\!\!2}} \, \right]^{-\frac{1}{2}} \!\!\!,
\end{equation}
and replace $\dot{Z}(r)$ with $\dot{Z}(\widehat{r})$.  The integration
limits remain unaffected by this change.

\begin{table*}[t] 
\vspace{0.15cm}
\hspace{-0.15cm}
\centerline{
\scalebox{1}{
\includegraphics{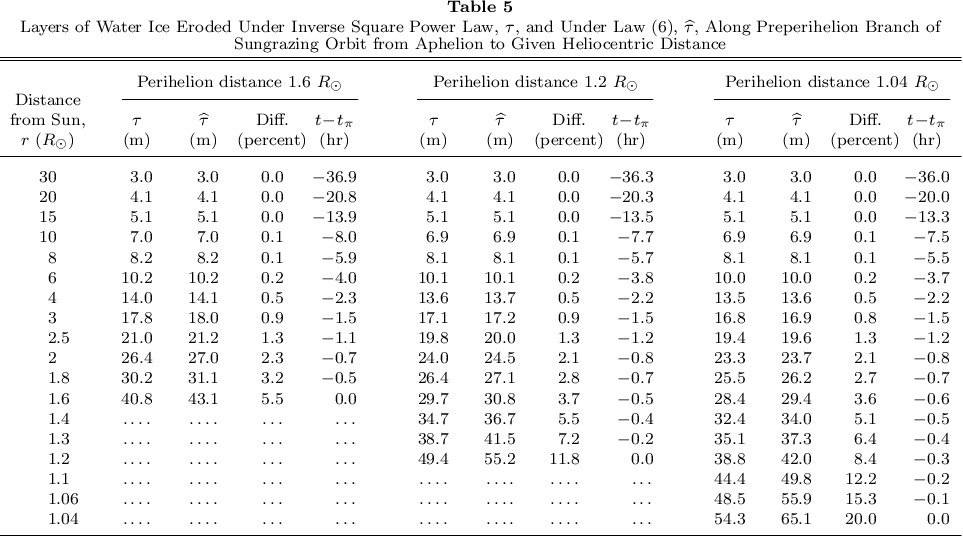}}}
\vspace{0.6cm}
\end{table*}

The increasing erosion of the layer on the surface of a water-ice
sphere along the preperihelion branch of a sungrazing orbit is for
three perihelion distances between 1.04 $R_\odot$ and 1.6 $R_\odot$
displayed in Table~5.  The~\mbox{numbers} are near a lower end of the usual
range, as they apply to the surface averaged case of the sublimation
regime, the so-called rapidly rotating comet.~The~approximate, widely
used model based on the inverse square power law for the incident
solar radiation, yielding the thickness $\tau$, is compared with the
preferred model governed~by~Equation~(6), increasing the thickness
to $\widehat{\tau}$.  The relative difference is given in percent,
showing that it could reach up to 20~percent only in orbits that
penetrate very deep, nearly to the very surface of the photosphere.

Most SOHO Kreutz sungrazers disintegrate \mbox{6--8 hr} before
perihelion (measured by extrapolation along the fitted parabolic
orbit --- see Part I of this investigation).  Table~5 shows that
by that time a layer of water ice 7--8 meters thick would erode.
Thus, an icy boulder 15~meters in diameter would by then sublimate
away completely.  The smallest Kreutz sungrazers detected by SOHO
are believed to be \mbox{5--10}~meters in diameter at the time they
begin their final approach (e.g., Battams \& Knight 2017).

\section{Modeling the Fragmentation Process}  
Failure of the SOHO Kreutz sungrazers to survive perihelion is powerful
evidence for an argument that the dimensions of the objects that did
survive as products of a parent comet's perihelion breakup to begin
their journey about the Sun must have been much larger.  Subjected
to the process of {\it cascading fragmentation\/}, many of them ended
up a revolution later as boulders in an estimated meter-size range.
The consensus is that the stream contains even larger numbers of still
smaller fragments.  Discovery of more than a dozen sungrazers with the
camera on board the Parker Solar Probe, some not seen by SOHO, appears
to be the case in point.

In the following, I describe the assumptions and constraints I employ
to incorporate cascading fragmentation in the computations.  But
first I address a hypothetical~yet~key~issue:\ {\it
How an arrival-rate curve of a stream of SOHO-like meter-sized
boulders would look like in the absence of fragmentation,
that is, if the boulders were directly stripped from the surface of
a parent sungrazer at perihelion and {\sf miracuously survived}
until next perihelion?}

\subsection{Stream's Hypothetical Fragmentation-Free Model} 
Let a massive sungrazer shed an enormous number of boulders at the
time of perihelion passage.  Allowed to survive, they end up in orbits
whose periods are governed by Equation~(1) and return to perihelion
at different times as a stream, presumably similar to that observed
in the coronagraphs on board SOHO.  What would be a range of the
boulders' orbital periods and the stream's timeline upon return
to perihelion?

I approximate the shape of the sungrazer's nucleus by a prolate
spheroid and assume that at the time the boulders were removed, the
long axis of the nucleus was aligned with the Sun direction.  Let
$U_{\rm frg}$ be a distance from the center of the spheroidal
nucleus along the Sun-comet line, reckoned positive away from the Sun,
and let the mass of the fraction of the boulders removed between the
radial distances of $U_{\rm frg}$ and \mbox{$U_{\rm frg} \!+\!
dU_{\rm frg}$} be $dm$ and their number, $d{\cal N}$, proportional
to $dm$.

Elsewhere I showed (Sekanina 2021a) that the normalized volume (or
mass), $\Im(\eta)$, of a prolate spheroid's cap of a normalized
height, $\eta$, measured in units of the spheroid's long diameter,
$\cal D$, from its sunward pole to a plane perpendicular to the
long axis equals
\begin{equation} 
\Im(\eta) = 3 \eta^2 \! \left( 1 \!-\! {\textstyle \frac{2}{3}} \eta \right).
\end{equation}
For \mbox{$\eta = 1$} this formula gives \mbox{$\Im = 1$}, confirming
that the spheroid cap's height has indeed been normalized.  Similarly,
as expected, \mbox{$\eta = {\textstyle \frac{1}{2}}$} implies \mbox{$\Im
= {\textstyle \frac{1}{2}}$}.  The proportionality
\mbox{$d{\cal N} \sim dm$} now becomes
\begin{equation} 
d{\cal N} = {\cal N}_0 \, d\Im = 6 {\cal N}_0 \eta (1 \!-\! \eta) \, d\eta,
\end{equation}
where ${\cal N}_0$ equals the total number of the boulders that were
stripped from the sungrazer's surface.  Replacing the normalized distance
of a boulder's position from the sunward end of the prolate-spheroidal
nucleus, $\eta$, with its normalized distance from the nucleus' center
of mass, $\upsilon$, along the radial axis, one has
\begin{equation} 
\upsilon = \eta - {\textstyle \frac{1}{2}}
\end{equation}
and Equation (11) becomes
\begin{equation} 
d{\cal N} = {\textstyle \frac{3}{2}} {\cal N}_0 \! \left( 1 - \upsilon^2
 \right) d\upsilon.
\end{equation}
Each fragment is handled as a dimensionless point, whose normalized radial
deviation from the heliocentric distance of the parent's{\vspace{-0.05cm}}
center of mass, $r_{\rm frg}$, is \mbox{$\upsilon = U_{\rm frg}/{\cal D}$},
where \mbox{$-{\textstyle \frac{1}{2}} \leq \upsilon \leq +{\textstyle
\frac{1}{2}}$}.  Since from Equation~(2)
\begin{equation} 
dU_{\rm frg} = {\textstyle \frac{1}{3}} r_{\rm frg}^2 P_{\rm
 frg}^{-\frac{5}{3}} dP_{\rm frg},
\end{equation}
and \mbox{${\cal D} \,d\upsilon = dU_{\rm frg}$}, I find by inserting
from Equations~(2) and (14) into Equation~(13):
\begin{equation} 
\dot{\cal N}(t) = {\textstyle \frac{1}{2}} {\cal N}_0 \chi P_{\rm frg}^
 {-\frac{5}{3}} \!\left[ 1 \!-\! \chi^2 \! \left(\! P_{\rm par}^{-\frac{2}{3}}
 \!-\! P_{\rm frg}^{-\frac{2}{3}} \right)^{\!2} \right]\!,
\end{equation}
where \mbox{$\dot{\cal N}(t) = d{\cal N}/dt = d{\cal N}/dP_{\rm
frg}$}, \mbox{$t = t_\pi \!+\! P_{\rm frg}$}, $t_\pi$ is the time
of perihelion, when the boulders were removed, and $\chi$ (in AU) equals
\begin{equation} 
\chi = \frac{r_{\rm frg}^2}{\cal D}.
\end{equation}

Equation (15) serves to determine the total number of boulders,
${\cal N}_0$, from{\vspace{-0.06cm}}  their observed annual arrival
rate, $\dot{\cal N}(t_{\rm obs})$.  The orbital period of the observed
boulders is \mbox{$P_{\rm frg} = P_{\rm obs} = t_{\rm obs} \!-\!
t_\pi$}, while the sungrazer itself returns at a time
\mbox{$t_{\rm par} = t_\pi \!+\! P_{\rm par}$}.

The annual arrival rate $\dot{\cal N}(t)$ varies with time, depending
on the distance $U_{\rm frg}$ of an element{\vspace{-0.04cm}}
of the surface from which the boulders were stripped
(\mbox{$-{\textstyle \frac{1}{2}}{\cal D} \leq U_{\rm frg}
\leq +{\textstyle \frac{1}{2}}{\cal D}$}).  The earliest ones
arrive at their perihelion well before does the returning sungrazer,
at time \mbox{$t_{\rm min} = t_\pi \!+\! P_{\rm min}$}, where
\begin{equation} 
P_{\rm min} = P_{\rm par} \!\left( \!1 \!+\! \frac{1}{\chi} P_{\rm
 par}^{\frac{2}{3}} \!\right)^{\!\!-\frac{3}{2}} \!\!.
\end{equation}
On the other hand, the last boulders arrive long after
the returning sungrazer, at time \mbox{$t_{\rm max} = t_\pi \!+\!
P_{\rm max}$}, where
\begin{equation} 
P_{\rm max} = P_{\rm par} \!\left( 1 \!-\! \frac{1}{\chi} P_{\rm
 par}^{\frac{2}{3}} \!\right)^{\!\!-\frac{3}{2}}\!\!.
\end{equation}
This equation makes sense only when
\begin{equation} 
P_{\rm par} < \chi^{\frac{3}{2}}.
\end{equation}
If $P_{\rm par}$ does not satisfy this condition, some of
the boulders get into hyperbolic orbits, ending up
as interstellar objects.

The annual arrival rate $\dot{\cal N}(t)$ reaches a maximum at time
\mbox{$t_{\rm peak} = t_\pi \!+\! P_{\rm peak}$}, where
\begin{equation} 
P_{\rm peak} = \frac{27 \sqrt{7}}{49} P_{\rm par} \!\left(\! 1 \!+\!
 \frac{2}{7} \sqrt{1 \!+\! \Phi} \right)^{\!\!-\frac{3}{2}}
\end{equation}
and
\begin{equation} 
\Phi = \frac{45}{4\chi^2} P_{\rm par}^{\frac{4}{3}}. 
\end{equation}
Because \mbox{$\Phi > 0$}, it is always \mbox{$P_{\rm peak} \!<\! P_{\rm
par}$} and $\dot{\cal N}$ reaches its peak before the parent sungrazer
{\vspace{-0.06cm}}returns to perihelion.  The peak rate \mbox{$\dot{\cal
N}_{\rm peak} = \dot{\cal N}(t_{\rm peak})$} equals
\begin{eqnarray} 
\dot{\cal N}_{\rm peak} & = & \frac{1}{2} \!\left(\!\frac{\sqrt{7}}{3}
 \right)^{\:\!\!\!\!5} \!{\cal N}_0 \chi P_{\rm par}^{-\frac{5}{3}} \!\!\left(
 \!1\!+\! \frac{2}{7} \sqrt{1\!+\! \Phi} \right)^{\!\!\frac{5}{2}} \nonumber \\
 & & \times \!\left[ 1 \!-\! \frac{5}{9 \Phi} \!\left( \!\sqrt{1 \!+\! \Phi}
 \!-\! 1 \!\right)^{\!2} \right] \!.
\end{eqnarray}

This hypothetical, fragmentation-free model predicts the rate of perihelion
arrival of boulders as a function of time via Equation~(15), in terms of
the orbital period, a measure of the location of their removal from
the surface of the parent sungrazer.  Since no consideration has been
given to the possibility that any of the initially stripped chunks could
further fragment along the orbit, their total number remains constant.

\begin{figure*} 
\vspace{0.19cm}
\hspace{-0.18cm}
\centerline{
\scalebox{0.68}{
\includegraphics{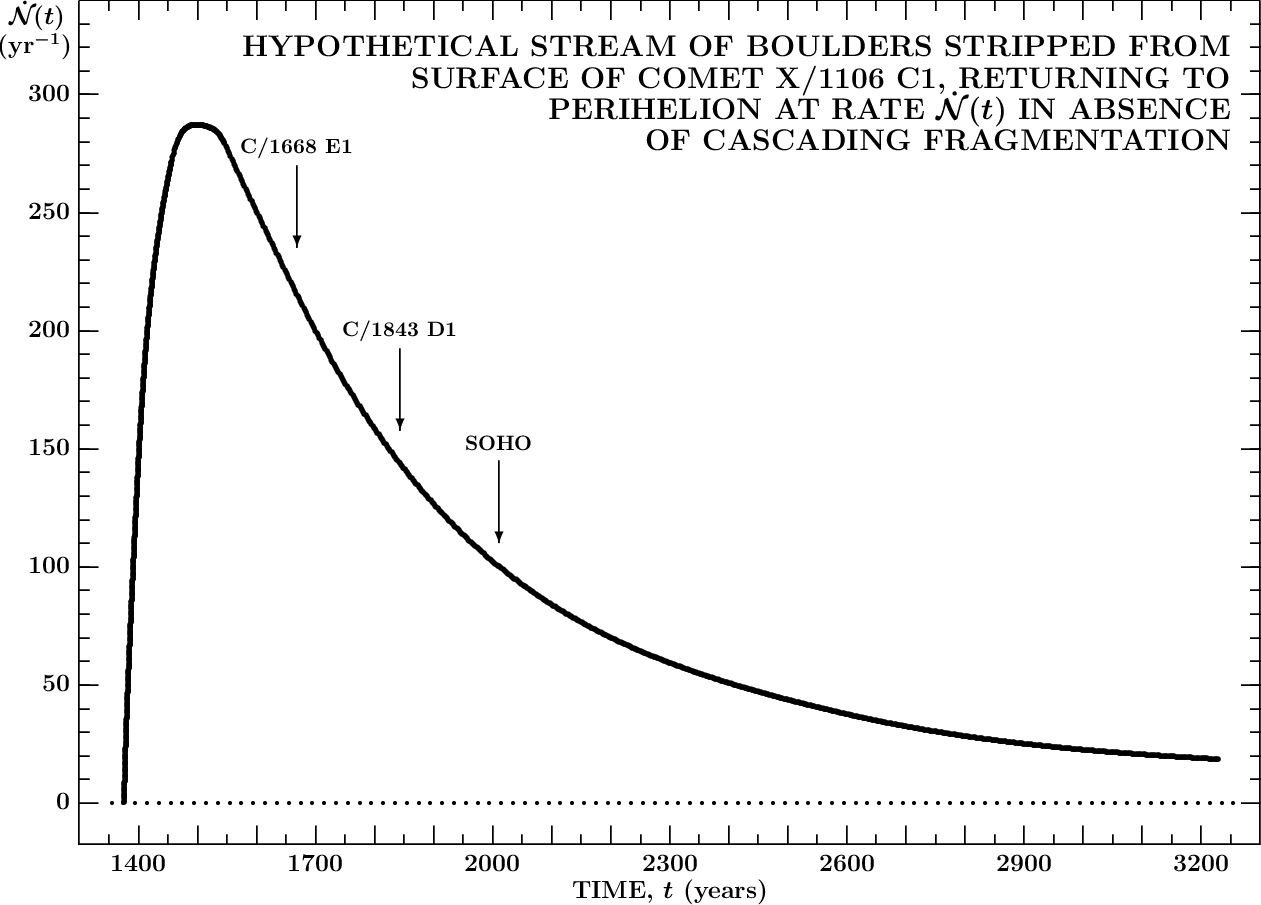}}}
\vspace{-0.1cm}
\caption{Curve of arrival rates of Population~I boulders in a hypothetical
 case of absence of cascading fragmentation.  The boulders are assumed to
 have directly been stripped from the surface of a 50-km long nucleus of
 comet X/1106~C1, ending up in independent orbits determined by their
 location and surviving their motion through the Sun's corona to return
 back to perihelion.  Note the sharp peak in the arrival rate near the
 year 1500.  An arrival rate of 100~yr$^{-1}$ in 2010
 requires that a total of 200,000 boulders be removed.  The time of the
 SOHO operations as well as the perihelion times of C/1843~D1 and
 C/1668~E1 are shown for reference.{\vspace{0.6cm}}}
\end{figure*}

For the sake of curiosity, the curve of the arrival rate of Population~I
boulders is in this unrealistic scenario displayed in Figure~3.  The
assumptions are that the boulders were stripped from the parent comet
X/1106~C1~at perihelion,~\mbox{$r_{\rm frg} = q$}; that the length of
its{\vspace{-0.03cm}} nucleus~along the~Sun-comet line{\vspace{-0.08cm}}
was \mbox{${\cal D} = 50$ km = 0.334$\times 10^{-6}$\,AU}; and that an
observed arrival rate was \mbox{$\dot{\cal N}\!\!\:(t_{\rm obs}) =
100$ yr$^{-1}$} at \mbox{$t_{\rm obs} = 2010$}, when \mbox{$P_{\rm frg} =
904$ yr}.  The parameters of the parent sungrazer, which returned to
perihelion as the Great March Comet of 1843, are taken from Table~3.

The results of this numerical exercise are most interesting.  The
earliest returning boulders, stripped from the sunward end of the
nuclear surface, would never get beyond 84~AU from the Sun and
would have reached perihelion in 1376 (period \mbox{$P_{\rm min}
= 270$ yr}); the last ones, from the antisunward end, would not
return for nearly 80,000~yr (\mbox{$P_{\rm max} = 77,100$ yr}),
moving in orbits reaching 3600~AU from the Sun at aphelion!  This
enormous range of arrival times is a result of the large assumed
dimensions of the comet's nucleus along the Sun-comet line at the
time of boulder removal.  If the length of the nucleus made an angle
of 45$^\circ$ with the Sun-comet line, the periods would be restricted
to a much tighter range from 342~yr to 3860~yr.

Because the arrival-rate curve has an extended tail toward the long
periods, the stream would effectively be limited to just a few thousand
years anyway; the arrival rates at later times would be extremely low.
The curve in Figure~3 displays a sharp peak near the year 1500, almost
400~yr after perihelion, when the arrival rate would reach nearly three
times the rate in 2010, followed by a gradual drop.  In the year 3100
the rate would be only 20~percent of the one in 2010.  The total number
of boulders stripped from the surface of X/1106~C1 would amount to
almost exactly 200,000.

\subsection{Seeds for Cascading Fragmentation} 
A straightforward manner to incorporate the process of cascading
fragmentation into a model for the stream of the SOHO Kreutz
sungrazers is to begin with a seed object that breaks up into two
equal fragments of the first generation, each of which
subsequently breaks up into two equal fragments of the second
generation, etc.  If the mass of the seed object is ${\cal M}_0$,
the mass of each fragment of the first generation is \mbox{${\cal
M}_1 = \frac{1}{2} {\cal M}_0$}, etc., and the mass of each
fragment of an arbitrary $k$-th generation, is
\begin{equation}
{\cal M}_k = \left( {\textstyle \frac{1}{2}} \right)^{\!k}
 \!{\cal M}_0. \\[0.6cm]
\end{equation}
Based on evidence offered by observations of the massive Kreutz sungrazers,
discussed in Section~2, a more~realistic scenario emerges that tidal
splitting~of~a~single~parent in close proximity of perihelion gives
rise in general to a multitude of seed objects.  The implication~is
that instead of one source, cascading fragmentation should begin
from several discrete sources that rapidly spread~over time.
However, regardless~of~the~number of seeds,~the~age of a SOHO-imaged
fragment seen to disintegrate suddenly upon return to perihelion
is known with high accuracy, once the identity of
the parent and the time of its tidal fragmentation,
essentially equal to the perihelion time, are available.

\subsection{Breakup by Rotation Instability:\ Spin-up and\\Separation
 Velocity} 
Strong observational evidence suggests that, subsequent to tidal splitting
of the parent sungrazer, the seed objects and their fragments continue to
fracture throughout the orbit.  Although the disruption mechanism is not known
with certainty and a variety of potential causes for splitting of comets has
over the years been proposed (e.g., Whipple \& Stefanik 1966; Boehnhardt
2002), rotational instability (breakup by runaway spin-up) has recently
become the primary suspect as a result of torques exerted by anisotropic
outgassing (e.g., Jewitt 2021, 2022; Ye et al.\ 2021).

To get insight into a spin-up scenario, I approximate an irregularly-shaped
object by two equal hemispheres attached to one another, each half
having its own center of mass.  If the radius equals~$\Re_{\rm frg}$,
the distance between the two centers of mass is $2\alpha \Re_{\rm frg}$,
where $\alpha \simeq 0.35$.  Let the hemispheres rotate with a period of
$P_{\rm rot}$.  The rotation velocity of the center of mass of each is
$2\pi \alpha \Re_{\rm frg}/P_{\rm rot}$ and the outward acceleration due
to{\vspace{-0.01cm}} the centrifugal force amounts to $4 \pi^2 \alpha
\Re_{\rm frg}/P_{\rm rot}^2$.  If $\rho_{\rm frg}$ is the{\vspace{-0.06cm}}
bulk density, the mass of each{\vspace{-0.06cm}} hemisphere comes to
\mbox{$\frac{2}{3} \pi \rho_{\rm frg} \Re_{\rm frg}^3$} and the
hemispheres are pulled together gravitationally by a force equal
to $\frac{1}{9}\pi^2 G \rho_{\rm frg}^2 \Re_{\rm frg}^4/\alpha^2$,
where $G$ is the universal{\vspace{-0.05cm} constant of gravitation.
The force generated by the rotation amounts to{\vspace{-0.055cm}}
\mbox{$\frac{16}{3} \pi^3 \alpha \rho_{\rm frg} \Re_{\rm frg}^4/P_{\rm
rot}^2$} and acts in the opposite direction, aiming to separate the
hemispheres from one another along their joint surface.  I let the
hemispheres spin up, until their tensile strength, $\sigma_{\rm T}$,
is overcome.  The pull per unit{\vspace{-0.02cm}} surface area,
$\pi \Re_{\rm frg}^2$, that exceeds the limit is
\begin{equation}
{\textstyle \frac{16}{3}} \pi^2 \alpha \rho_{\rm frg}
 \frac{\Re_{\rm frg}^2}{P_{\rm rot}^2} - {\textstyle
 \frac{1}{9}} \pi G \rho_{\rm frg}^2 \frac{\Re_{\rm frg}^2}{\alpha^2}
 > \sigma_{\rm T}.
\end{equation}
The condition for $P_{\rm rot}$ becomes
\begin{equation}
P_{\rm rot} < 4 \pi \Re_{\rm frg} \sqrt{\frac{\alpha \rho_{\rm frg}}{3
	\sigma_{\rm T}}} \frac{1}{\sqrt{1 \!+\! \kappa}},
\end{equation}
where
\begin{equation}
\kappa = {\textstyle \frac{1}{9}} \pi G \rho_{\rm frg}^2 \,
 \frac{\Re_{\rm frg}^2}{\alpha^2 \sigma_{\rm T}}.
\end{equation}
When the hemispheres break apart, the separation velocity of each
center of mass, $v_{\rm sep}$, amounts to
\begin{equation}
v_{\rm sep} = \frac{2 \pi \alpha \Re_{\rm frg}}{P_{\rm rot}} >
 \sqrt{\frac{3 \alpha \sigma_{\rm T}}{4 \rho_{\rm frg}}} \sqrt{1
 \!+\! \kappa}.
\end{equation}
Should \mbox{$\kappa \ll 1$}, a solution that satisfies the condition
of hemispheric separation implies $v_{\rm sep}$ whose lower limit
depends on the fragment's tensile strength and bulk density but not
on its size.  Of course, it is the lower limit of $v_{\rm sep}$ that
counts because that is when the separation begins. For a bulk density
of \mbox{$\rho_{\rm frg} = 0.5$ g cm$^{-3}$}, $\kappa$ equals
\begin{equation}
\kappa = 0.000048 \frac{\Re_{\rm frg}^2}{\sigma_{\rm T}},
\end{equation}
where $\Re_{\rm frg}$ is in meters and $\sigma_{\rm T}$ in pascals.

It is possible to investigate the maximum possible size $\Re_{\rm frg}$
for which $v_{\rm sep}$ stays within required limits of its value for
\mbox{$\kappa = 0$}.  The maximum size refers to the seed objects, with
which cascading fragmentation begins, so this exercise inquires about
how large the seed objects would have to be to satisfy the condition
(27) within any particular limit.  As an example, suppose I do not want
$v_{\rm sep}$ to vary by more than 10~percent of the magnitude given
by \mbox{$\kappa = 0$}.  In this case the radius of the seed object,
$\Re_{\rm seed}$, would have to be such that \mbox{$\sqrt{1 \!+\!
\kappa_{\rm seed}} = 1.1$} and therefore \mbox{$\kappa_{\rm seed} =
0.21$}.  From Equation~(26) this radius is equal to
\begin{equation}
\Re_{\rm seed} = \frac{3 \alpha}{\rho_{\rm frg}} \sqrt{\frac{\kappa_{\rm
 seed} \, \sigma_{\rm T}}{\pi G}}.
\end{equation}

Now, at spin periods{\vspace{-0.06cm}} exceeding $3.3 \,\rho_{\rm
frg}^{-\frac{1}{2}}$~hr solid bodies do not break rotationally even
if their tensile strength is nil (e.g., Jewitt 1997).\footnote{Note
that by equating the left side of Equation~(24) with~zero (rather
than making it larger than $\sigma_{\rm T}$), one{\vspace{-0.09cm}}
would readily obtain \mbox{$P_{\rm rot} = (48 \pi \alpha^3\!/G)^{1/2}
\rho_{\rm frg}^{\raisebox{-0.15cm}{\tiny {-1/2}}}$}, which is a relevant
{\vspace{-0.08cm}}condition for $P_{\rm rot}$.  To equate the constant
with 3.3~hr, one would need \mbox{$\alpha = 0.397$}, the small difference
from the value used being the product of the applied formalism.}  At a
bulk density of 0.5~g~cm$^{-3}$ this critical spin period equals
\mbox{$P_{\rm crit} = 4.67$ hr}.  In the search for the maximum dimensions
of the seeds that would satisfy the condition (27) (with \mbox{$\kappa =
\kappa_{\rm seed}$}), one obviously should require that at the same
time their radius be smaller than the critical value $\Re_{\rm crit}$,
given by a condition\footnote{Note that while the spin-period condition
is independent of the body's size, the parallel introduction into the
consideration of the constraint (27) makes it size-dependent.}
\begin{equation}
v_{\rm sep} = \frac{2 \pi \alpha \Re_{\rm crit}}{P_{\rm crit}},
\end{equation}
that is,
\begin{equation}
\Re_{\rm crit} = \frac{P_{\rm crit}}{4 \pi} \sqrt{\frac{3 \sigma_{\rm
 T}}{\alpha \rho_{\rm frg}}} \sqrt{1 \!+\! \kappa_{\rm peak}}.
\end{equation}
For the ratio I find, again with \mbox{$\rho_{\rm frg} \!=\! 0.5$ g cm$^{-3}$}
\begin{equation}
\frac{\Re_{\rm seed}}{\Re_{\rm crit}} = \frac{4}{P_{\rm crit}}
 \sqrt{\frac{3 \pi \alpha^3 \kappa_{\rm seed}}{G \rho_{\rm frg}
 (1 \!+\! \kappa_{\rm seed})}} = 0.345.
\end{equation}
\begin{table}[t] 
\vspace{0.18cm}
\hspace{-0.15cm}
\centerline{
\scalebox{1}{
\includegraphics{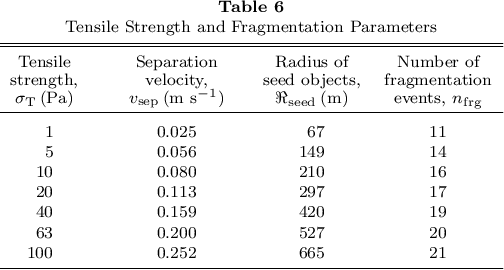}}}
\vspace{0.55cm}
\end{table}

The size of the seed bodies under consideration is thus clearly not
in conflict with the implied constraint.  Even though this does not
by itself determine what the actual size of the seeds is, the result
(32) is encouraging in that Equation~(27) with a reasonably low value
of $\kappa_{\rm seed}$, that is, with an approximately constant separation
velocity, is a plausible condition that could be applied when
incorporating the process of cascading fragmentation into the model.

The next issue concerns the magnitude of the tensile strength of the
Kreutz sungrazers.  I rely here on a few independent sources.  Attree
et al.\ (2018) estimated that a {\it minimum\/} tensile strength needed
to support overhanging cliffs on the surface of comet
67P/Churyumov-Gerasimenko was extremely low, on the order of 1~Pa.
Tensile strengths between 1 and 150~Pa were previously proposed from
similar investigations by Groussin et al.\ (2015) and by Vincent et
al.\ (2017).  On the other hand, from his research of tidal splitting
and rotational breakup Davidsson (2001) concluded that a strength of
$\!$\lapeq$\!$100~Pa was consistent with the data.

Table 6 presents the separation velocity and a possible radius of
the seeds, computed conservatively as a function of the tensile
strength.  The entries tabulated near the bottom appear to
be more realistic in terms of both $v_{\rm sep}$ and $\Re_{\rm
seed}$.  The (nearly) constant separation velocity implies
a gradual spin-up on account of a systematically decreasing size of
the boulders with every fragmentation event.  Also, because of the
action of torques, the boulders must tumble, so that the direction of the
separation velocity is subject to chaotic behavior from one fracture
to the next.  The number of breakup events, which each fragment of each
seed undergoes before reaching the estimated 5-meter radius of the
smallest SOHO Kreutz sungrazers, is tabulated as $n_{\rm frg}$ in the
last column.  I return to issues involving the seeds in the following
sections.

An important implication is that the seeds needed for the formation
of the stream of SOHO Kreutz comets, are not to be confused with
the primary products of the parent sungrazer's tidal splitting at
perihelion.  The seeds --- apparently highly-active, irregularly-shaped,
subkilometer-sized fragments --- represent only a minor fraction of the
mass lost in this process.  This conclusion is not surprising because
it is supported very obviously by the observed return to perihelion
of one or more of the parent's main surviving fragment(s), the
Great March Comet of 1843 and possibly the Great Comet of 1668 as remains
of the Great Comet of 1106 in the case of Population~I.  Likewise
corroborating is an apparent consensus that the mass of the SOHO stream
is orders of magnitude smaller than the mass of a large sungrazer
(e.g., Sekanina 2003;{\nopagebreak} Knight et al.\ 2010).

\begin{figure*}[t] 
\vspace{0.18cm}
\hspace{-0.15cm}
\centerline{
\scalebox{0.8395}{
\includegraphics{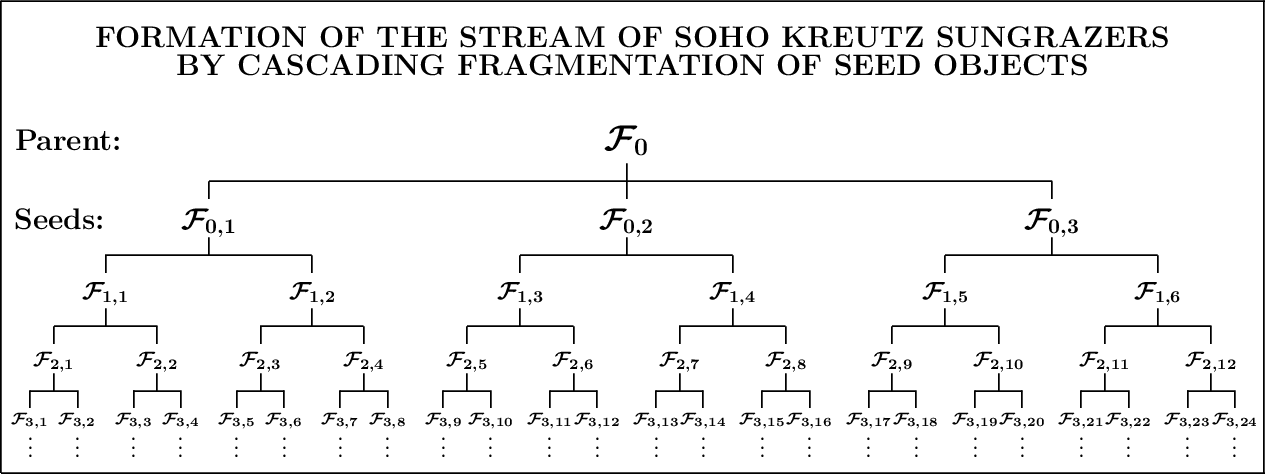}}}
\vspace{0cm}
\caption{Schematic picture of the formation of the stream of SOHO
Kreutz sungrazers by the process of cascading fragmentation of the
seed objects, which effectively are fragments of generation zero.
The depicted scenario shows the parent comet, ${\cal F}_0$; three
seeds --- ${\cal F}_{0,1}$, ${\cal F}_{0,2}$, and ${\cal F}_{0,3}$;
six fragments of the first generation; 12~fragments of the second
generation; and 24~fragments of the third generation.{\vspace{0.6cm}}}
\end{figure*}

\section{Fragmentation Events' Spacing Along Orbit}
I argued in Section 3.2 that the seeds that give rise to the stream of
SOHO sungrazers begin their existence at the time of tidal breakup of
the parent comet.  I designate the time of their birth as $t_0$; it
may or may not coincide with the comet's perihelion time $t_\pi$,
but the two times are never far apart.  In line with the narrative of
Section~3.2, the seeds are classified as fragments of generation zero;
their number is $n_{\rm seed}$; they are referred to in the following by
the symbols ${\cal F}_{0,1}$, ${\cal F}_{0,2}$, \ldots, ${\cal F}_{0,n_{\rm
seed}}$; and their masses amount to ${\cal M}_{0,1}$, ${\cal M}_{0,2}$,
\ldots, ${\cal M}_{0,n_{\rm seed}}$.

As stipulated in Section 3.2, the fragments of the first generation are
born by a breakup of each seed into two halves, so that their total number
equals $2n_{\rm seed}$.  A seed ${\cal F}_{0,k}$ fragments into ${\cal
F}_{1,2k-1}$ and ${\cal F}_{1,2k}$, each then acquiring an extra momentum;
if the seed's orbital velocity vector at the time is $\dot{\bf r}$ and
its components in the RTN coordinate system\footnote{The RTN right-handed
ortogonal coordinate system is referred to the center and orbital plane
of the body and rotates with it about the Sun.  The R coordinate is in
the direction away from the Sun, the N axis points to the north orbital
pole, and the T coordinate is perpendicular to R in the plane.}
$|\dot{\bf r}|_{\rm R}$, $|\dot{\bf r}|_{\rm T}$, and $|\dot{\bf r}|_{\rm N}$,
one fragment is released with an orbital{\vspace{-0.02cm}} velocity
whose components are \mbox{$|\dot{\bf r}|_{\rm R} \!+\! v_{\rm R}$},
\mbox{$|\dot{\bf r}|_{\rm T} \!+\! v_{\rm T}$}, and \mbox{$|\dot{\bf
r}|_{\rm N} \!+\! v_{\rm N}$}, the other with a velocity whose components
are \mbox{$|\dot{\bf r}|_{\rm R} \!-\! v_{\rm R}$}, \mbox{$|\dot{\bf r}|_{\rm
T} \!-\! v_{\rm T}$}, and \mbox{$|\dot{\bf r}|_{\rm N} \!-\! v_{\rm N}$}.
Because the seed is expected to tumble, $v_{\rm R}$, $v_{\rm T}$, and
$v_{\rm N}$ are modeled by means of a random number generator, yet they
satisfy{\vspace{-0.055cm}} the condition \mbox{$v_{\rm R}^2 \!+\! v_{\rm T}^2
\!+\! v_{\rm N}^2 = v_{\rm sep}^2$} (Section~3.3).  The same rules apply
to fragments of the higher generations, except that the number of
fragments of a $k$-th generation is $2^k n_{\rm seed}$ and its $m$-th
fragment, ${\cal F}_{k,m}$, breaks up into fragments ${\cal F}_{k+1,2m-1}$
and ${\cal F}_{k+1,2m}$ of the $(k\!+\!1)$-st generation (Figure~4).

The number and orbital positions of fragments depend strongly on the
number of the seeds and on the locations in the orbit at which the
seeds and fragments of the previous generations fractured.  The spacing
of the fragmentation events is thus another important factor that
governs the structure of the SOHO sungrazers' stream.

I had previously addressed this issue (Sekanina 2002)~in connection with the
progressive fragmentation~of~the ill-fated comet D/1993~F2 (Shoemaker-Levy)
before it impacted Jupiter in July 1994.  The comet split tidally at the time
of its close encounter with the planet in 1992 and continued to fragment
nontidally along its final pre-impact jovicentric orbit.  Investigation of
this fragmentation sequence (Sekanina et al.~1998) showed that the rate of
breakup episodes in the nontidal phase was subsiding with time, possibly a
fundamental property of the process.  If so, a similar slowdown should be
expected for the Kreutz sungrazers.  Accordingly, I subscribe to the
fragmentation-chain law tested on comet Shoemaker-Levy and assume that a
progression of breakup events started with the birth of the seeds at time
$t_0$ and continued in a fashion, illustrated on a particular branch of
fragments in the following:

Let a seed ${\cal F}_{0,k}$ split into the first-generation fragments
${\cal F}_{1,2k-1}$ and ${\cal F}_{1,2k}$ at time \mbox{$t_1 = t_0 \!+\!
\Delta t_0$}.  Next,~let \mbox{$m = 2k \!-\! 1$} or \mbox{$m = 2k$} and
${\cal F}_{1,m}$ split into ${\cal F}_{2,2m-1}$~and ${\cal F}_{2,2m}$
at time \mbox{$t_2 = t_1 \!+\! \Delta t_1$}, where \mbox{$\Delta t_1 =
\Lambda \, \Delta t_0$}~and~the parameter \mbox{$\Lambda > 1$} describes
the slowdown rate.  Similarly, let all fragments of the second generation
split~into~frag\-ments of the third generation at time \mbox{$t_3 = t_2 \!+\!
\Delta t_2$},~where \mbox{$\Delta t_2 = \Lambda \, \Delta t_1 =
\Lambda^{\!2} \Delta t_0$},~etc.~Generally, fragments~of~a~\mbox{$k$-th}
generation are produced by fragments~of~a~\mbox{$(k \!-\! 1)$-st}
generation at a time
\begin{equation}
t_k = t_{k-1} + \Lambda^{\!k-1} \Delta t_0.
\end{equation}
The rate of increase in the lengths of intervals $\Delta t_k$ follows
a geometric progression, so that the fragmentation times $t_k$ make up
a sequence
\begin{equation}
t_k = t_0 + \frac{\Lambda^{\!k} \!-\!1}{\Lambda \!-\! 1} \, \Delta t_0.
\end{equation}
In the limit, using L'Hospital's rule,
\begin{equation}
\lim_{\Lambda \rightarrow 1} (t_k \!-\! t_0) = \lim_{\Lambda \rightarrow
 1} (k \Lambda^{k-1} \Delta t_0) = k \, \Delta t_0, \\[0.02cm]
\end{equation}
as expected.

Equation (34) becomes an important boundary condition when written for
the arrival time of a seed or a fragment to perihelion, \mbox{$t_k = t_{\rm
ref}$}.  Then one can write \mbox{$t_{\rm ref} \!-\! t_0 = P_{\rm frg}$},
the orbital period, the exponent $k$ in Equation~(34) becoming
the number of fragmentation events $n_{\rm frg}$ that the seed and its
fragments undergo in order that an initial diameter $\Re_{\rm seed}$
is reduced to a diameter $\Re_{\rm min}$ of the smallest dwarf sungrazers
detectable by the SOHO coronagraphs, adopted in Section~3.3 to amount to
10~meters; $n_{\rm frg}$ also equals the number of fragment generations.
Listed in Table~6, it is derived from Equation~(23) on the assumption
that a fragment's mean dimension varies as a cube root of its mass,
\begin{equation}
n_{\rm frg} = \frac{\log {\cal N}_0}{\log 2} \simeq 10 \,
 \log \! \left( \! \frac{\Re_{\rm seed}}{\Re_{\rm min}} \! \right) \!,
\end{equation}
where
\begin{equation}
{\cal N}_0 = \!\left( \! \frac{\Re_{\rm seed}}{\Re_{\rm min}} \!
 \right)^{\!\!3}
\end{equation}
is the total number of SOHO-like sungrazers collected from a seed.~For
example, for \mbox{$\Re_{\rm seed} \simeq 40\;{\rm to}\;100 \; \Re_{\rm
min}$} one finds \mbox{$n_{\rm frg} = 16\;{\rm to}\;20$} and \mbox{${\cal
N}_0 \simeq 6.5 \!\times\!10^4\;{\rm to}\;10^6$}.

With $t_{\rm ref}$ replacing $t_k$, Equation~(34) reads
\begin{equation}
P_{\rm frg} = \frac{\Lambda^{n_{\rm frg}} \!-\! 1}{\Lambda \!-\! 1} \,
 \Delta t_0,
\end{equation}
linking four parameters.  But \mbox{$P_{\rm frg} \simeq 900$ yr} for
the sungrazer stream of Population~I in the early 21-st century and the
number of fragment generations $n_{\rm frg}$ cannot deviate much from 16
or 20, as the seeds cannot be much smaller or larger than adopted
above.  Equation~(38) then becomes a relationship between $\Lambda$
and $\Delta t_0$.  And for any such pair Equation~(34) allows one
to compute the fragmentation times for each of the $n_{\rm frg}$
generations of fragments, as exhibited in Table~7 ($\Delta t_0$
equaling by definition the interval of time between the births of
the seed and the fragments of the first generation) with $P_{\rm
frg}$ and $n_{\rm frg}$ held constant.

The reader should be aware of an ambiguity regarding the numbering
of the fragmentation cycles, depending on their perception.  In
particular, the simulation process is terminated at the time of
{\it birth\/} of the pair of fragments of the $n_{\rm frg}$-th
generation, which is the same as the time of {\it breakup\/} of
their parent, a fragment of the $(n_{\rm frg} \!-\! 1)$-st
generation.

\begin{table}[t] 
\vspace{0.15cm}
\hspace{-0.18cm}
\centerline{
\scalebox{1}{
\includegraphics{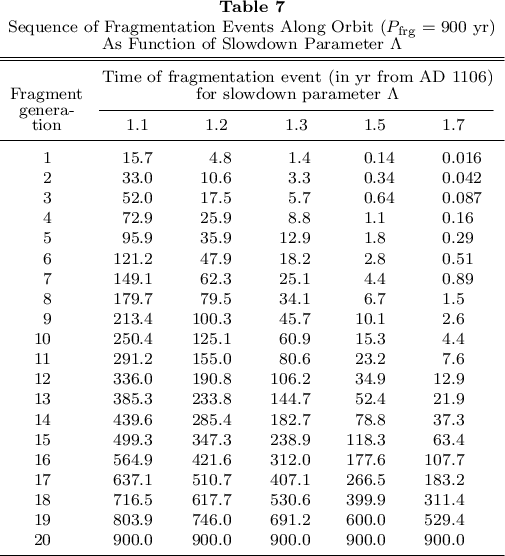}}}
\vspace{0.6cm}
\end{table}

Table 7 shows that an increase from 1.1 to 1.7 in the parameter $\Lambda$
results in a drop by three orders of magnitude in the parameter $\Delta
t_0$.  The higher the value of $\Lambda$, the more crowded the
fragmentation events are in the early post-perihelion period of time.
The major significance of this crowding for the eventual distribution of
SOHO-like sungrazers is apparent from Section~5.

Table 7 also exhibits the increasing asymmetry between the numbers
of pre- and post-aphelion fragmentation episodes as $\Lambda$
increases.  For \mbox{$\Lambda = 1.1$} the first 14 events
occur before aphelion, whereas for \mbox{$\Lambda = 1.5$} to 1.7 the
number increases to 18.~The condition for the number of
pre-aphelion fragmentation events, $n_{\rm pre}$,~is~given~by
\begin{equation}
n_{\rm pre} = {\rm trunc}\!\left\{ \!\frac{\log \!\left[ 1 \!+\! {\textstyle
 \frac{1}{2}} (\Lambda \!-\! 1) P_{\rm frg}/\Delta t_0 \right]}{\log \Lambda}
 \!\right\}. \\[0.01cm]
\end{equation}

It is straightforward to set the condition for $\Lambda$ to signal the
absence of fragmentation events along the post-aphelion leg of the
orbit.  Writing Equation~(38) for the pre-aphelion half of the orbit as
\begin{equation}
{\textstyle \frac{1}{2}} P_{\rm frg} \geq \frac{\Lambda_{\rm pre}^{\!n_{\rm
 frg}-1} \!-\! 1}{\Lambda_{\rm pre} \!-\! 1} \, \Delta t_0
\end{equation}
and dividing this expression by Equation~(38), one finds a condition
for $\Lambda_{\rm pre}$:
\begin{equation}
\Lambda_{\rm pre} \geq 2 \!-\! \Lambda_{\rm pre}^{\!1-n_{\rm frg}}.
\end{equation}
For large values of $n_{\rm frg}$}, which I consider in this paper,
the post-aphelion half of the orbit is free from fragmentation events
when $\Lambda_{\rm pre}$ is nearly exactly 2 or larger.

In Part~I, I mentioned occasional observations of close pairs of SOHO
sungrazers that provide evidence on relatively recent and definitely
post-aphelion events of fragmentation.  Such scenarios require values
of $\Lambda$ substantially lower than 2.  In general, fragmentation
histories of the seed objects clearly vary widely from case to case.
One of the tasks of the proposed simulation computations is to examine
a range of such variations.

\section{Orbital Perturbations of Fragments}  
In a breakup of a parent fragment, equal orbital momenta of opposite
signs are assumed to be transferred to the two generated subfragments
of equal mass, a basic scenario adopted here to incorporate the effects
of cascading fragmentation into a model for the SOHO stream formation.
The orbital-velocity vector for one of the new fragments is simulated
by adding a separation velocity vector to the orbital-velocity vector
of the parent fragment, while for the other new fragment its
orbital-velocity vector is modeled by subtracting the same separation
velocity vector from the parent's orbital-velocity vector.  The
separation velocity vector is oriented at random, a product of the
fragments' expected tumbling, whereas its magnitude is tentatively
estimated at being close to 0.2~m~s$^{-1}$ (Table~6).

The components of the separation velocity affect the fragments' orbital
elements, which differ, usually only slightly, from the orbital elements
of the parent.  In a sense, one can say that the products of a sequence
of fragmentation events move in {\it perturbed\/} orbits.  The degree
of orbital similarity is measured by a {\it dispersion\/} in each orbital
element, a method extensively used in Part~I of this investigation to
describe the observed structure of the stream of SOHO sungrazers.  To
understand the relationships among the dispersions of the orbital
elements, it is desirable to examine the magnitudes of these separation
velocity driven orbital {\it perturbations\/} as a function of not only
the separation velocity vectors but the orbital positions of the parent
at the fragmentation times as well.  This examination requires the
computation of the perturbed orbital motion of each simulated fragment,
a topic that I am addressing next.

\subsection{Computing Elements of a Fragment's Orbit} 
Let a Kreutz sungrazer move about the Sun unaffected by the planetary
perturbations and let its orbit be determined by the argument of perihelion
$\omega$, the longitude of the ascending node $\Omega$, the inclination
$i$, the perihelion distance $q$, and the orbital period $P$.  Let the
comet pass through perihelion at time $t_\pi^0$ and{\vspace{-0.03cm}}
break up at time $t_{\rm frg}$, where \mbox{$t_\pi^0 \leq t_{\rm frg}
< t_\pi$} and $t_\pi$ is the time of{\vspace{-0.05cm}} return to perihelion,
\mbox{$t_\pi = t_\pi^0 + P$}.  Let the comet's{\vspace{-0.01cm}}
heliocentric distance and true anomaly at $t_{\rm frg}$ be, respectively,
\mbox{$r(t_{\rm frg}) = r_{\rm frg}$} and \mbox{$u(t_{\rm frg}) = u_{\rm
frg}$}.  An outcome~of~the breakup event is the birth of a fragment that
separates at a rate of $v_{\rm sep}$ relative to the main, parent mass.

An RTN right-handed orthogonal coordinate system, whose origin is in the
parent's center of mass, the R axis pointing radially away from the Sun, the
T axis in the orbital plane and perpendicular to R, and the N axis normal to
this plane, has already been introduced in Section~4.  Let the components of
the fragment's separation velocity vector {\boldmath $v_{\bf sep}$} in the
cardinal directions of the RTN system be $v_{\rm R}$, $v_{\rm T}$, and
$v_{\rm N}$.  This velocity vector prompts the fragment to end up in a new
orbit, described by the asterisk-labeled elements
\begin{eqnarray}
\omega^{\displaystyle \ast}\! & = & \omega + \Delta \omega(t_{\rm frg},
 \mbox{\boldmath $v_{\bf sep}$}), \nonumber \\
\Omega^{\displaystyle \ast}\! & = & \Omega + \Delta \Omega(t_{\rm frg},
 \mbox{\boldmath $v_{\bf sep}$}), \nonumber \\
i^{\displaystyle \ast}\! & = & i + \Delta i(t_{\rm frg}, \mbox{\boldmath
 $v_{\bf sep}$}), \nonumber \\[-0.02cm]
q^{\displaystyle \ast}\! & = & q + \Delta q(t_{\rm frg}, \mbox{\boldmath
 $v_{\bf sep}$}), \nonumber \\[0.02cm]
P^{\displaystyle \ast}\! & = & P + \Delta P(t_{\rm frg}, \mbox{\boldmath
 $v_{\bf sep}$}), \nonumber \\[-0.02cm]
t_\pi^{\displaystyle \ast} & = & \, t_\pi + \Delta t_\pi(t_{\rm frg},
 \mbox{\boldmath $v_{\bf sep}$}).
\end{eqnarray}

The aim of this exercise is to derive $\Delta \omega$, \ldots, $\Delta t_\pi$
as a function of the time, $t_{\rm frg}$, of the fragmentation event (which
determines the parent comet's orbital state vectors) and the fragment's
separation velocity vector, {\boldmath $v_{\bf sep}$}.\,\,

To begin, I assume that at the fragmentation time the
parent comet's position vector and orbital-velocity vector in the
ecliptic coordinate system equal {\boldmath $S_{\bf frg}$} and
{\boldmath $V_{\bf frg}$}, respectively:
\begin{equation}
{\hspace{-0.2cm}}\mbox{\boldmath $S_{\bf frg}$} = (x_{\rm frg}, y_{\rm frg},
 z_{\rm frg}), \;\;
 \mbox{\boldmath $V_{\bf frg}$} = (\dot{x}_{\rm frg}, \dot{y}_{\rm frg},
 \dot{z}_{\rm frg}).
\end{equation}
They are given by the expressions:
\begin{equation}
\left( \!\!
\begin{array}{c}
x_{\rm frg} \\
y_{\rm frg} \\
z_{\rm frg}
\end{array}
\!\! \right) \!= r_{\rm frg} \!\left( \!\!
\begin{array}{cc}
P_x & Q_x \\
P_y & Q_y \\
P_z & Q_z
\end{array}
\!\! \right) \!\! \times \!\!\left( \!\!
\begin{array}{c}
\cos u_{\rm frg} \\
\sin u_{\rm frg}
\end{array}
\!\! \right),
\end{equation}
and
\begin{equation}
\left( \!\!
\begin{array}{c}
\dot{x}_{\rm frg} \\
\dot{y}_{\rm frg} \\
\dot{z}_{\rm frg}
\end{array}
\!\! \right) \!=\! \frac{k_0}{\sqrt{p}} \!\left( \!\!
\begin{array}{cc}
P_x & Q_x \\
P_y & Q_y \\
P_z & Q_z
\end{array}
\!\! \right) \!\!\times \!\!\left( \!\!
\begin{array}{c}
-\sin u_{\rm frg} \\
e \!+\! \cos u_{\rm frg}
\end{array}
\!\! \right),
\end{equation}
where $p$ is the orbit parameter in AU and $k_0$ the{\vspace{-0.05cm}}
Gaussian gravitational constant in \mbox{AU$^\frac{3}{2}$ day$^{-1}$}.  The
directional cosines $P_x$, \ldots, $Q_z$, are, together with $R_x$, $R_y$,
and $R_z$ (used below), the ecliptic components of the unit vectors {\boldmath
$P$}, {\boldmath $Q$}, and {\boldmath $R$} in the orthogonal coordinate
system tied to the orbital plane and the line of apsides:
\begin{eqnarray}
\left( \!
\begin{array}{ccc}
P_x & \,P_y & \,P_z \\
Q_x & \,Q_y & \,Q_z \\
R_x & \,R_y & \,R_z
\end{array}
\! \right) & = & \left( \!\!
\begin{array}{ccc}
\;\;\:\cos \omega & \;\sin \omega & \;0 \\
-\!\sin \omega    & \;\cos \omega & \;0 \\
        0         &         0     & \;1
\end{array}
\right) \nonumber \\[0.15cm]
& \times & \! \left(
\begin{array}{ccc}
1 & \:    0        & \;   0   \\
0 & \;\;\;\:\cos i & \;\sin i \\
0 & \ -\!\sin i    & \;\cos i
\end{array}
\right) \!\!\times \!\! \left( \!\!
\begin{array}{ccc}
\;\;\:\cos \Omega & \;\sin \Omega & \;0 \\
   -\!\sin \Omega & \;\cos \Omega & \;0 \\
       0          & \;     0      & \;1
\end{array}
	\right) \!. \nonumber \\[-0.1cm]
\end{eqnarray}

At the fragmentation time, $t_{\rm frg}$, the position{\vspace{-0.04cm}}
vector of the fragment, {\boldmath $S$}$_{\bf frg}^{\displaystyle \ast}$, is
taken to coincide with the{\vspace{-0.04cm}} parent's position vector, while
the fragment's orbital-velocity vector, {\boldmath $V$}$_{\bf
frg}^{\displaystyle \ast}$, is the sum{\vspace{-0.06cm}} of the parent's
orbital-velocity vector and the fragment's separation velocity vector,
{\boldmath $v_{\bf sep}$}, whose ecliptic components are $v_x$, $v_y$,
and $v_z$,
\begin{equation}
\left( \!\!
\begin{array}{c}
x_{\rm frg}^{\displaystyle \ast} \\
y_{\rm frg}^{\displaystyle \ast} \\
z_{\rm frg}^{\displaystyle \ast}
\end{array}
\!\! \right) \!=\! \left( \!\!
\begin{array}{c}
x_{\rm frg} \\
y_{\rm frg} \\
z_{\rm frg}
\end{array}
\!\! \right) \!, \;\;
\left( \!\!
\begin{array}{c}
\dot{x}_{\rm frg}^{\displaystyle \ast} \\
\dot{y}_{\rm frg}^{\displaystyle \ast} \\
\dot{z}_{\rm frg}^{\displaystyle \ast}
\end{array}
\!\! \right) \!=\! \left( \!\!
\begin{array}{c}
\dot{x}_{\rm frg} \\
\dot{y}_{\rm frg} \\
\dot{z}_{\rm frg}
\end{array}
\!\! \right) \!+\! \left( \!\!
\begin{array}{c}
v_x \\
v_y \\
v_z
\end{array}
\!\! \right)\! .
\end{equation}
The ecliptic components of the separation velocity vector are related to its
components in the RTN coordinate system by
\begin{equation}
\left( \!\!
\begin{array}{c}
v_x \\
v_y \\
v_z
\end{array}
\!\! \right) \!= \! \left( \!\!
\begin{array}{ccc}
P_x & Q_x & R_x \\
P_y & Q_y & R_y \\
P_z & Q_z & R_z
\end{array}
\!\! \right) \!\times \! \left( \!\!
\begin{array}{ccc}
\cos u_{\rm frg} &  \!-\!\sin u_{\rm frg} & 0 \\
\sin u_{\rm frg} & \;\:\,\cos u_{\rm frg} & 0 \\
      0          &         0          & 1
\end{array}
\! \right) \!\times \!\left( \!\!
\begin{array}{c}
v_{\rm R} \\
v_{\rm T} \\
v_{\rm N}
\end{array}
\!\! \right) \!.
\end{equation}
The determination of the fragment's orbital elements can now proceed by
introducing the angular-momentum vector components:
\begin{eqnarray}
\Im_{xy} & = & \left|\,
\begin{array}{cc}
x_{\rm frg} & \; y_{\rm frg} \\[0.1cm]
\dot{x}_{\rm frg}^{\displaystyle \ast} & \; \dot{y}_{\rm frg}^{\displaystyle
 \ast}
\end{array}
\, \right| \!, \nonumber \\[0.2cm]
\Im_{yz} & = & \left| \,
\begin{array}{cc}
y_{\rm frg} & \; z_{\rm frg} \\[0.1cm]
\dot{y}_{\rm frg}^{\displaystyle \ast} & \;\dot{z}_{\rm frg}^{\displaystyle
 \ast}
\end{array}
 \,\right| \!, \nonumber \\[0.2cm]
\Im_{zx} & = & \left|\,
\begin{array}{cc}
z_{\rm frg} & \; x_{\rm frg} \\[0.1cm]
\dot{z}_{\rm frg}^{\displaystyle \ast} & \;\dot{x}_{\rm frg}^{\displaystyle
 \ast}
\end{array}
 \,\right| \! .
\end{eqnarray}
%
%

The following computations allow to incorporate, if deemed desirable, a
nongravitational acceleration into the orbital motion of the fragment on the
assumptions that it points in the antisolar direction and varies inversely
as the square of heliocentric distance, the constraints employed by Hamid
\& Whipple (1953) in their investigation and likewise integrated into the
standard model for the split comets (Sekanina 1982).  Let $\gamma_0$ be
a dimensionless parameter describing the fragment's nongravitational
acceleration in units of 10$^{-5}$\,the Sun's gravitational acceleration.
The fragment's {\it effective\/} Gaussian gravitational constant
$k_0^{\displaystyle \ast}$, which replaces $k_0$ below, equals
\begin{equation}
k_0^{\displaystyle \ast} \!= \sqrt{k_0^2 - 10^{-5} \gamma_0 k_0^2} =
 k_0 \sqrt{1 \!-\! 10^{-5} \gamma_0}.
\end{equation}

The fragment's longitude of the ascending node, $\Omega^{\displaystyle
\ast}\!$, inclination, $i^{\displaystyle \ast}\!$, and orbit parameter,
$p^{\displaystyle \ast}\!$, related to the perihelion distance,
$q^{\displaystyle \ast}\!$, via the orbital eccentricity, $e^{\displaystyle
\ast}\!$, by \mbox{$p^{\displaystyle \ast} = q^{\displaystyle \ast}
(1 \!+\! e^{\displaystyle \ast}\:\!\!)$}, follow from the relations, 
\begin{equation}
\left( \!\!
\begin{array}{c}
\widehat{\Im}_{xy} \\
\widehat{\Im}_{yz} \\
\widehat{\Im}_{zx}
\end{array}
\!\! \right) = k_0^{\displaystyle \ast} \sqrt{p^{\displaystyle
 \ast}} \left( \!\!
\begin{array}{c}
\cos i^{\displaystyle \ast} \\
\sin \Omega^{\displaystyle \ast} \sin i^{\displaystyle \ast} \\
-\cos \Omega^{\displaystyle \ast} \sin i^{\displaystyle \ast}
\end{array}
\!\!\! \right) \!.
\end{equation}
For the longitude of the ascending node one finds
\begin{equation}
\tan \Omega^{\displaystyle \ast} \!=
 -\frac{\widehat{\Im}_{yz}}{\widehat{\Im}_{zx}}, \;\;\;
 {\rm sign}(\sin \Omega^{\displaystyle \ast}\:\!\!) =
 {\rm sign}(\widehat{\Im}_{yz});
\end{equation}
for the inclination
\begin{equation}
\tan i^{\displaystyle \ast} \!= \frac{\sqrt{\widehat{\Im}_{yz}^2 \!+\!
 \widehat{\Im}_{zx}^2}}{\widehat{\Im}_{xy}},
\end{equation}
where the sign of the denominator determines the quadrant of the inclination;
and for the parameter
\begin{equation}
p^{\displaystyle \ast} = \frac{\widehat{\Im}_{xy}^2 \!+\! \widehat{\Im}_{yz}^2
 \!+\! \widehat{\Im}_{zx}^2}{(k_0^{\displaystyle \ast\:\!\!})^2} .
\end{equation}
Next, the fragment's orbital eccentricity is
\begin{equation}
e^{\displaystyle \ast} \! = \sqrt{1 + \! p^{\displaystyle \ast} \! \left\{ \:\!\!
 \frac{1}{(k_0^{\displaystyle \ast\:\!\!})^2} \!\left[ \! \left( \! \dot{x}_{\rm
 frg}^{\displaystyle \ast} \!\right)^{\!2} \!\!+\! \left(\! \dot{y}_{\rm
 frg}^{\displaystyle \ast} \!\right)^{\!2} \!\!+\! \left(\! \dot{z}_{\rm
 frg}^{\displaystyle \ast} \!\right)^{\!2} \right] \!-\!  \frac{2}{r_{\rm frg}}
 \!\right\}}\, ,
\end{equation}
so that the perihelion distance comes out from\\[-0.15cm]
\begin{equation}
q^{\displaystyle \ast} \!= \frac{p^{\displaystyle \ast}}{1 \!+\!
 e^{\displaystyle \ast}}
\end{equation}
and the orbital period from
\begin{equation}
P^{\displaystyle \ast} \!= \frac{2\pi}{k_0^{\displaystyle \ast}}
 \! \left(\!\frac{q^{\displaystyle \ast}}{1 \!-\! e^{\displaystyle
 \ast}} \!\! \right)^{\!\!\frac{3}{2}} \!\!.
\end{equation}
The fragment's true anomaly at the time of separation from the primary
is given by
\begin{equation}
\sin u_{\rm frg}^{\displaystyle \ast} = \frac{\sqrt{p^{\displaystyle
 \ast}}}{k_0^{\displaystyle \ast} e^{\displaystyle \ast}
 r_{\rm frg}} \! \left( x_{\rm frg} \dot{x}_{\rm frg}^{\displaystyle
 \ast} \!+\! y_{\rm frg} \dot{y}_{\rm frg}^{\displaystyle \ast}
 \!+\! z_{\rm frg} \dot{z}_{\rm frg}^{\displaystyle \ast} \right)
\end{equation}
with \mbox{${\rm sign}(\cos u_{\rm frg}^{\displaystyle \ast}) ={\rm
sign}(p^{\displaystyle \ast} \!\!-\! r_{\rm frg})$}.  The{\vspace{-0.02cm}
sum of the argument of perihelion and the true anomaly at separation~are
calculated from
\begin{equation}
\cos (\omega^{\displaystyle \ast} \!\!+\! u_{\rm frg}^{\displaystyle
 \ast}) = \frac{x_{\rm frg} \cos \Omega^{\displaystyle
 \ast} \!+\! y_{\rm frg} \sin \Omega^{\displaystyle \ast}}{r_{\rm frg}}
\end{equation}
with \mbox{${\rm sign}[\sin (\omega^{\displaystyle \ast} \!\!+\! u_{\rm
frg}^{\displaystyle \ast})] = {\rm sign}(z_{\rm frg})$}.  Equations~(58)
{\vspace{-0.06cm}}and (59) isolate the argument of perihelion.  Finally,
to{\vspace{-0.03cm}} derive{\nopagebreak} the time of the fragment's
return to perihelion, $t_\pi^{\displaystyle \ast}$, one{\pagebreak} first
gets the eccentric anomaly at separation, $\epsilon_{\rm frg}^{\displaystyle
\ast}$,
\begin{equation}
\epsilon_{\rm frg}^{\displaystyle \ast} = 2 \arctan \! \left( \! \sqrt{ \frac{1
 \!-\! e^{\displaystyle \ast}}{1 \!+\! e^{\displaystyle \ast}}}
 \tan {\textstyle \frac{1}{2}} u_{\rm frg}^{\displaystyle \ast} \! \right) \!,
\end{equation}
which provides the following relation for the perihelion time:
\begin{equation}
t_\pi^{\displaystyle \ast} = t_{\rm frg} - \frac{\epsilon_{\rm
 frg}^{\displaystyle \ast} \!-\! e^{\displaystyle \ast} \! \sin
 \epsilon_{\rm frg}^{\displaystyle \ast}}{k_0^{\displaystyle \ast}}
 \! \left( \! \frac{q^{\displaystyle \ast}}{1 \!-\! e^{\displaystyle
 \ast}} \:\!\!\! \right)^{\!\!\frac{3}{2}} \!\!.
\end{equation}
The eccentric anomaly $\epsilon_{\rm frg}^{\displaystyle \ast}$ is here
{\vspace{-0.08cm}}in radians and its range for fragmentation
{\vspace{-0.05cm}}times $t_{\rm frg}$ between $t_\pi^0$ and
$t_\pi^{\displaystyle \ast}$ is \mbox{$-2\pi \leq
\epsilon_{\rm frg}^{\displaystyle \ast} \leq 0$}.  Inserting{\vspace{-0.04cm}}
$\omega^{\displaystyle \ast}\!$, \ldots, $t_\pi^{\displaystyle \ast}$ into
Equation~(42) completes the derivation of effects of the orbital location
of the fragmentation event and the separation velocity of the fragment on its
orbital elements relative to those of the parent, $\Delta \omega(t_{\rm frg},
\mbox{\boldmath $v_{\bf sep}$})$, \ldots, $\Delta t_\pi(t_{\rm frg},
\mbox{\boldmath $v_{\bf sep}$})$.

\begin{table*} 
\vspace{-0.15cm}
\hspace{-0.15cm}
\centerline{
\scalebox{1}{
\includegraphics{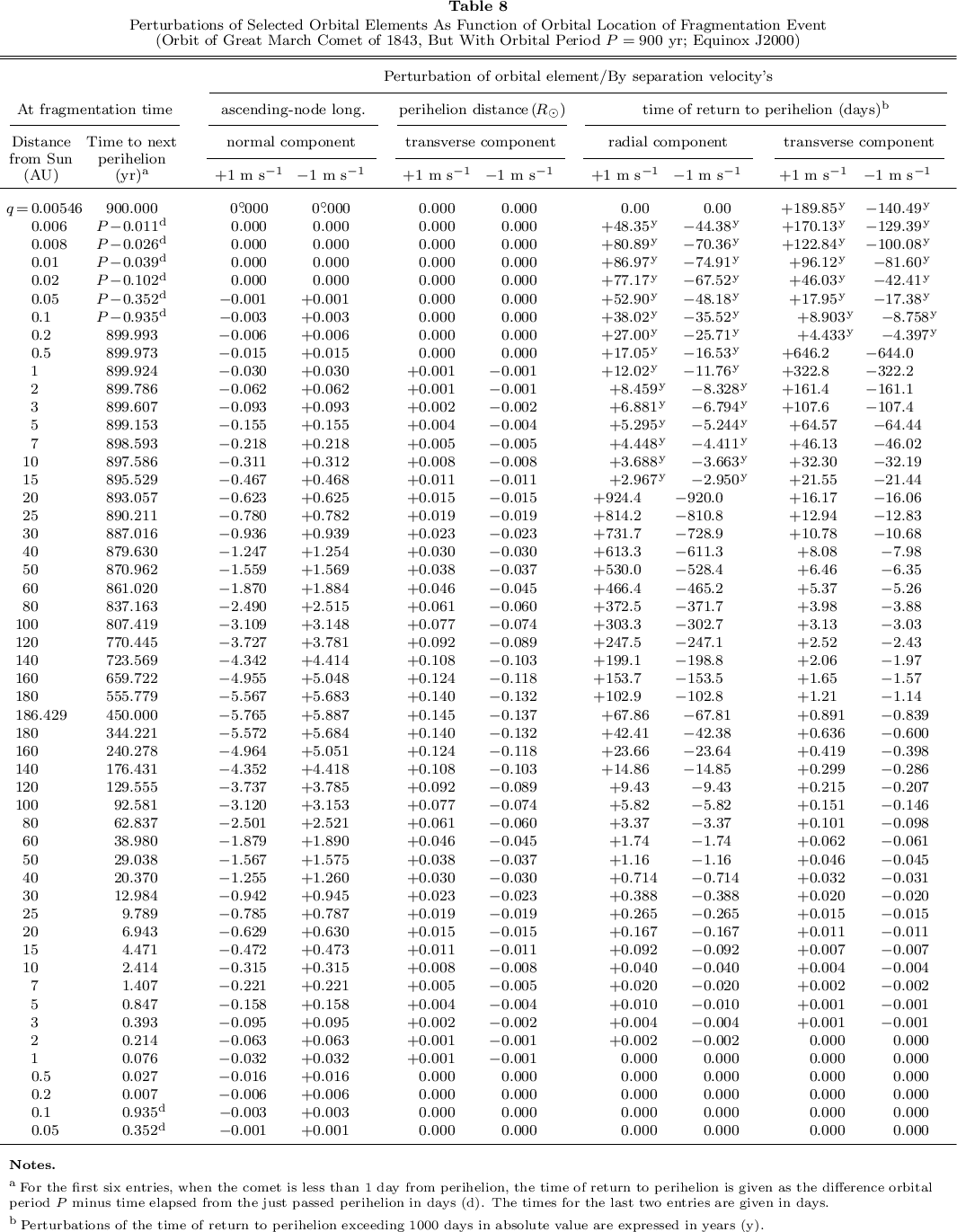}}}
\vspace{-0.01cm}
\end{table*}

\subsection{Perturbations of Fragment's Motion As Function of\\Location
 of Fragmentation Event in Orbit} 
Equations (52) to (61) provide a means to map out the structure of the
stream of SOHO Kreutz sungrazers in general and examine their
orbital-dispersion relationships in particular.  Focusing on Population~I
again, I summarize in Table~8 the separation velocity driven perturbations
of the key orbital elements of a fragment as a function of a breakup event's
location in the orbit.  The separation velocity in one of six cardinal
directions is normalized to 1~m~s$^{-1}$ and the parent (which could be
a seed or a fragment of lower generation) is assumed to move in the orbit
of the Great March Comet of 1843, but with a period of 900~years, returning
to perihelion in the year 2006.  The orbit of the 1843 sungrazer is believed
to provide {\vspace{0.01cm}}an appropriate approximation to the orbit
of X/1106~C1 at the time of interest.

I tabulate the perturbations of three elements of fragments' orbits that
have been deemed especially important:\ (i)~those of the longitude of the
ascending node, triggered off by the normal component of the separation
velocity; (ii)~those of the perihelion distance, precipitated by the
velocity's transverse component; and (iii)~those of the time of return to
perihelion, prompted by both the radial and transverse components.

I do not tabulate the perturbations of the orbit inclination, triggered off
by the normal component and strongly correlated with the more prominent
perturbations of the longitude of the ascending node; the perturbations
of the argument of perihelion, due primarily to the normal component,
but with a minor contribution from the transverse component as well; and the
perturbations of the orbit eccentricity, due to the transverse and radial
components, and correlated with the perturbations of the time of the
fragment's return to perihelion via its orbital period.

The table documents the correlations between each of the three components
of the separation velocity and the elements affected.  Likewise, the
extreme properties of a sungrazing orbit illustrate dramatically the
dependence of a perturbation on the orbital location of the comet at
the time of fragmentation.  In particular, a breakup near perihelion
may affect considerably the orbital period and thus the time of return
to perihelion but practically has no effect on the longitude of
the ascending node (and the other angular elements) or the perihelion
distance.  However, the reader should be aware that this scenario of a
major perturbation of the orbital period is fundamentally different from
the effect addressed in Section~2.1 and the following, where fragments
were assumed to move at the time of separation at {\it exactly\/}
the same orbital velocities; the driver of very different future
motions was a slight difference between the heliocentric distances
of the centers of mass of the fragments at the instant of breakup and
no separation velocity was involved.  On the other hand, here I assume
that fragments were at {\it exactly\/} the same heliocentric distance at
the time of breakup and they ended up in diverse orbits on account of
their separation velocities.  As is plainly demonstrated, fragmentation
effects on the perihelion time could in either scenario be enormous.

In contrast to near-perihelion breakups, fragmentation near aphelion
--- and generally at large heliocentric distance --- has only a very
minor effect on the orbital period and on the time of a fragment's
return to perihelion.  On the other hand, the events of disruption
at these locations may greatly affect a fragment's perihelion distance
and/or the angular elements, the longitude of the ascending node in
particular.

Interesting aspects of the perturbation variations are symmetries,
of which there are two kinds.  One of them, forward vs backward,
is seen in Table~8 to be universal --- columns~3 and 4 for the
longitude of the ascending node; columns~5 and 6 for the perihelion
distance; etc.\ --- but always only approximate.  The other
symmetry, pre-aphelion vs post-aphelion, does not apply, obviously,
to the time of return to perihelion.  It applies approximately to
the longitude of the ascending node, but for the perihelion distance
the match is apparently perfect to better than 10$^{-4}\,R_\odot$.

From Table 8 one can estimate a crude upper limit for the
range of each element inherent to~the~stream of Kreutz sungrazers,
which is consistent with the fragmentation parameters considered in
Section~3.3, namely, a separation velocity of \mbox{$\sim\:\!$0.2 m
s$^{-1}$} per event and $\sim$20~generations of fragments, equivalent
to $\sim$4~m~s$^{-1}$. One finds $\sim$20$^\circ\!${\footnotesize +}
in the nodal longitude, 0.5 to 0.6~$R_\odot$ in the perihelion distance,
and nearly 1000~years in the perihelion time, on top of the adopted
900~yr orbital period.  These numbers do by no means look excessive,
given that the realistic ranges should amount to only a small
fraction of the upper limits.  They also suggest that the orbits
of the seeds may be exhibiting a fairly wide range of periods
that the subsequent process of cascading fragmentation is merely
extending ever further.

\subsection{Dispersions of SOHO Sungrazer Swarms As\\Fragments'
 Orbital Perturbations}  
A striking result in Part~I was the{\vspace{-0.08cm}} contradiction
between the correlations of {\sf disp}($\widehat{\Omega}$) with
{\sf disp}($\widehat{t}_\pi$) among the swarms of the SOHO Kreutz
sungrazers in narrow intervals of the nodal longitude on the one
hand and in time on the other hand.  The two dispersions correlated
inversely among the swarms of the first kind, when they ranged from
0$^\circ\!$.01 to 0$^\circ\!$.1 in the nodal longitude and from 2 to
7~years in the perihelion time.  The dispersions varied in the same
direction among the swarms of the second kind, when they averaged
about 2$^\circ$ in the nodal longitude and about 1.3~days in the
perihelion time.

Inspection of Table 8 shows that the {\it swarms in the nodal longitude\/},
which include sungrazers with nearly coinciding nodal longitudes,
that is, with near-zero perturbations of this element, refer to
objects sharing the early fragmentation history, mostly the first
days and weeks after the seeds had separated from the parent comet.
The table shows that in this period of time the perturbations
of the nodal longitude and the perturbations of the perihelion time
do vary in opposite directions.  The magnitudes of the dispersions
in time are also of the correct order of magnitude.  For example,
about 20 weeks after perihelion, when the tabulated perturbation of
the nodal longitude is 0$^\circ\!$.093, the perturbation of the
perihelion time is about 6.8~years, or $\sim$2500~days; for
comparable nodal-longitude dispersions the first entries of Table~3
in Part~I yield the perihelion-time dispersions near 2000~days, which
actually is better agreement than one would expect.

For the SOHO sungrazers returning to perihelion nearly simultaneously
(within, say, a few days of one another), which I referred to in
Part~I as the \it swarms in time\/}, and whose dispersion in the nodal
longitude equaled on the average a few degrees, expectation from
Table~8 is that these objects share a history of breakups some tens of
years before perihelion, when they are tens of AU from the Sun.  The
table suggests that at 1~m~s$^{-1}$ the perturbations of the nodal
longitude and perihelion time both decline with progressing time of
fragmentation.  For example, a tabulated perturbation of the
nodal line, equaling $\sim$2$^\circ$ and implying a perturbation
of $\sim$2~days in the arrival time, could be achieved by a
fragmentation event some 40~years before perihelion.  At
the same separation velocity, the perturbations are only 1$^\circ$
and 0.5~day, respectively, with an event about 25~years later.

The seemingly bizarre relationships between the dispersions in
the SOHO sungrazers' nodal longitudes and arrival times
detected in Part~I of this investigations are thus readily
explained as orbital perturbations of the fragments' motions
exerted by the separation velocities acquired at breakup of
their parent bodies, the fragments of the preceding generation.

The SOHO sungrazers are not the only Kreutz fragments that fit the
presented perturbation scheme.  The four nuclear fragments of the
Great September~Comet~of 1882, for which Kreutz (1891) computed
separate sets of orbital elements, appear to comply with the scheme
as well, regardless of the nature of the mechanism that fragmented
the comet's parent nucleus.  Using Kreutz's results I derive in
Table~9 the differences, in the longitude of the ascending node,
in the perihelion distance, and in the time of return to perihelion
(coinciding with the future orbital period taken{\vspace{-0.1cm}} from
Table~1), between the individual nuclei and compute the{\vspace{-0.07cm}}
dispersions {\sf disp}($\widehat{\Omega}$), {\sf disp}($\widehat{q}$),
and {\sf disp}($\widehat{t}_\pi$).  Unfortunately, the dispersions in
the nodal longitude and in the perihelion distance are so small that
the mean errors (converted from Kreutz's probable errors) exceed
their values.  Fair expectation is that the dispersion in the nodal
longitude is less than 0$^\circ\!$.01 and the dispersion{\vspace{-0.01cm}}
in the perihelion distance less than 0.001~$R_\odot$.  From
Table~8 both conditions give very soft limits on the fragmentation
time, which nonetheless are consistent with the result from the
dispersion in time --- expectation that the fragmentation event
took place{\nopagebreak} within 30~minutes of perihelion.{\vspace{-0.05cm}}

\begin{table}[t] 
\vspace{0.12cm}
\hspace{-0.24cm}
\centerline{
\scalebox{1}{
\includegraphics{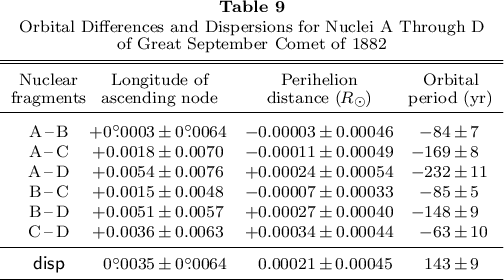}}}
\vspace{0.6cm}
\end{table}

\section{Monte Carlo Simulation of the SOHO Stream:\\Putting Things Together}
If the Kreutz system consists of nine discrete populations (Sekanina 2022b),
it makes sense to perceive the observed stream of SOHO sungrazers as a sum
of nine independent streams.  It is hoped that a Monte Carlo simulation
of the stream of Population~I --- by far the most prominent one --- will be
beneficial, at least to a point, to our understanding of the fundamental
features of the other ones as well, even though significant variations in the
arrival rates with time were revealed in Part~I to exist among them.

\subsection{The Objectives and the Algorithm} 
A major objective of this project is to fit the observed annual-rate curve
of perihelion arrival for the stream of SOHO Population~I sungrazers as
closely as possible.  The peak in 2006, seen in the histogram in the
upper part of Figure~5 of Part~I, is considered a fluke, which does not
need to be computer simulated.  On the other hand, the steady increase in
the rate, supported by the data over the period of ten years, is deemed
genetically significant.  The total number of the Population~I sungrazers
in the stream, detected in the C2 coronagraph, is estimated to vary from
about 60 per year in 2000 to 100 per year in 2009, with the rate of
increase gradually diminishing with time, as indicated in Part~I.  The
{\vspace{-0.055cm}}annual arrival rate of Population~I sungrazers,
$\dot{\cal N}$, observed in C2 is approximately expressed by
\begin{equation}
\dot{\cal N}(Y) = 60 + 5.1 \, (Y \!-\! 2000) - 0.12 \, (Y \!-\! 2000)^2,
\end{equation}
where $Y$ is the year (taken as an integer).  Since~as~much as a half of
the sungrazer population avoids the field~of the C2 coronagraph because of
seasonal effects, the absolute term on the right-hand side of Equation~(62)
may in fact be up to a factor of two higher.  However, there is no evidence
that the year-to-year trend should be affected.  Although the Marsden orbits
are unavailable after mid-2010, it appears that the annual arrival rate of
Kreutz sungrazers stabilized more recently (Battams \& Knight 2017) and I
accept that this conclusion applies to Population~I as well.

Another objective is to simulate the range of dispersion in the orbital
elements as the number of fragment generations increases with time. In
line with the results of Part~I, I focus on three orbital elements:\
the perihelion times of the returning fragmets, their longitudes of
the ascending node, as well as their perihelion distances.

The reader will note that the model offers the user a breathtaking variety
of options and a virtually endless range of choices in his effort
to fit the stream's observed structure.  In the rest of this paper
I examine an array of plausible scenarios and constraints on their
parameters, search for an optimum solution to the annual arrival
rate of the SOHO sungrazer stream, discuss the problem of dispersion
in the orbital elements, and contemplate potential avenues for
further work in the future.

The algorithm of the computer code follows closely the narrative in the
preceding sections.  The code allows the process to begin with a {\it
single\/} seed, presumably a subkilometer-sized fragment of  X/1106~C1,
which is assumed, as one of countless products of the tidal
fragmentation event at perihelion, to have separated from the parent on
1106 February~1.0~TT.  This is a plausible date, whose uncertainty of a
few days has no effect on the outcome.

To simplify the computations, the indirect planetary perturbations (the
only ones of any impact on the problem) have been ignored, the orbits
of the seed and its fragments taken as ideal ellipses.  Given this
constraint, an appropriate set of orbital elements (other than the
starting perihelion time and orbital period) to use for the seed is that
of the Great March Comet of 1843.  And since the seed most obviously
responsible for the stream of Population~I sungrazers observed by SOHO
is that in an orbit that would bring it back to perihelion at about
this time, the appropriate orbital-period choice is approximately
900~years.  This completes the problem of the seed's initial orbit
in the algorithm.

The pyramidal construct, illustrated schematically in Figure~4, has at
this point been introduced with any one of the three marked ${\cal F}_{0,1}$,
\ldots, ${\cal F}_{0,3}$ at the top.  The process of nontidal, cascading
fragmentation has thus been set in motion, the seed breaking up
rotationally into two equal halves, as described in Section~3.3.  Each
half has subsequenly broken again into two equal halves, etc., until
the fragments have reached average dimensions of the faint SOHO
sungrazers, their size distribution ignored.  There is nothing that
could prevent the user to run the code as many times as he wishes,
to complete the computations for $n_{\rm seed}$ different seeds, and
to combine the outcomes into a single output.

On certain assumptions, a typical separation velocity acquired by fragments
at breakup has, in terms of the tensile strength, been estimated at
0.2~m~s$^{-1}$, but this may require much adjustment.  A postulate of
fragments tumbling out of control calls for use of a random-number
generator to determine the separation velocity components in the radial,
transverse, and normal directions of an orthogonal coordinate system
(Sections~4 and 6.5).  The formalism of breaking up into two halves of
equal mass has required no additional assumptions to describe the
fragments' motions in the breakup's aftermath, as the separation velocities
gained by the pair should be of the same magnitude in opposite directions.
When summed up with the parent's orbital-velocity vector, the opposite
separation velocity vectors make the two fragments end up in new
orbits, thus contributing to the stream's increased orbital scatter.
The elements of these orbits depend not only on the separation
velocity vectors, but substantially also on the location of the
fragmentation episode in the orbit (Section~5 and Table~8).
 
The topic of orbital locations at which generations~of fragments
have been breaking up parallels the issue of separation velocity in
the overall problem of cascading fragmentation.  The history of the
law proposed to describe a sequence of fragmentation events along
the orbit has briefly been described in Section~4.  I discuss the
issue of incorporating the law into the model in greater detail in
Section~6.3.  Here I only note that its parameter, $\Lambda$, could be
allowed to become a random quantity in an interval of up to, say,
\mbox{$1 < \Lambda < 2$}, which excludes scenarios with fragmentation
events limited to pre-aphelion locations (Section~4).  Two additional
parameters of the law have been kept constant (Section 6.3), so
that the incorporation of the fragmentation process into the
computer code exhibits a blending of features of random nature
with deterministic ones.  The process extends over a number of
generations of fragments, as discussed in Section~4.

The outcome is a computer-simulated stream of members of Population~I.
In order to be ready for direct comparison with the observed stream
of relevant SOHO sungrazers, the computer-generated list still has
to be sorted by the quantity of interest (e.g., the perihelion arrival
time), using a sorting code, and then prepared for graphics either
as a cumulative distribution or in the form of a histogram.  In an
effort to fit genetically significant features of the observed stream,
the entire procedure of computer simulation has to be iterated by
trial and error to optimize the resulting solution.

\subsection{Incorporation of the Seeds} 
The optimization is a complex process that involves every part of
the algorithm.  Of major importance are the dimensions and number
of seeds as well as their locations in the chain of debris from the
tidal fragmentation event at perihelion, all of which greatly affects
the makeup and properties of the simulated stream of SOHO-like
sun\-grazers.  

Seed dimensions are the only of the three parameters that is subject
to obvious constraints.  The observed behavior of the SOHO sungrazers
offers an estimate for the minimum size of a seed to survive the
hostile environment.  Even the brightest SOHO sungrazers undetectable
from the ground, such as the pair of C/1998~K10 and K11, C/2003~K7,
C/2008~K4, C/2010~G4, and others, whose peak brightness
was near magnitude 0 (Knight et al.\ 2010, Sekanina \& Kracht 2013),
were not massive enough, disintegrating several hours before perihelion.
And all sungrazers discovered with the coronagraphs on board
the Solwind satellite (Michels et al.\ 1982, Sheeley et al.\ 1982)
and Solar Maximum Mission (MacQueen \& St.\ Cyr 1991), most of
them in the same brightness category, met the same fate.

On the other hand, the sungrazer C/2011~W3, already mentioned in Section
2.4, had a peak brightness not fainter than magnitude $-$3 (Green 2011;
estimates~by~two independent observers), survived perihelion but its
nucleus perished less than 2~days after perihelion (Sekanina \& Chodas
2012).  The effective nuclear diameter at times well before the demise
has by independent techniques been estimated at about 400~meters
(Sekanina \& Chodas 2012, McCauley et al.\ 2013, Raymond et al.\
2018).  The object was not a member of Population~I, but it does not
make much difference, because this was not the only case of its kind
observed.  The ``headless wonder'' --- a long dust tail emanating
out of nothing --- observed as a sungrazing comet C/1887~B1, which
did happen to be a member of Population~I, was another example of
the same phenomenon (Section~2.4).

One cannot rule out that the event experienced by Lovejoy is of the
kind that a seed needs to undergo to initiate the process of formation
of its own stream.  The dimensions of the largest debris left at the
location of the former nucleus are unknown but some sizable boulders
are always likely to survive intact or nearly intact; likewise, one
certainly cannot exclude the possibility that the above description
refers not to a single event but to a {\it sequence\/} of episodes,
whose combined effect could from a distance of the terrestrial observer
appear --- and be interpreted --- in a simplified manner.

Now, as described by Sekanina \& Chodas (2012), a peculiar trait of the
Lovejoy post-perihelion event was that, contrary to a typical comet
disintegration episode, {\it no major flare-up} was observed, only
a few relatively minor outbursts, the timing of the last one just about
coinciding with the disintegration event, as determined from the variations
in the spine-tail orientation.  Also highly unusual (if not unique) was
the sudden, dramatic, day-to-day transformation of the comet's appearance
(on December 19--20) and low dust{\vspace{-0.03cm}} velocities in the
spine tail, estimated at \mbox{20--30 m s$^{-1}$}.  These observations
prompted the authors to remark that, contemplated in its entirety, the
event was reminiscent of a collapse of the nucleus rather than its
explosion.  One has a good reason to believe that --- its seemingly
healthy appearance notwithstanding --- the comet was leaving perihelion
already severely damaged and the 40-or-so hours were needed to
complete the dismantling of the sick nucleus.

If these ideas are correct, one may have to admit that there could be
two classes of seeds that generate streams of SOHO-like sungrazers:\
besides those released at perihelion from a massive sungrazer in the
course of its nuclear splitting by tidal forces (such as the Great
September Comet of 1882), there may exist solitary seeds (such as Lovejoy). 
In either case, the seeds appear to be recruited only from objects of a
particular makeup and dimensions.  A seed should be sizable and resilient
enough to survive the perihelion environment, but small and weak enough
to get its fabric damaged enough in the process to subsequently become
prey to the forces of cascading fragmentation.

If one accepts that an average faint SOHO sungrazer is approximately
\mbox{$\Re_{\rm min} = 10$}~meters across, then the inference that
a seed should be about the size of the nucleus of comet Lovejoy would
imply some 60,000 to 70,000 fragments per seed, if no major fraction
of mass has been lost to microscopic dust.  With the adopted formalism,
a convenient analytical tool to approximate the actual process, these
numbers would seem to call for 16 generations of fragments, as seen
directly from Equation~(36) or equivalently from
\begin{equation}
\Re_{\rm seed} = \Re_{\rm min} \, 2^{\frac{1}{3}n_{\rm frg}} =
 \Re_{\rm min} \exp\left( 0.231 \, n_{\rm frg} \right).
\end{equation}
Because the stream of SOHO sungrazers demonstrates the existence of
major scatter in both the arrival time and other elements, it is obvious
that seeds fragment gradually, throughout the orbit.

\subsection{Incorporation of the Orbital Distribution Law\\of Fragmentation
Events} 
Direct use of Equation~(33) in the Monte Carlo simulation code would
be inconvenient because two random numbers would have to be assigned,
one to $\Lambda$ and a second to $\Delta t_0$, whose value ought to
be confined to some prescribed range.  It is preferable to employ
instead Equation~(34) and relate the time scale to the fragmentation
number, $n_{\rm frg}$, via $\Lambda$, as in Equation~(38).

In practice, the reference time $t_{\rm ref}$ in Equation~(34)~---
following the text at the top of page 12 --- is kept
fixed, corresponding generally to a time by when the continuing process
of cascading fragmentation reduces the dimensions of simulated fragments
{\it below\/} the estimated sizes of the SOHO sungrazers.  The time scale
over which cascading fragmenation has been going on is about 900~years,
unless the observed stream of SOHO Population~I sungrazers is a product
of a solitary Lovejoy-like Population~I fragment that had separated from
the parent comet to X/1106~C1 in AD~363 and barely survived the following
perihelion passage.  I deem this scenario too speculative to consider
seriously and prefer to stay with X/1106~C1 as the source of Population~I
seeds.

As \mbox{$\Delta t_0 = t_1 \!-\! t_0$}, Equation~(34) can be written
thus:
\begin{equation}
t_1 = t_0 +\frac{\Lambda \!-\! 1}{\Lambda^{n_{\rm frg}} \!-\! 1}
 (t_{\rm ref} \!-\! t_0).~
\end{equation}
This is an expression for the period of time elapsed between the
breakups of a seed and a first-generation fragment, $t_1 \!-\! t_0$,
as a function of the parameter $\Lambda$, the fragmentation number,
$n_{\rm frg}$, and the reference time, $t_{\rm ref}$.  If in
Equation~(34) for $t_{\rm ref}$ I next insert the identities,
\mbox{$(t_2 \!-\! t_1)/\Lambda$} for $\Delta t_0$, and
\mbox{$[(\Lambda \!+\! 1) \:\! t_1 \!-\! t_2]/\Lambda$} for $t_0$,
I obtain
\begin{equation}
t_2 = t_1 + \frac{\Lambda \!-\! 1}{\Lambda^{n_{\rm frg}-1} \!-\! 1}
 (t_{\rm ref} \!-\! t_1),
\end{equation}
an expression for the period of time elapsed between the breakups of
a fragment of the first generation and a fragment of the second
generation.  Generally, the time between the breakups of a fragment
of a $k$-th generation and a fragment of a $(k \!+\! 1)$-st generation
is given by
\begin{equation}
t_{k+1} = t_k + \frac{\Lambda \!-\! 1}{\Lambda^{n_{\rm frg}-k} \!-\! 1}
 (t_{\rm ref} \!-\! t_k).
\end{equation}
This formalism can conveniently be incorporated into the Monte Carlo
simulation code, because it warrants that \mbox{$t_{k+1} > t_k$}
for any \mbox{$k < n_{\rm frg}$} and any \mbox{$t_k < t_{\rm ref}$}.

\begin{table}[b] 
\vspace{0.6cm}
\hspace{-0.2cm}
\centerline{
\scalebox{1}{
\includegraphics{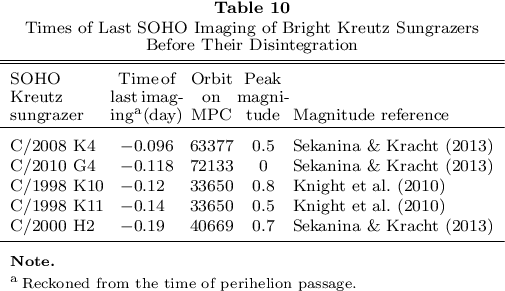}}}
\vspace{-0.08cm}
\end{table}
\begin{table}[t] 
\vspace{0.15cm}
\hspace{-0.21cm}
\centerline{
\scalebox{1}{
\includegraphics{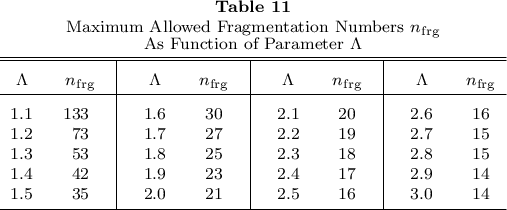}}}
\vspace{0.6cm}
\end{table}

There are however two caveats.  The first one is obvious:\ Equation~(66)
is not defined for \mbox{$k = n_{\rm frg}$}, whereby the meaningless
simultaneous arrival of all fragments to perihelion (Table~7) is
automatically eliminated.

The second caveat concerns environmental limits on the time of the
seed's {\it initial fragmentation event\/}, $t_1$.  This event must
not take place too close to the Sun too soon after the seed's birth,
in order to avoid the complete sublimation of the fragments of the
first generation before the process of cascading fragmentation takes
over.  Denoting
\begin{equation}
\nu = \frac{t_{\rm ref} \!-\! t_0}{t_1 \!-\! t_0},
\end{equation}
the condition requires that the time interval between the two
breakup events, \mbox{$t_1 \!-\! t_0$}, exceed a particular minimum,
thereby implying a maximum value of $\nu_{\rm max}$.  A conservative
estimate for \mbox{$(t_1 \!-\! t_0)_{\rm min}$} is obtained from the
last SOHO observations of the brightest Kreutz sungrazers, of peak
magnitudes between 0 and +1, before they disintegrated.  The five
with the known orbits, which the SOHO's C2 coronagraph was imaging
down to the smallest heliocentric distances, are listed in Table~10.
The suggested limit of \mbox{$(t_1 \!-\! t_0)_{\rm min} \simeq 0.1$
day}, equivalent to a distance of a little over 4~\Rsun \,from the Sun,
{\vspace{-0.05cm}}together with \mbox{$t_{\rm ref} \!-\! t_0 \simeq
900$ yr}, implies \mbox{$\nu_{\rm max} \simeq 3.3 \times \! 10^6$}.
From Equation~(64) one gets the constraint on $n_{\rm frg}$ as follows:
\begin{equation}
n_{\rm frg} < \frac{\log \left[1 \!+\!(\Lambda \!-\! 1) \:\! \nu_{\rm
 max}\right]}{\log \Lambda}.
\end{equation}
The highest allowed values of $n_{\rm frg}$ as a function of $\Lambda$
are listed in Table~11.  In a limit, L'Hospital's rule gives
\begin{equation}
\lim_{\Lambda \rightarrow 1} n_{\rm frg} < \lim_{\Lambda \rightarrow 1} \,
 \frac{\Lambda \nu_{\rm max}}{1 \!+\! (\Lambda \!-\! 1) \:\!\nu_{\rm max}}
 = \nu_{\rm max}.
\end{equation}

The way the computer code was written, the standard output is a set of
lists, sorted sequentially by three orbital elements, and histograms of
all simulated fragments of a fixed \mbox{$n_{\rm gen}$-th} generation,
where \mbox{$n_{\rm gen} < n_{\rm frg}$}.~It~is~now $n_{\rm gen}$ that
controls the size ratio of the simulated SOHO sungrazers and seeds, as
well as the total number of test fragments:\ rising $n_{\rm gen}$ by one
increases the ratio of $\Re_{\rm seed}/\Re_{\rm min}$ by a factor of
1.26 and ${\cal N}$ by a factor of two.\,\,\,\,

The model's simulation computations began on the assumption that
\mbox{$n_{\rm gen} = 16$} and \mbox{$n_{\rm frg} = 20$}, which
required that $\Lambda$ not exceed 2.13 according to Equation~(68).
To explain the existence of close pairs of SOHO sungrazers implying
post-aphelion fragmentation, one should allow a limited number of
fragments of generations between $n_{\rm gen}$ and $n_{\rm frg}$ or,
alternatively, subjected to \mbox{$\Lambda < 1.3$}, if $\Lambda$ is
a random quantity (within limits).

\subsection{Constants and Options} 
One of the tasks of this investigation is to find~out~how is the difference
between $n_{\rm gen}$ and $n_{\rm frg}$ going to affect the simulated stream
of sungrazers.  This procedure could further be manipulated by changing the
reference time $t_{\rm ref}$, tied to the seed's orbit.  This shows that the
two free constants, $n_{\rm frg}$ and $t_{\rm ref}$, have a significant
impact on the outcome of the simulation.

The other constant that affects the simulated orbital evolution of
fragments is their separation velocity.  Nominally, its ``recommended''
magnitude, based on the considerations in Section~3.3, is 0.2~m~s$^{-1}$,
independent of fragment size and therefore implying spin-up.  Since
the overall dispersion of the orbital elements clearly correlates with
the magnitude of the separation velocity, comparison of the simulated
orbital sets with the observations should provide a valuable test of
the constants. 

The last topic that involves decisions of the optional nature is the
problem of whether or not to incorporate a nongravitational acceleration
in the orbital computations.  While these effects on the motions of
the SOHO Kreutz sungrazers in the final hours of their life is enormous
(Sekanina \& Kracht 2015b), I believe that --- given other uncertainties
--- their overall contribution is not significant enough for inclusion.

\begin{table}[b]  
\vspace{0.6cm}
\hspace{-0.205cm}
\centerline{
\scalebox{1}{
\includegraphics{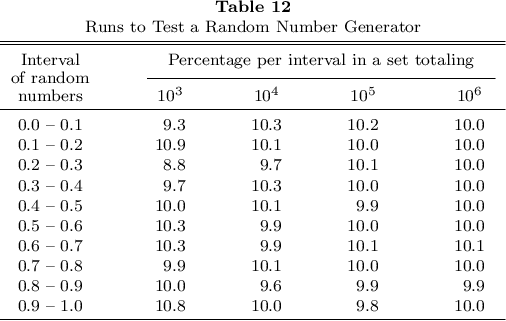}}}
\vspace{-0.03cm}
\end{table}
\subsection{Random Number Generator and Applications}  
A random-number generator routine supplies~\mbox{10-digit} random numbers
{\boldmath ${\cal R}$} in the range \mbox{0\,$<$\,{\boldmath ${\cal
R}$}\,$<$\,1}, working with several positive integers.  Three of them
are constants:\ $A_0$ and $B_0$ are on the order of 10$^{13}$ to 10$^{14}$,
while $Z_0$ is an odd number on the order of 10$^2$.  Two additional
integers, ${\cal A}$ and ${\cal B}$, are variable. 

To begin the generation of random numbers, I choose ${\cal A}$ such that
it satisfies a condition
\begin{equation}
{\cal B} = {\cal A} Z_0 \!+\! A_0 \gg B_0
\end{equation}
and then compute
\begin{equation}
{\cal A}^\ast\! = {\cal B} \: {\rm mod} \: B_0.
\end{equation}
The first random number {\boldmath ${\cal R}$} is given by
\begin{equation}
{\mbox{\boldmath ${\cal R}$}} = {\cal A}^\ast\!/B_0.
\end{equation}
Inserting ${\cal A}^\ast$ for ${\cal A}$ into Equation~(70) and
executing successively Equations~(70) through (72), I get a second
random number {\boldmath ${\cal R}$}, etc.

Next I have generated sets of random numbers, totaling a thousand,
ten thousand, hundred thousand, and a million entries, respectively.
Each set has been divided into ten equal intervals between 0 and 1,
and a percentage of the total in each interval computed to test the
degree of randomness.  The results in Table~12 show, as expected,
that the relative variations in the number of entries per interval
drop with increasing size of the set.\,\,\,\,\,

Four random numbers are used in the computer simulations per
fragmentation event:\ one for the parameter $\Lambda$ and
three for the separation velocity components.  To complete the
computations down to fragments of an \mbox{$n_{\rm gen}$-th} generation
requires a file of \mbox{$4 \! \cdot \! (2^{n_{\rm gen}} \!-\!1)$}
random numbers.  A preferred range for $\Lambda$, \mbox{$\Lambda_{\rm
min} \!<\! \Lambda \!<\! \Lambda_{\rm max}$}, where \mbox{$\Lambda_{\rm
min} \!\geq\! 1$} and \mbox{$\Lambda_{\rm max} \!\leq\! 2$}, is readily
satisfied by
\begin{equation}
\Lambda = \Lambda_{\rm min} \!+\! (\Lambda_{\rm max} \!-\! \Lambda_{\rm
 min}) \:\!\mbox{\boldmath ${\cal R}$}\,.
\end{equation}
When \mbox{$\Lambda_{\rm min} \!=\! \Lambda_{\rm max}$}, $\Lambda$
becomes a parametric constant.

The separation velocity imposes a different condition on random
numbers.  The radial, transverse, and normal components should
each be allowed to vary in extreme cases from $-v_{\rm sep}$ to
$+v_{\rm sep}$, so that a component{\vspace{-0.065cm}} equaling
zero is given by a random number of $\frac{1}{2}$.  One can
write
\begin{eqnarray}
v_{\rm R} & = & v_{\rm sep} (\mbox{\boldmath ${\cal R}$}_1 \!-\!
 {\textstyle \frac{1}{2}}) \, \lambda, \nonumber \\
v_{\rm T} & = & v_{\rm sep} (\mbox{\boldmath ${\cal R}$}_2 \!-\!
 {\textstyle \frac{1}{2}}) \, \lambda, \nonumber \\
v_{\rm N} & = & v_{\rm sep} (\mbox{\boldmath ${\cal R}$}_3 \!-\!
 {\textstyle \frac{1}{2}}) \, \lambda,
\end{eqnarray}
where $\lambda$ is a normalization constant.  The condition that
the sum of squares of the components equal the square of the
separation velocity is equivalent to a condition
\begin{equation}
\lambda^2 \!\!\!\: \left[ \left( \mbox{\boldmath ${\cal R}$}_1 \!-\!
 {\textstyle \frac{1}{2}} \right)^{\!2} \!+\! \left( \mbox{\boldmath 
 ${\cal R}$}_2 \!-\! {\textstyle \frac{1}{2}} \right)^{\!2} \!+\!
 \left( \mbox{\boldmath ${\cal R}$}_3 \!-\! {\textstyle \frac{1}{2}}
 \right)^{\!2} \right] = 1,
\end{equation}
from which
\begin{equation}
\lambda = \frac{2}{\sqrt{3 \!-\! 4 \sum_{k=1}^{3} \! {\mbox{\boldmath
 ${\cal R}$}}_k (1 \!-\! {\mbox{\boldmath ${\cal R}$}}_k)}} \,.
\end{equation}
The normalization constant attains values of \mbox{$\lambda > \frac{2}{3}
\sqrt{3}$}.

\section{Results for Fragments of 16th Generation} 
The results of Monte Carlo computations reported below present
simulations of a semi-stochastic process that was to generate a stream of
the SOHO Kreutz sungrazers of Population~I as products of cascading
fragmentation of a single seed, a subkilometer-sized object, born in
the course of tidal fragmentation at perihelion.  The seed was moving
essentially in the orbit of the Great Comet of 1106, the seed's parent,
but with an orbital period of $\sim$900~yr, by $\sim$163~years longer
than the orbital period of C/1843~D1, the 1106 comet's principal fragment.
Deemed appropriate as an approximation to the 1106 comet's orbit at
the beginning of the 21st century (Section~6.1), an orbital solution
for the 1843 comet derived by Sekanina \& Chodas (2008) was adopted.
The 1106 comet was assumed to have passed its perihelion on February~1.0~TT,
the date from which time was reckoned.

In reality, the Population I stream is a sum of fragmentation products
of a number of seeds from the 1106 parent comet that may or may not
partially or completely overlap one another in time.  In addition,
the observed SOHO stream also contains minor contributions from
fragmented seeds of other Kreutz populations.

\subsection{Expanding Spans of Simulated Orbital Elements\\As a Function
 of Fragment Generation} 
The first task I undertook with the simulation code was an investigation
of the systematic propagation of perturbations of the seed's orbit,
driven by the separation velocities of fragments and manifested by
scatter among their orbits.  The perturbations were growing with ever
increasing number of fragmentation events, described in terms of the
number of fragment generations, $n_{\rm gen}$.  For the fragmentation
constants, I used --- somewhat arbitrarily, but still more or less in
line with the arguments in the preceding sections of this paper ---
the following:\ \mbox{$n_{\rm frg} = 20$}, \mbox{$\Lambda_{\rm min} =
1$}, \mbox{$\Lambda_{\rm max} = 2$}, \mbox{$v_{\rm sep} = 0.2$ m
s$^{-1}$}, and \mbox{$t_{\rm ref} \!-\! t_0 = 900$ yr}.

Because nontidal, cascading fragmentation began near the Sun,
continuing toward aphelion, the birth of the {\it early\/} fragment
generations was confined to relatively small heliocentric distances.
As seen from Figure~3, these generations contained a limited number
of objects larger than a typical SOHO sungrazer.  They have not been
included in Figure~5, which begins with the 9th fragment generation
and ends with the 16th.

I let the computer code run and sort~the~sets~of~the time of return to
perihelion, $t_\pi$, the longitude of the ascending node, $\Omega$, and the
perihelion distance, $q$. {\small \bf The difference between the absolutely
minimum and maximum values that a pair of simulated fragments of each
generation, {\boldmath $n$}, exhibited in each element defined its span\/},
{\small \sf Sp}$_n(t_\pi)$, {\small \sf Sp}$_n(\Omega)$, and {\small \sf
Sp}$_n(q)$, respectively.  The number of simulated fragments, equaling $2^n$,
was increasing from 512 at the 9th generation to 65,536 at the 16th generation.
The dimensions of the fragments in this highest generation were 1/40-th
part of the seed's dimensions.

Figure 5 confirms that scatter in the orbital elements was growing
with increasing fragment generation, as expected.  More importantly,
the figure also shows that the {\it rate of increase\/} in {\sf
Sp}$_n(t_\pi)$ was diminishing with increasing generation (that is, with
time and increasing heliocentric distance), while the parallel rates of
increase in {\sf Sp}$_n(\Omega)$ and, to a degree, in {\sf Sp}$_n(q)$
had a tendency to accelerate.  This behavior suggests that scatter
in each orbital element was essentially proportional to the magnitude
of the perturbation listed in Table~8.

Comparison of the numbers for the 16th generation with the data
in Part~I is clearly unsatisfactory.  The stream's observed
duration ever since 1996, now over a period of nearly 30~years, is
utterly incompatible with the predicted span of 19~years, from
the end of 1996 through early 2016.  On the other hand, the range
of the longitudes of the ascending node, which for Population~I is
estimated to cover about 7$^\circ$, is predicted to extend too wide,
over more than 11$^\circ$, from 358$^\circ$ to more than 9$^\circ$.

\subsection{Spans of Simulated Orbital Elements As a Function\\of
 Fragmentation Number, $n_{\rm frg}$} 
Throughout the rest of Section 7 the number of fragment generations is being
held {\vspace{-0.04cm}}constant at \mbox{$n_{\rm gen} = 16$}.  In a first
step, $v_{\rm sep}$ was kept at 0.2~m~s$^{-1}$ and \mbox{$t_{\rm ref} \!-\!
t_0$} at 900~years, while the parameter $\Lambda$ was varied again between
1 and 2 to plot the dependence of the spans of the three orbital elements
on the fragmentation number, $n_{\rm frg}$, whose allowed range was
\mbox{$17 \leq n_{\rm frg} \leq 21$}.  The results are exhibited in Figure~6.  

\begin{figure}[t] 
\vspace{0.17cm}
\hspace{-0.17cm}
\centerline{
\scalebox{0.675}{ 
\includegraphics{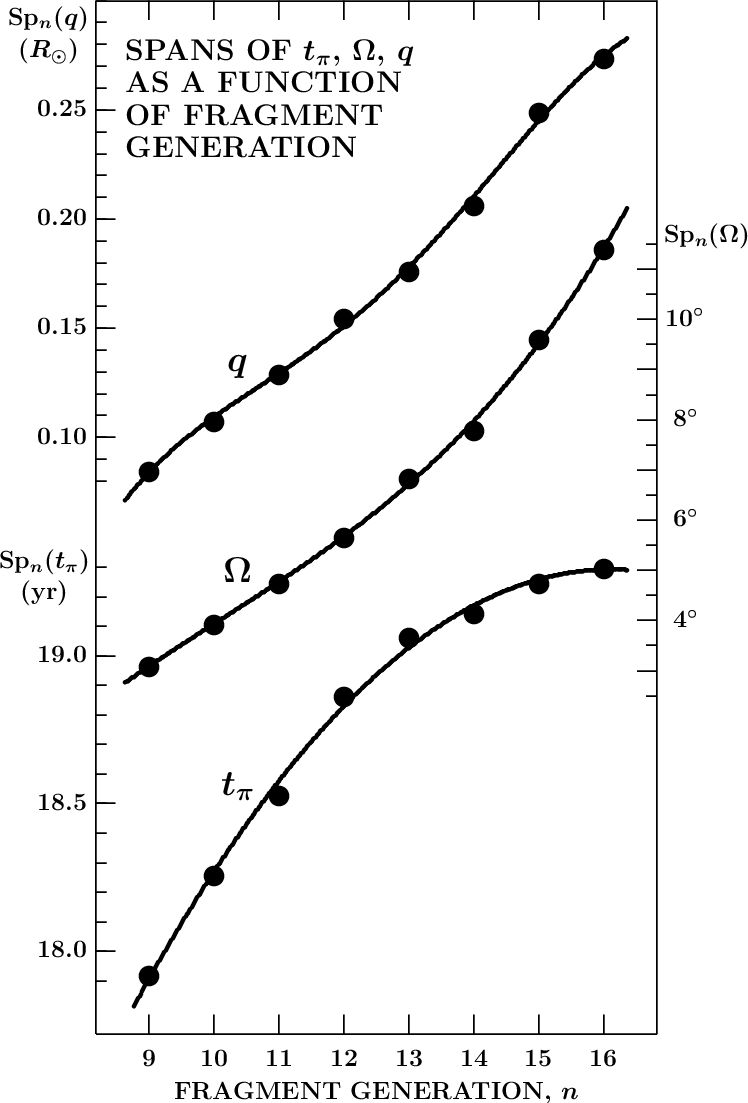}}}
\vspace{-0.08cm}
\caption{Spans of the times of return to perihelion, $t_\pi$, the longitudes
of the ascending node, $\Omega$, and the perihelion distances, $q$, as
a function of fragment generation, $n$, from computer simulation runs
based on the following parameters:\ \mbox{$n_{\rm frg} = 20$},
\mbox{$v_{\rm sep} = 0.2$ m s$^{-1}\!$}, \mbox{$t_{\rm ref}
\!-\! t_0 = 900$\,yr}, \mbox{$\Lambda_{\rm min} = 1$}, and
\mbox{$\Lambda_{\rm max} = 2$}.  Note that the {\it rate\/} of increase
in {\sf Sp}$_n(t_\pi)$ diminishes with time, as the fragmentation
process enters ever larger heliocentric distances, while the
parallel rates of increase in {\sf Sp}$_n(\Omega)$ and, less
prominently in {\sf Sp}$_n(q)$, continue increasing, in line
with the data in Table~8.{\vspace{0.6cm}}}
\end{figure}

The opposite trends between {\sf Sp}$_n(t_\pi)$ on the one hand and {\sf
Sp}$_n(\Omega)$ and {\sf Sp}$_n(q)$ on the other hand can be explained by
recognizing that an increase in the difference \mbox{$n_{\rm frg} \!-\!
n$} means increasing crowding of the fragmentation times
toward smaller heliocentric distance.  According to Table~8 the
perturbation of $t_\pi$ increases in that direction, while the
perturbations of $\Omega$ and $q$ decrease.  The spans thus vary
qualitatively in line with expectation.

\begin{figure}[t] 
\vspace{0.17cm}
\hspace{-0.17cm}
\centerline{
\scalebox{0.665}{
\includegraphics{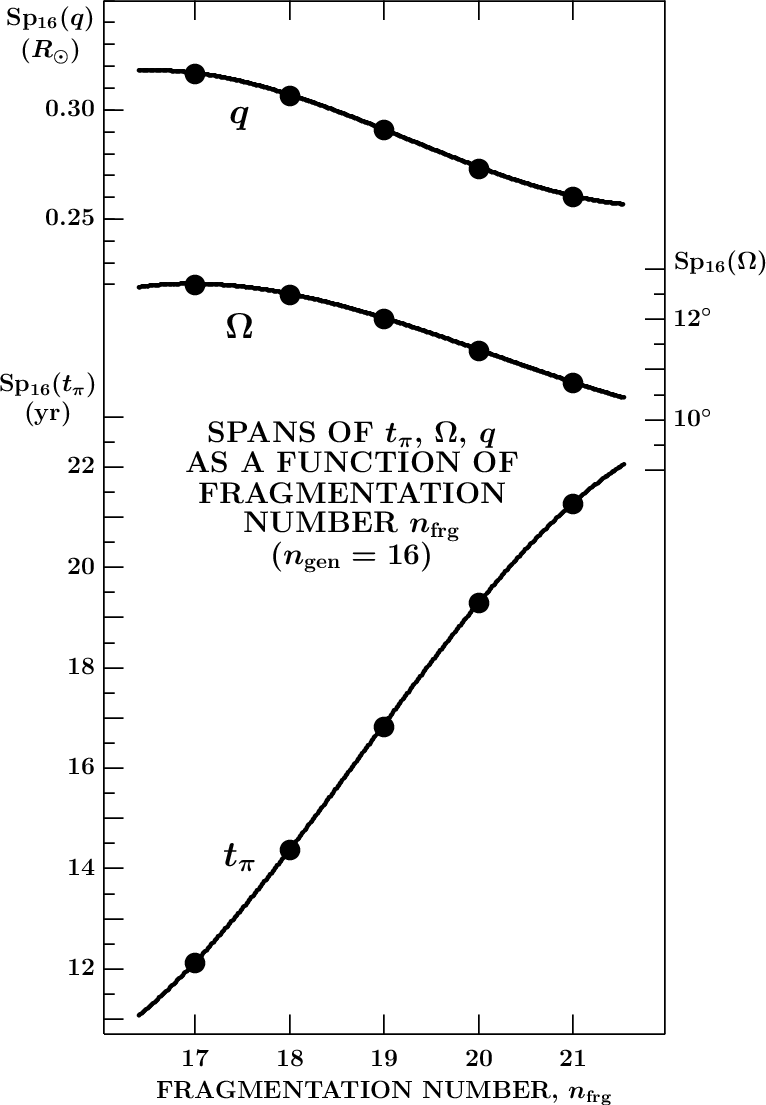}}}
\vspace{-0.06cm}
\caption{Spans of the times of return to perihelion, $t_\pi$, the longitudes
 of the ascending node, $\Omega$, and the perihelion distances, $q$, for
 simulated fragments of the 16th generation as a function of the
 fragmentation number, $n_{\rm frg}$, from computer{\vspace{-0.07cm}}
 runs based on the following parameters:\ \mbox{$n_{\rm gen} = 16$},
 \mbox{$v_{\rm sep} = 0.2$ m s$^{-1}$}, \mbox{$t_{\rm ref} \!-\! t_0 =
 900$ yr}, \mbox{$\Lambda_{\rm min} = 1$}, and \mbox{$\Lambda_{\rm max}
 = 2$}.  The curve of {\sf Sp}$_{16}(t_\pi)$ now behaves very differently
 in comparison with those of {\sf Sp}$_{16}(\Omega)$ and {\sf Sp}$_{16}(q)$,
 increasing sharply with $n_{\rm frg}$.{\vspace{0.6cm}}}
\end{figure}

Comparison of the trends in $n_{\rm frg}$ with the observed data from
Part~I suggests that the smaller the difference between $n_{\rm frg}$
and $n$, the poorer fit to the data.  For the continuing experimentation,
it was deemed appropriate to keep $n_{\rm frg}$ at 20.

\subsection{Spans of Simulated Orbital Elements As a Function\\of
 Parameter $\Lambda$} 
To examine the effect of the parameter $\Lambda$ on the spans of the
orbital elements, I turned off the random number generator for $\Lambda$
by putting \mbox{$\Lambda_{\rm min} = \Lambda_{\rm max}$}.  I continued
to adopt:\ \mbox{$n_{\rm frg} = 20$}, \mbox{$n = n_{\rm gen} = 16$},
\mbox{$v_{\rm sep} = 0.2$ m s$^{-1}$}, and \mbox{$t_{\rm ref} \!-\!
t_0 = 900$ yr}.  The runs for $\Lambda$ between 1.1 and 2.3 yielded the
results for {\sf Sp}$_{16}(t_\pi)$, {\sf Sp}$_{16}(\Omega)$, and {\sf
Sp}$_{16}(q)$ that are plotted in Figure~7.

\begin{figure}[t] 
\vspace{0.17cm}
\hspace{-0.17cm}
\centerline{
\scalebox{0.665}{
\includegraphics{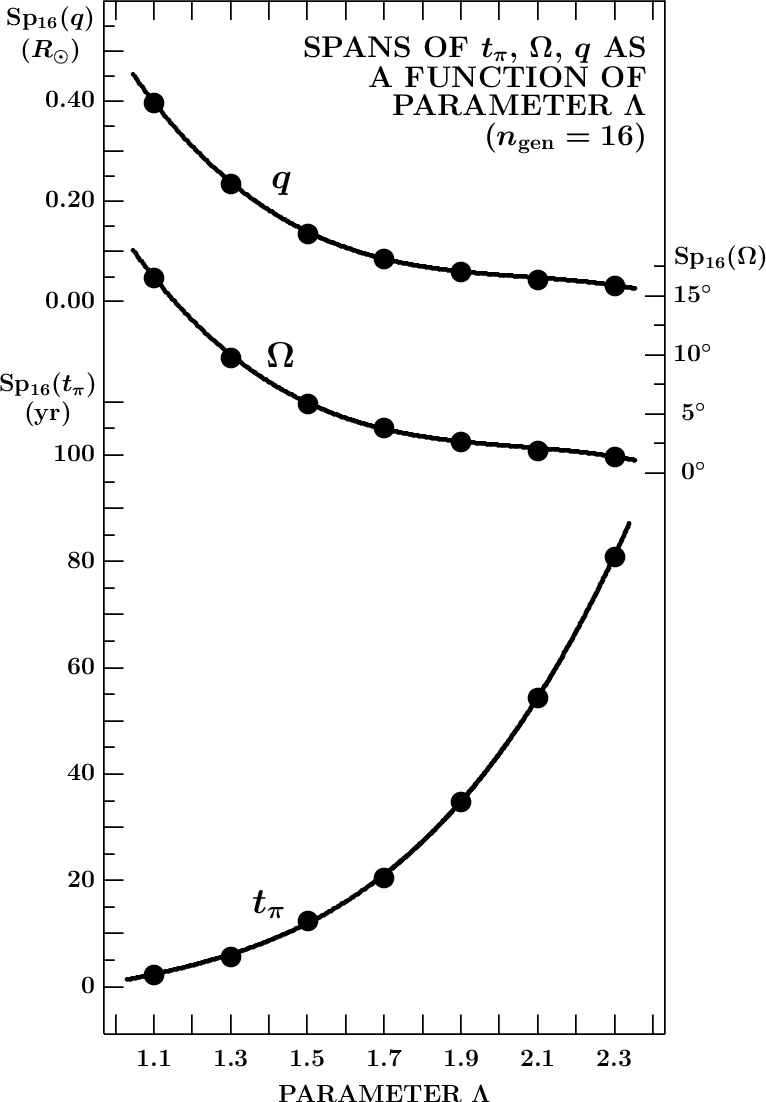}}}
\vspace{-0.01cm}
\caption{Spans of the times of return to perihelion, $t_\pi$, the longitudes
of the ascending node, $\Omega$, and the perihelion distances, $q$, for
simulated fragments of the 16th generation as a function of the parameter
$\Lambda$ from computer runs{\vspace{-0.01cm}} based on the following:\
\mbox{$n_{\rm frg} = 20$}, \mbox{$n_{\rm gen} = 16$}, \mbox{$v_{\rm sep}
= 0.2$ m s$^{-1}$}, and \mbox{$t_{\rm ref} \!-\! t_0 = 900$ yr}.  It is
noted that the general trend of the curves is somewhat similar to that
in Figure~6 except that the variations are now strongly nonlinear and
the spans are much wider.{\vspace{0.65cm}}}
\end{figure}

It is obvious that $\Lambda$ affects each of the three spans dramatically,
{\sf Sp}$_{16}(t_\pi)$ in particular.  However, the variations are now
strongly nonlinear, the highest rates being achieved at \mbox{$\Lambda
> 2$} in the time of return to perihelion, but at \mbox{$\Lambda
\rightarrow 1$} in the nodal longitude and perihelion distance.  The
observed range of  7$^\circ$ in the nodal longitude is matched most
closely by \mbox{$\Lambda \simeq 1.4\!-\!1.5$}, whereas the observed
scatter in time nominally requires a minimum \mbox{$\Lambda \simeq 1.9$}.
Interestingly, the set for \mbox{$\Lambda = 1.1$} includes orbits
with perihelion distances that are smaller than the Sun's radius.

\subsection{Spans of Simulated Orbital Elements As a Function\\of
 Separation Velocity} 
The dependence of the spans of the three elements on the simulated
separation velocities of the fragments~of the 16th generation is
rather straightforward. Computer runs, based on the set of
parameters as before (\mbox{$n_{\rm frg} = 20$}, \mbox{$n_{\rm gen}
= 16$}, \mbox{$\Lambda_{\rm min} = 1$}, \mbox{$\Lambda_{\rm max} =
2$}, \mbox{$t_{\rm ref} \!-\! t_0 = 900$ yr}), showed that the
relationships were nearly perfectly linear, {\sf Sp}$_{16}(t_\pi)$
varying at a rate of 97~yr per 1~m~s$^{-1}$, {\sf Sp}$_{16}(\Omega)$
at a rate of 56$^\circ$ per 1~m~s$^{-1}$, and {\sf Sp}$_{16}(q)$ at
about 1.4~$R_\odot$ per 1~m~s$^{-1}$.  Perihelion distances smaller
than 0.7~$R_\odot$ were obtained in extreme cases.  The results,
plotted in Figure~8, predict much too high a nodal-longitude rate
compared to the rate of arrival times.

\begin{figure}[t] 
\vspace{0.19cm}
\hspace{-0.19cm}
\centerline{
\scalebox{0.665}{
\includegraphics{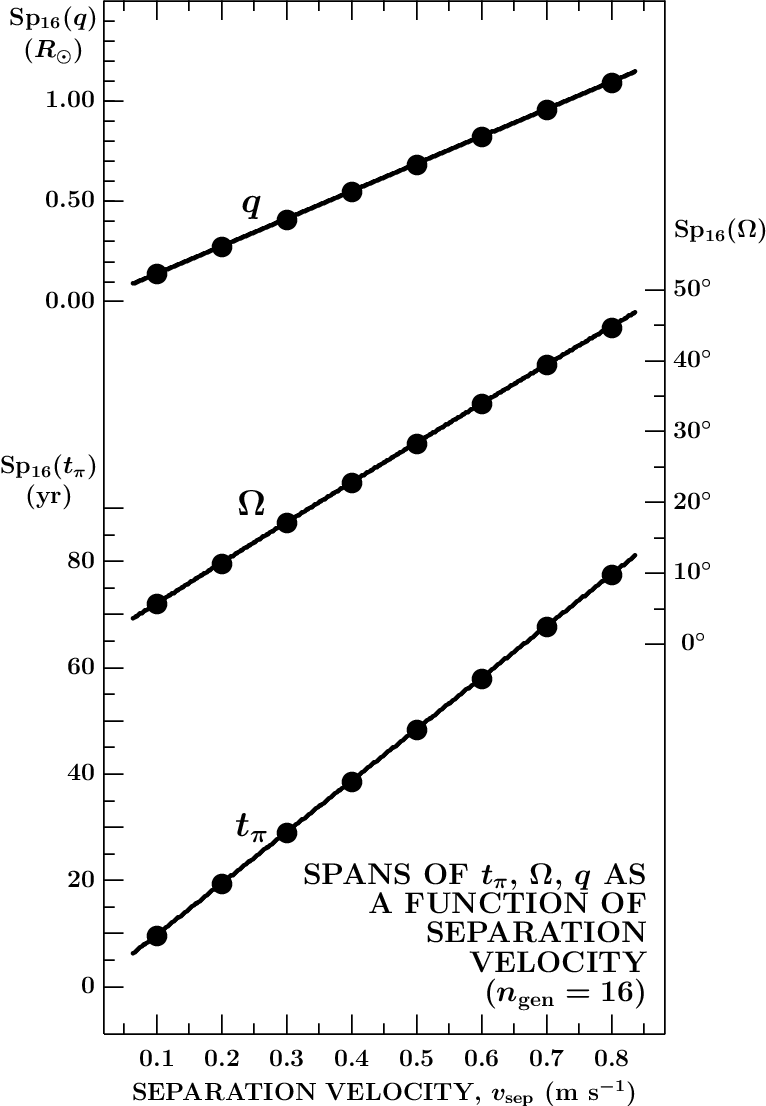}}}
\vspace{-0.075cm}
\caption{Spans of the times of return to perihelion, $t_\pi$, the
longitudes of the ascending node, $\Omega$, and the perihelion distances,
$q$, for simulated fragments of the 16th generation as a function of
the separation velocity from computer runs based on the following
parameters:\ \mbox{$n_{\rm frg} = 20$}, \mbox{$n_{\rm gen} = 16$},
\mbox{$\Lambda_{\rm min} = 1$}, \mbox{$\Lambda_{\rm max} = 2$},
and \mbox{$t_{\rm ref} \!-\! t_0 = 900$ yr}.  The relationships
are nearly perfectly linear.{\vspace{0.65cm}}}
\end{figure}

\subsection{Spans of Simulated Orbital Elements As a Function\\of
 Reference Time, $t_{\rm ref}$} 
This procedure examines the dependence of the spans of
the three elements on the orbital period of the seed, as $t_0$ is
expected to differ at most by a fraction of a day from the perihelion
time of the parent comet.  The computer runs were based again on
the standard parameters:\ \mbox{$n_{\rm frg} = 20$}, \mbox{$n_{\rm
gen} = 16$}, \mbox{$\Lambda_{\rm min} = 1$}, \mbox{$\Lambda_{\rm
max} = 2$}, {\vspace{-0.03cm}}and \mbox{$v_{\rm sep} = 0.2$ m
s$^{-1}$}.  The runs should provide information on the orbital
relationship between the seed and the related SOHO sungrazer stream.

\subsection{Spans of Simulated Orbital Elements As\\a Function of a
 Boundary Condition} 
Usual in problems of this kind is a dependence~of~results --- in
this case the spans {\sf Sp}$_{16}(t_\pi)$, {\sf Sp}$_{16}(\Omega)$,
and {\sf Sp}$_{16}(q)$ --- on boundary conditions.  Intuitively, one
would expect that the orbital position of the seed at the time of its
splitting into the two fragments of the first generation should be
instrumental in the process of cascading fragmentation and affect
the spans of orbital elements in a profound manner.
\begin{figure}[t] 
\vspace{0.19cm}
\hspace{-0.19cm}
\centerline{
\scalebox{0.665}{
\includegraphics{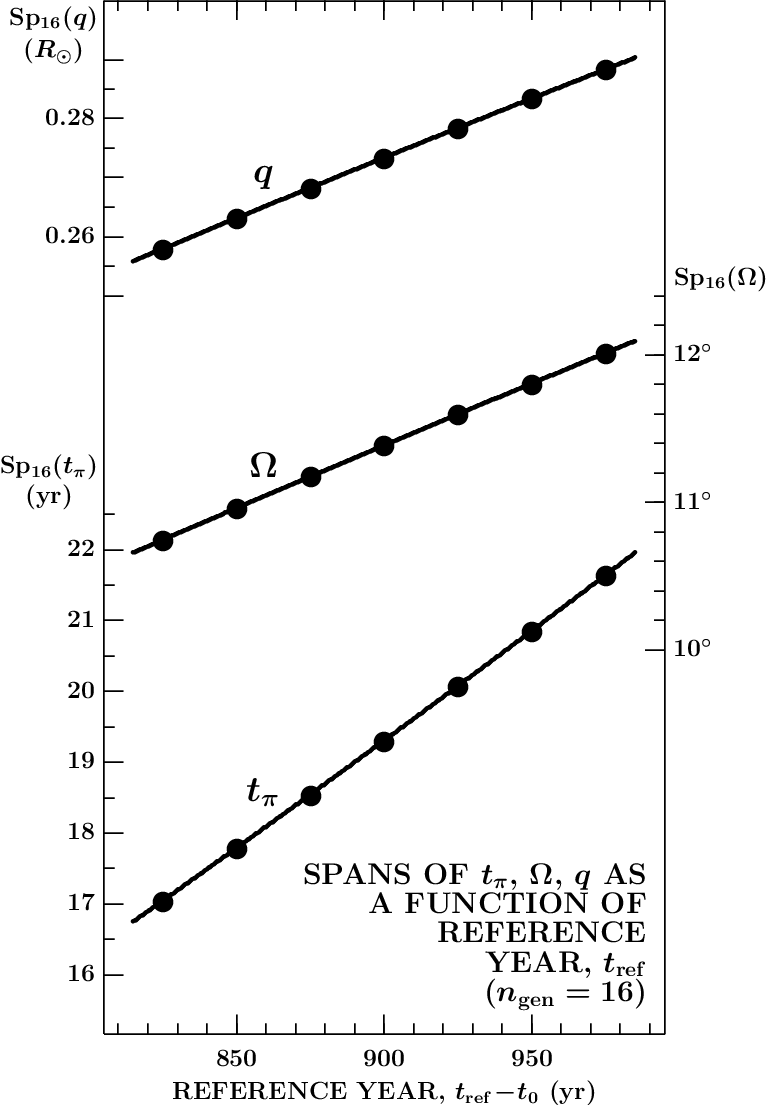}}}
\vspace{-0.075cm}
\caption{Spans of the times of return to perihelion, $t_\pi$, the
 longitudes of the ascending node, $\Omega$, and the perihelion
 distances, $q$, for simulated fragments of the 16th generation as
 a function of the reference time, $t_{\rm ref}$, showing an effect
 of the seed's orbital period.  The same parametric values
 {\vspace{-0.01cm}}have been used as before:\ \mbox{$n_{\rm frg} = 20$},
 \mbox{$n_{\rm gen} = 16$}, \mbox{$\Lambda_{\rm min} = 1$},
 \mbox{$\Lambda_{\rm max} = 2$}, and \mbox{$v_{\rm sep} = 0.2$ m
 s$^{-1}$}.{\vspace{0.7cm}}}
\end{figure}

Figure 10 shows this expectation to be fully confirmed for the span of
the times of return to perihelion, validated to a modest degree for the
span of the longitudes of the ascending node, but not corroborated
by the span of the perihelion distances, which is independent of the
boundary condition.  This peculiar outcome of examination of this
issue calls for its more extensive investigation.

\begin{figure}[t] 
\vspace{0.19cm}
\hspace{-0.19cm}
\centerline{
\scalebox{0.665}{
\includegraphics{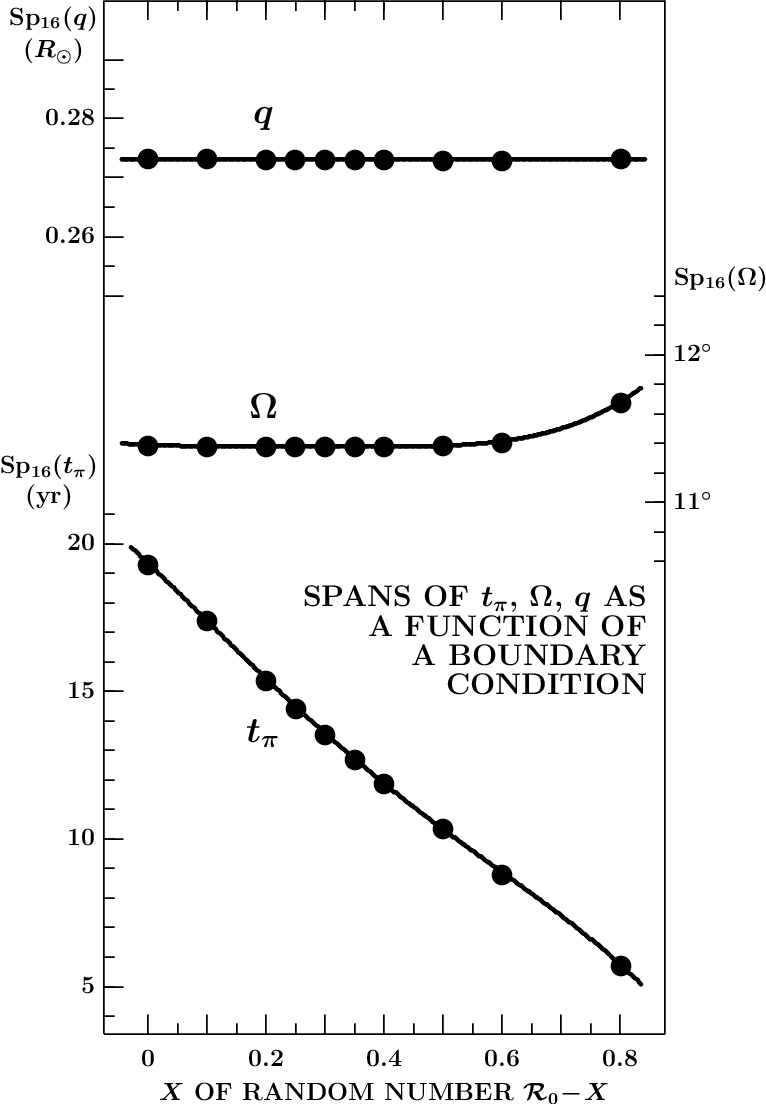}}}
\vspace{-0.08cm}
\caption{Spans of the times of return to perihelion, $t_\pi$, the longitudes
 of the ascending node, $\Omega$, and the perihelion distances, $q$, for
 simulated fragments of the 16th generation as a function of a boundary
 condition, represented by the first random number \mbox{{\boldmath
 ${\cal R}_0$} = 0.9923017714} and its forced variations.  They controled
 the values of $\Lambda$ and the times of the seed's initial fragmentation.
 The parameters have remained{\vspace{-0.07cm}} unchanged:\ \mbox{$n_{\rm
 frg} = 20$}, \mbox{$n_{\rm gen} = 16$}, \mbox{$\Lambda_{\rm min} = 1$},
 \mbox{$\Lambda_{\rm max} = 2$}, \mbox{$v_{\rm sep} = 0.2$ m s$^{-1}$},
 and \mbox{$t_{\rm ref} \!-\! t_0 = 900$ yr}.{\vspace{0.7cm}}}
\end{figure}

The nominal results presented in Figures 5 through 9 were based on a
few constants (Sections~6.4, \mbox{7.1--7.5}) and two variables, the
parameter $\Lambda$ and the {\it vector\/} of the separation velocity,
{\boldmath $v_{\rm sep}$}, while the {\it magnitude\/}, $v_{\rm sep}$,
was kept constant.  The variables were controled by random numbers, the
parameter $\Lambda$ over a wide range.  The random numbers \mbox{$0 \,<\:
${\boldmath $\cal R$}$ \:<\, 1$} were read sequentially from a file of one
million entries produced by a random-number generator (Section~6.5).  The
first entry, determining the starting parameter $\Lambda$ happened to have
an extreme value of \mbox{{\boldmath ${\cal R}_0$}$\,= 0.9923017714$} (i.e.,
\mbox{$\Lambda = 1.9923017714$}), implying that the seed split into the
first-generation fragments as early as 0.34~day after the parent comet's
tidal fragmentation event, at a heliocentric distance{\vspace{-0.0cm}}
of 10.5~\Rsun.  The submeter separation velocities in the opposite
directions were enough to get the two fragments into orbits whose
periods differed by nearly 11~years:\ one would have returned to
perihelion on 2000 September 29, the other on 2011 August 2.  The
continuing randomization of the orbital elements of higher-generation
fragments led eventually to a span of times of return to perihelion of
nearly 20~years, as shown in the upper left panel of Figure~11 (for
\mbox{$X = 0$}).  In scenarios in which $X$ was forced to grow, so was
the heliocentric distance at the time of the seed's initial breakup,
so that the gap between the two peaks in Figure~11, associated with the
first-generation fragments, was progressively narrowing down.  Accordingly,
the span of the times of return to perihelion was getting ever shorter.
On the other hand, the nodal longitudes of the first-generation fragments
essentially coincided, so that the span of the nodal longitudes was
unaffected by the increasing values of $X$ until it exceeded $\sim$0.5.
For the same reason, the perihelion distances remained independent of
the boundary condition for up to $X$ nearing unity ({\boldmath
${\cal R}$}$\, < 0.2$).

\begin{figure}[ht]  
\vspace{0.19cm}
\hspace{-0.185cm}
\centerline{
\scalebox{0.887}{
\includegraphics{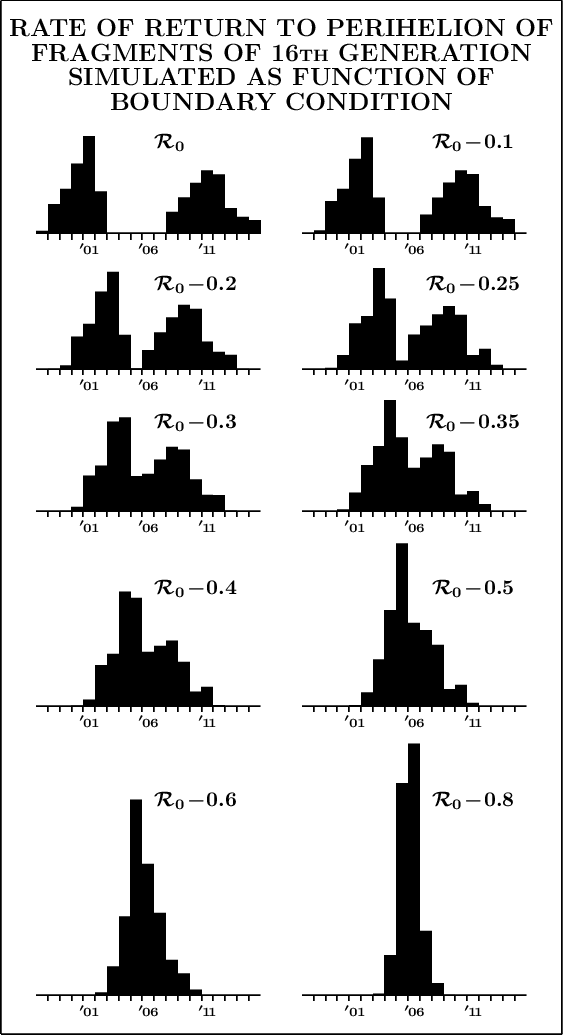}}}
\vspace{-0.02cm}
\caption{Effects of a boundary condition:~Histograms~of~a~rate of return
 to perihelion of fragments of the 16th generation simulated as a function
 of the time of the seed's initial breakup into two fragments of the first
 generation, expressed via the parameter $\Lambda$, selected by a random
 number \mbox{{\boldmath ${\cal R}_0$} = 0.9923017714} and its sequential
 variations.  The times of the seed's initial fragmentation event were,
 in days following the near-perihelion tidal disruption of the Great Comet
 of 1106:\ 0.34 for {\boldmath ${\cal R}_0$}; 0.85 for \mbox{\boldmath
 ${\cal R}_0$\,--\,0.1 = 0.8923\,\ldots}; 2.2 for \mbox{\boldmath ${\cal
 R}_0$\,--\,0.2}; 3.7 for \mbox{\boldmath ${\cal R}_0$\,--\,0.25}; 6.1 for
 \mbox{\boldmath ${\cal R}_0$\,--\,0.3}; 10.3 for \mbox{\boldmath ${\cal
 R}_0$\,--\,0.35}; 17.8 for \mbox{\boldmath ${\cal R}_0$\,--\,0.4}; 54 for
 \mbox{\boldmath ${\cal R}_0$\,--\,0.5}; 172 for \mbox{\boldmath ${\cal
 R}_0$\,--\,0.6}; and 1933 (or 5.3~yr) for \mbox{\boldmath ${\cal
 R}_0$\,--\,0.8 = 0.1923\,\ldots}\,.  The abscissa shows time in years
 ($^\prime 01$ stands for 2001, etc.); the ordinate, uniform throughout,
 displays the annual rate of return to perihelion of simulated 16th~generation
 fragments; for example, the highest rate at the upper left panel is
 11,041 fragments per year (in 2001), the highest rate at the bottom
 right panel is 28,364 fragments per year (in 2006).  It is noted that,
 especially in the upper four histograms, the distributions are prominently
 double-peaked because the seed split into two first-generation fragments
 of very different orbital periods, one returning to perihelion in September
 2000, the other in August 2011, at times that the peaks are centered on.
 The projected return of the seed, in early 2006, approximately coincides
 with the center of the wide gap.  On the other hand, the distributions in
 the last three histograms are increasingly sharp-peaked on the year of
 2006, as at these values of $\Lambda$ the projected perihelion returns
 of the first generation fragments and their successive products nearly
 coincided with the projected return of the seed.{\vspace{-0.35cm}}}
\end{figure}

\subsection{Trends, Constraints, and Conditions} 
In a way, it was fortunate that the initial random~number nearly coincided
with the upper boundary, as it led to an unexpected scenario that otherwise
may not have been discovered:\ the first-generation fragments of the seed
moved in orbits differing from one another enough to have lasting effect
of the orbital evolution of fragments of the 16th generation, causing a
two-peaked distribution of their times of return to perihelion (centered
on 2001 and 2011/2012).  The peaks were separated by a huge gap with no
fragments, centered on the projected time of return of the seed
(in 2006), as is amply demonstrated by the panels in the two upper rows
of Figure~11.  However, when I forced the initial random number {\boldmath
${\cal R}_0$} to attain progressively lower values, the two peaks gradually
became less prominent, the gap between them shrinking and getting more
shallow.  At \mbox{{\boldmath ${\cal R}_0$}$-0.4$} one peak was on the verge
of diappearance, the span of the times of return to perihelion continuing
to diminish.  Eventually, the single peak near the seed's projected time
of return to perihelion became prominent and narrow as {\boldmath ${\cal
R}_0$} continued to decrease toward its lower boundary of zero.  This
was a shape of the distribution of perihelion arrival times that better
met one's expectation.

The peak arrival rates that I was getting are seen from Figure~11 to reach
about 10$^4$ SOHO-like sungrazers per year (sic!), or {\it two orders of
magnitude\/} too high in comparison with the observations that were mentioned
in Part~I [and here is Section~6.1 and Equation~(62)].  Furthermore, to
be in line with the data, the span of the times of return to perihelion in
Figure~10 should be wider by a factor of at least 2 to 5 (for the stream
to last an absolute minimum of 30~years).  On the other hand, the span
of the longitudes of the ascending node should be narrower to better fit the
width of 7$^\circ$, estimated for Population~I in Part~I.  It is obvious
that while the exercises performed in Sections~7.1 through 7.6 were useful
in terms of learning the effects of the various parameters on the spans
of the three orbital elements, the values used were apparently nowhere
near those needed to fit the observed stream of SOHO sungrazers.

\subsubsection{Attempting Corrective Actions} 
For a given $n_{\rm gen}$ and a number of objects, the annual arrival rate
must drop when a span of the times of return to perihelion broadens.  This
was one of the rules followed in an effort to rectify the problems with
the high rates.  To find appropriate values of the parameters, I focused
on two key issues:\ (i)~to fit the slope and end points of the arrival
rate of Population~I sungrazers observed in the C2 coronagraph in the
years \mbox{2000--2009} --- I opted for the rates corrected for the missed
data, 120~objects in 2000 and 200~objects in 2009 (Section~6.1); and
(ii)~to match the range of the corrected longitudes of the ascending
node of these objects.  Also taken into account were other well-known
properties of the stream of SOHO sungrazers, such as (iii)~an essentially
constant rate of arrivals after 2009 (Battams \& Knight 2017),\footnote{An
anomaly was the arrival of a swarm of bright Population~I sungrazers,
detected in the years 2009 through 2012.  The peak rates were 4.6 per
year in late 2010 for the objects brighter at maximum light than
magnitude~3 and 4.3 per year in early 2011 for the objects brighter
than magnitude~2 (Sekanina \& Kracht 2013).} a conclusion believed to
be in general valid for Population~I as well, even though its membership
was less clear in the absence of orbital information after 2010; and
(iv)~evidence of post-aphelion fragmentation events from detection of
close pairs of sungrazers (Sekanina 2000).

Given the degree of stochasticity, the search for the ``best'' solution
had to be conducted by trial and error, with hardly any interpolation
possible.  The constraints provided by the simulated distribution of the
longitudes of the ascending node proved decisive, as they ruled out
separation velocities higher than 0.7~m~s$^{-1}$, which would have led
to an unacceptably wide range of these nodal longitudes.

The limits dictated by the nodal-longitude conditions did unfortunately
have a very detrimental effect on the options left open to modeling the
annual arrival rates of the stream sungrazers.  The extremely high rates
would have been brought down at least to a degree by higher separation
velocities, because they would have stretched arrivals over wider periods
of time.  With this option now unavailable, the only way left to bring
the arrival rates in \mbox{2000--2009} down to \mbox{120--200} objects
per year was to accept that these years were near the beginning of the
span of the times of return to perihelion.

The best bet was to elevate $\Lambda$ to values near 2 to curtail the
number of fragmentation events at larger heliocentric distances, where
the nodal longitudes get perturbed more strongly, and to move the seed's
projected arrival time far into the future by choosing a proper value of
\mbox{$t_{\rm ref} \!-\! t_0$}.  In spite of the two caveats regarding
\mbox{$\Lambda > 2$} [one in Section 4 and Equation~(41), the other in
Section~6.3 and Table~11], neither objection ruled out maximum values
of $\Lambda$ near 2.1, so long as $\Lambda$ remained a random variable
with a lower limit below 2.

\subsubsection{Rejection of SOHO Sungrazer Stream Models\\Based on
 16 Fragment Generations} 

These efforts notwithstanding, it was not possible to avoid a continuing
rise of the simulated arrival rates at times following the tested period
of \mbox{2000--2009}.  The rates reached absurdly high levels, exceeding
900~objects per year (sic!) at maximum in the mid-21st century and they
topped 350~objects per year in the early 2020s, which we now know for
sure to be much too high.

\begin{figure*}[t] 
\vspace{0.19cm}
\hspace{-0.19cm}
\centerline{
\scalebox{0.68}{
\includegraphics{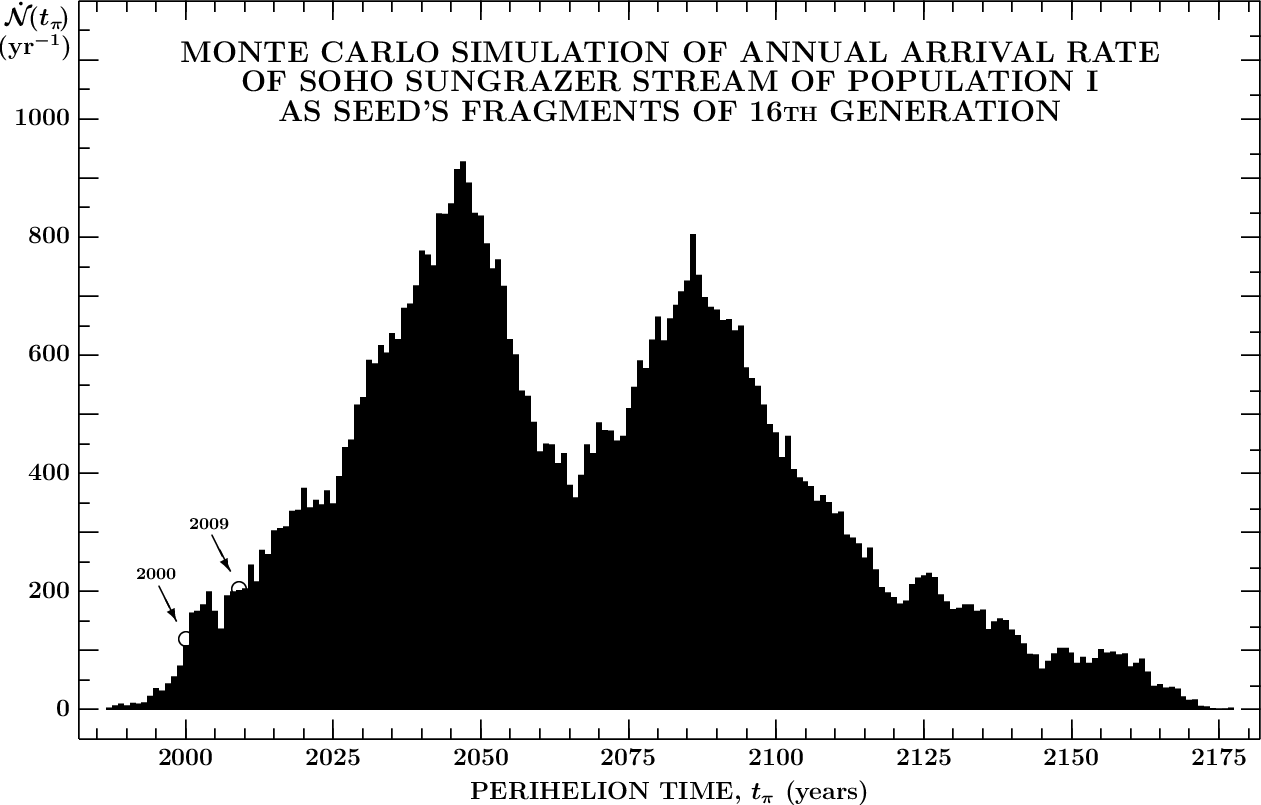}}}
\vspace{-0.17cm}
\caption{Simulation of an arrival rate $\dot{\cal N}$ (per year) of the stream
 of SOHO sungrazers of Population~I as fragments of the~16th~generation of
 a seed born at near-perihelion tidal breakup of comet X/1106~C1.
 {\vspace{-0.09cm}}Other parameters of the simulation curve:\ \mbox{$n_{\rm
 frg} = 20$}, \mbox{$\Lambda_{\rm min} = 1.93$}, \mbox{$\Lambda_{\rm max}
 = 2.13$}, \mbox{$v_{\rm sep} = 0.69$ m s$^{-1}$}, \mbox{$t_{\rm ref} \!-
 \! t_0 = 962$ yr}.  The two peaks measure the difference between the
 orbital periods of the two fragments of the first generation.  The shapes
 of these simulation curves are largely independent of the boundary condition.
 {\vspace{-0.065cm}}The simulated arrival rates, \mbox{$\dot{\cal N}(2000)
 = 110$ yr$^{-1}$} and \mbox{$\dot{\cal N}(2009) = 203$ yr$^{-1}$}, are
 compared with the corrected rates derived from the data, 120 and
 200~yr$^{-1}$, respectively, depicted by open circles.  The arrival-rate
 curve is grossly inconsistent with the observed arrival rates outside the
 interval \mbox{2000--2009}, implying 350 arrivals per year in the early
 2020s; all models simulating the stream of SOHO sungrazers as fragments
 of the 16th generation failed miserably.{\vspace{0.6cm}}}
\end{figure*}

An example of the arrival rate of the stream of SOHO sungrazers of
Population~I, simulated as consisting of 65,500 16th generation
fragments of a seed born at near-perihelion tidal breakup of the
parent comet X/1106~C1, is displayed in Figure~12.  The fragments
stretched over about 190~years, thus averaging about 350~objects
per year.  I came up with a number of alternative scenarioss, all
of which rather closely resembled the plotted one.  To bring the
rate down to less than 200~objects to get it in line with the
observations would have required that the stream stretched over
more than four centuries, an unacceptable condition.

The treatment of the SOHO sungrazers as fragments of the 16th generation
was based on a probable seed-to-sungrazer size ratio of about
40 --- a seed of 400~meters in diameter and a SOHO
sungrazer 10 meters across.  This ratio, which up to this point was the
most fundamental constant of these computations, could not be sustained
dynamically, leading for Population~I to a stream about
four times as dense as is the observed stream.  Accordingly, in the
framework of the methodology employed in this investigation, {\it all
sungrazer stream models based on the premise that the SOHO sungrazers
are fragments of the 16th generation must be rejected\/}.  The degree of
disparity suggests that the seed-to-sungrazer{\vspace{-0.055cm}} size ratio
has to be reduced by a factor of about \mbox{$4^{\frac{1}{3}} \simeq 1.6$},
from 40 down to 25; the SOHO sungrazers of Population~I should be
treated as {\it fragments of the 14th generation\/}, whose total number
from a single seed is \mbox{$2^{14} \simeq 16,400$}.  Considering that,
{\it after corrections\/}, the number of ``known'' members of this population
should amount to more than 4000, at least a quarter of the stream is
already on the books, not including the part before 1996.

One should add two caveats.  Contemplating the reduction of the
size ratio between the seed and a SOHO-like sungrazer, one should
remember that all fragments of a given generation are assumed to be of
equal size.  Accordingly, by the size of a SOHO sungrazer one should
understand the dimensions of an {\it average\/} member of this category,
not of the most common (smallest detected) member.  It very well may be this
difference that takes care of the discrepancy; one only needs to replace
the diameter of 10~meters with 16~meters, a difference too subtle
to argue about.

The other caveat concerns the seeds.  There is no doubt that
a single tidal breakup of the parent comet generates a number ---
probably a large number --- of seeds, whose orbital periods differ to
various degrees.  If the SOHO-like sungrazer debris from a single
seed gets scattered over an arc of the orbits equivalent to a span of
two centuries, the debris from any two seeds may or may not overlap,
scenarios that could further complicate the interpretation of the sungrazer
stream data.

\section{Results for Fragments of 14th Generation:\\A Success} 
Already the earliest computer runs for \mbox{$n_{\rm gen} = 14$} suggested
that the problems with high arrival rates were gone.  The trial and error
approach led to major differences in some parameters in comparison with
runs for \mbox{$n_{\rm gen} = 16$}.  Probably the most significant was
the shift in \mbox{$t_{\rm ref} \!-\! t_0$}.  While runs for the 16th
fragment generation predicted values near 960~years, indicating that its
original orbit would bring the seed back to perihelion in the late 2060s,
runs for the 14th generation led to times about 30~years earlier.
This shift moved the main peak in Figure~12 from the late 2040s by just
about the same amount of time, while the secondary peak in Figure~12 now
disappeared.

\begin{figure*}[t] 
\vspace{0.19cm}
\hspace{-0.19cm}
\centerline{
\scalebox{0.68}{
\includegraphics{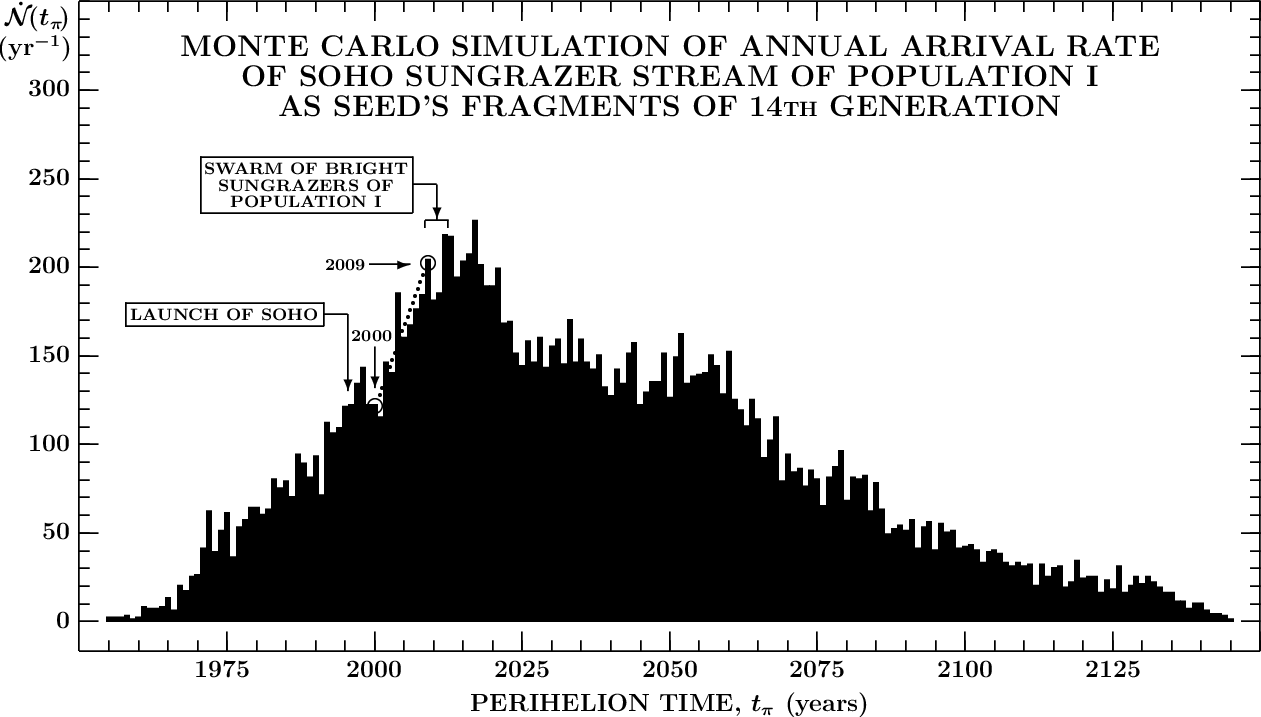}}}
\vspace{-0.17cm}
\caption{Successful simulation of an arrival rate $\dot{\cal N}$ (per
 year) of the stream of SOHO sungrazers of Population I as fragments of
 the 14th generation of a seed born at near-perihelion tidal breakup of
 comet X/1106~C1.  Other parameters{\vspace{-0.065cm}} of the simulation
 curve:\ \mbox{$n_{\rm frg} = 20$}, \mbox{$\Lambda_{\rm min} = 1.6$},
 \mbox{$\Lambda_{\rm max} = 2.1$}, \mbox{$v_{\rm sep} = 1.20$ m s$^{-1}$},
 \mbox{$t_{\rm ref} \!-\! t_0 = 930$ yr}.  The peak measures the orbital
 period of one of the two fragments of the first generation, which would
 have passed perihelion in June 2011.  The shapes of these simulation
 {\vspace{-0.055cm}}curves are largely independent of the boundary
 condition.  The simulated arrival rates, \mbox{$\dot{\cal N}(2000) =
 123$ yr$^{-1}$} and \mbox{$\dot{\cal N}(2009) = 205$ yr$^{-1}$}, are
 {\vspace{-0.045cm}}compared with the corrected rates derived from the
 data, 120 and 200~yr$^{-1}$, respectively, depicted by the open circles.
 Also shown are the launch of the SOHO observatory and the detection of
 a swarm of bright SOHO sungrazers (of peak magnitude not fainter than 3)
 in \mbox{2009--2012} (Sekanina \& Kracht 2013), a coincidence that may
 corroborate the predicted time of the stream's maximum arrival
 rate.{\vspace{0.6cm}}}
\end{figure*}

An example of the arrival-rate simulation curve for the stream of SOHO
sungrazers of Population~I, established from an excellent Monte Carlo
run based on fragments of the 14th generation, is displayed in Figure~13
and its parameters are listed in Table~13.  The plot offers several
predictions:\ (i)~the stream should survive for almost exactly
200~years, from about 1950 to 2150, even though over the last 50~years
the arrival rate is expected to average less than 1~sungrazer per
week and in the last decade or so less than 1~sungrazer per month;
(ii)~the stream should have peaked in the 2010s, when the arrival
rate of the sungrazers detected in the C2 coronagraph (and corrected
for the missed ones) should have attained about 220 per year; the
peak appears to be a product of fragmentation of one of the two parts
of the seed that ended up in an orbit that would have brought it back
to perihelion in June 2011; (iii)~perhaps significantly, the timing
of the peak essentially coincides with the reported arrival of a swarm
of bright SOHO sungrazers (Sekanina \& Kracht 2013), of magnitude not
fainter than 3 at maximum light; these could be larger fragments of
the same part of the seed, whose fragmentation had stopped or was slow;
and (iv)~values of the parameters of the simulated arrival-rate curve
were searched for by trial and error to approximately{\vspace{-0.04cm}}
fit the corrected rates of \mbox{$\dot{\cal N}(2000) = 120$ per year}
and \mbox{$\dot{\cal N}(2009) = 200$ per year}, as established for
Population~I in Part~I; this was the only constraint used from the
stream's arrival rates.

\begin{table}[b] 
\vspace{0.7cm}
\hspace{-0.19cm}
\centerline{
\scalebox{1}{
\includegraphics{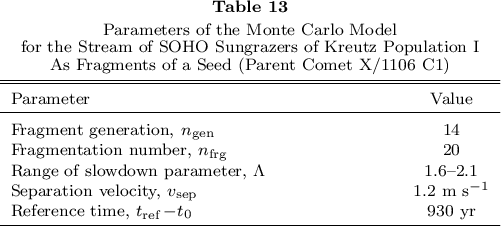}}}
\vspace{0.15cm}
\end{table}

The separation velocity in Table 13 is surprisingly high compared to the
range of values contemplated in Section 3.3 and Table~6.  The disparity
has two important implications.  One is an effect on the relationship
between the separation velocity and what I took to be a {\it tensile\/}
strength of the fragments.  In retrospect, the involved quantity of
$\sigma_{\rm T}$ should better be called a {\it cohesion\/} strength,
for which $v_{\rm sep}$ from Table~13 implies a value of about 2700~Pa,
more than one order of magnitude higher than the tensile strength of
comets is believed to average.  The second implication concerns the
rotation period of the minor sungrazers, which appears to be on the
order of {\it 1 minute\/} if not shorter.

Another constraint was provided by the span of nodal longitudes.  For
the period from 1996 through mid-2010 the distribution of 390 members of
Population~I from the corrected Marsden data was plotted in Figure~3 of
Part~I, showing the nodal longitudes in a range of about 7$^\circ$, from
359$^\circ\!$.5 to 6$^\circ\!$.5.  The sungrazer stream simulated here
by fragments of the 14th generation offers a very good match; for the
period from 1996 through mid-2010 the histogram is plotted in Figure~14,
for the \mbox{2000--2009} period in Figure~15.  Both plots show a clear
peak at the longitude of the ascending node of the parent comet (and the
parent seed), as expected.  The range of the longitudes is slightly
shifted to higher values, but the contamination by the excess beyond
6$^\circ\!$.5 is less than 2~percent.  The three potential swarms are
not seen in the simulation, but this is not surprising given that the
observations provided only an average of about 11~data points per
0$^\circ\!$.2 wide interval of longitude, thus failing to prevent
significant scatter.

For the sake of interest, I also provide, in Figure~16, a histogram of
the distribution of perihelion distances among simulated fragments of
the 14th generation in the \mbox{2000--2009} period.  As already noted
in Part~I, unlike the longitude of the ascending node, the perihelion
distance of the Marsden gravitational orbits for the SOHO sungrazers
cannot be corrected for effects of the nongravitational acceleration
and, as a result, can provide no additional orbital information on the
fragmentation process.  This point is supported by Figure~16, which
predicts the simulated distribution of fragments' perihelion distances
to be confined to between 1.08 and 1.28~\Rsun, while the orbital data
show about 10~percent of the Population~I sungrazers from the
\mbox{2000--2009} period to have perihelia beyond 1.5~\Rsun.

\begin{figure*}[t] 
\vspace{0.19cm}
\hspace{-0.19cm}
\centerline{
\scalebox{0.8}{
\includegraphics{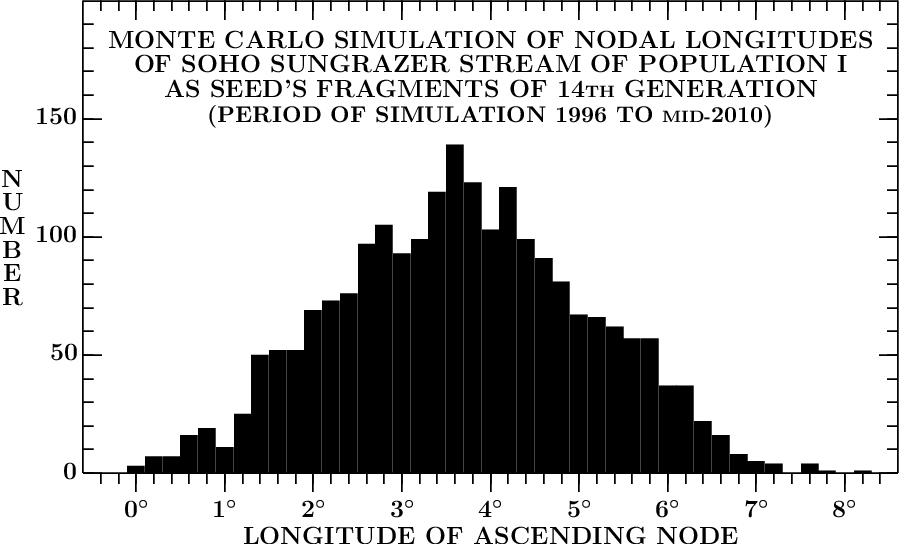}}}
\vspace{0cm}
\caption{Histogram of the Monte Carlo simulated longitudes of the ascending
node for the stream of SOHO sungrazers of Population~I, modeled as 14th
generation fragments of a seed.  The selected period is 1996 to mid-2010,
the time interval for which Marsden's gravitational orbits were available.
Note a good match to the observed range of nodal longitudes of 7$^\circ$,
from 359$^\circ\!$.5 to 6$^\circ\!$.5, presented in Part~I.{\vspace{0.55cm}}}
\end{figure*}

\section{Summary and Conclusions}
This Part II completes the first serious effort aimed at providing a physical
model for the stream of SOHO sungrazers\footnote{I keep referring to the
{\it SOHO sungrazer stream\/}, but the imagers on board the {\it STEREO\/}
spacecraft (and, very recently, the {\it Parker Solar Probe\/}) should also
be acknowledged for contributing to the stream's observation.  And to the
extent that one could speak of streams made up of fewer than two dozen comets,
the same applies to the imagers on board the {\it Solwind\/} (P78-1) and {\it
Solar Maximum Mission\/} satellites in \mbox{1979--1989}. My reference to the
SOHO stream is merely a short-cut term that recognizes SOHO's data and
dominant position in this field of scientific activity.{\vspace{0.1cm}}} 
of the Kreutz system.  Even though the first SOHO sungrazers, which have for
nearly 30~years been dominated by the Kreutz comets, were not detected
until more than two years after the space observatory's launch, the pace
has been accelerating over the years thanks to spontaneous enthusiasm of
participating amateur comet hunters, and a celebration of the discovery
of the incredible 5000th object is still in progress as these lines are
being written!

The stream of Kreutz sungrazers is complex because the Kreutz system is
complex, consisting of a number of populations of genetic significance.
This investigation deals with the stream of only one of them ---
Population~I, which was introduced by Marsden (1967) as Subgroup~I and
which has ever since 1996 been the main contributor to the SOHO sungrazer
stream.  One can take a view that the observed stream consists of a number
of individual streams, each population having its own.  Whether or to what
extent is a model for one stream useful to formulating models for the
others remains to be determined.

While the sections of this paper have been organized thematically, the
summary is arranged chronologically, which the reader may prefer.  One
point concerning the stream of SOHO sungrazers that is unlikely to invite
any controversy is a declaration that the stream consists of fragments
of a larger, parent body or bodies.  Speaking of the Kreutz system as
a whole, the ultimate parent was the progenitor, which my recent
contact-binary model (Sekanina 2021b) identifies with Aristotle's
comet of 372~BC.  Speaking more specifically of Population~I, as I do
in this investigation, a more direct parent was this population's
principal mass, which in my model was the Great Comet of 1106 (X/1106~C1).
But I propose that the {\it immediate\/} parent of the sungrazer stream
that the SOHO's coronagraphs see was a subkilometer-sized fragment of the
1106 comet, which separated from it (with many other objects of different
sizes, both larger and smaller) in the course of an event of tidal
fragmentation in close proximity of perihelion, at the beginning of
February 1106.  I refer to this immediate parent as a {\it seed\/}.
I believe that the seed of the SOHO sungrazer stream of Population~I was a
single object, even though it is virtually certain that there were many
seeds, each generating its own stream at some time.  Streams of different
seeds may, but do not have to, overlap one another.  From the limited
data available I have as yet seen no evidence that would contradict the
premise of a single seed.

Not every fragment of the parent comet can become a seed.  A seed has
to be resilent and large enough to survive on its own the first hours
after birth in the harsh environment of the solar corona, yet in the long
run it should be brittle enough to be prone to progressive fragmentation.
Objects like comet Lovejoy (C/2011~W3), which was not a member of
Population~I, or the Great Southern Comet of 1887 (C/1887~B1), which was,
may be appropriate seed prototypes.  Their existence suggests that the
formation of a SOHO-like stream does not have to be necessarily linked
to the perihelion passage of a major Kreutz sungrazer, even though in
most cases it probably is.

\begin{figure*}[t] 
\vspace{0.19cm}
\hspace{-0.19cm}
\centerline{
\scalebox{0.8}{
\includegraphics{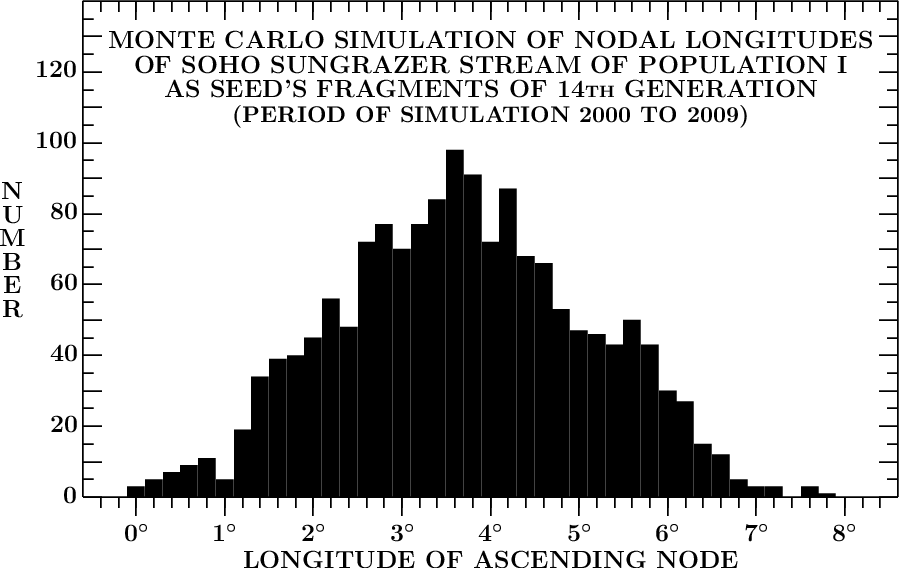}}}
\vspace{-0.05cm}
\caption{Histogram of the Monte Carlo simulated longitudes of the ascending
 node for the stream of SOHO sungrazers of Population~I, modeled as 14th
 generation fragments of a seed.  The selected period is 2000 to 2009,
 used in Figure~13 to fit the arrival-rate curve.  Note a good match to
 the observed range of nodal longitudes of 7$^\circ$, from 359$^\circ\!$.5
 to 6$^\circ\!$.5, presented in Part~I.{\vspace{0.7cm}}}
\end{figure*}
\begin{figure}[b] 
\vspace{0cm}
\hspace{-0.19cm}
\centerline{
\scalebox{0.933}{
\includegraphics{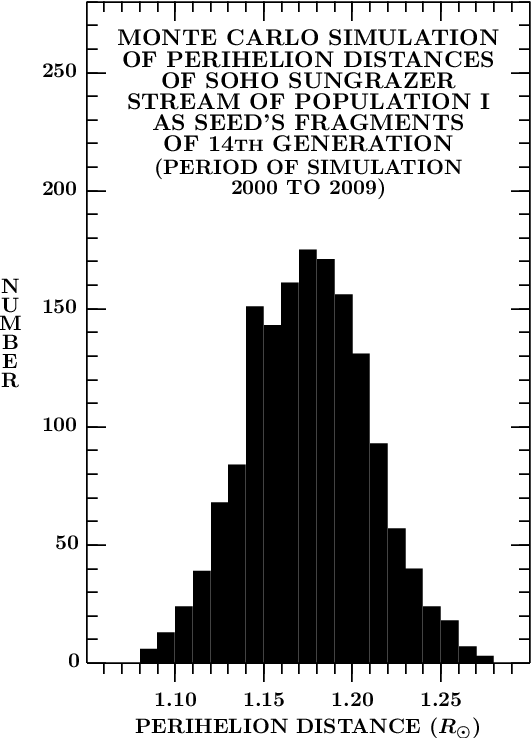}}}
\vspace{-0.1cm}
\caption{Histogram of the Monte Carlo simulated perihelion distances for
the stream of SOHO sungrazers of Population~I, modeled as 14th generation
fragments if a seed.  The selected period is 2000 to 2009, used in
Figure~13 to fit the arrival-rate curve.  Unfortunately, this distribution
cannot be tested on a set of observed perihelion distances, because the
available data are not even approximately corrected for major nongravitational
effects.{\vspace{0cm}}}
\end{figure}

Modeling of a seed's fragmentation process leading to a SOHO-like stream
consists of handling two different issues:\ (i)~repetitive scaling down
of a rotationally-induced fragmentation event whose product is assumed
to be self-similar and (ii)~a distribution law for the sequence of such
fragmentation events along the orbit.  Since all fragments are irregular
in shape, their figure is ignored when developing the algorithm for
solving the first problem.  At the same time, the fragment size is
linked to the number of events by postulating that, in each fragmentation
episode, a spherical object breaks into two equal halves.  This procedure
is easy to handle mathematically, introducing a dependence on the
material's tensile or, rather, cohesive strength against rotational
fragmentation.  The value of this quantity is determined indirectly,
via a separation velocity.  Its {\it magnitude\/} is assumed to be
constant, which implies a spinup as the fragments' dimensions diminish
with time.  On the other hand, as the fragments are expected to tumble
uncontrolably, the separation velocity {\it vector\/} is allowed to
vary at random.   In practice, its RTN components are determined by
a random-number generator.

The separation velocity vector serves as a perturber of fragments' paths,
which is ultimately responsible for orbital scatter of the sungrazer
stream.  The section on the perturbations demonstrates in some detail,
how the contributions to the radial, transverse, and normal components
of the separation velocity project as variations in the orbital period,
the perihelion distance, and the longitude of the ascending node.  The
orbital position at the time of fragmentation plays an important role
in determining the amounts of change in the various orbital elements.  

%
These effects are in the paper described by a law, whose beginnings date
to my early investigation of cascading fragmentation (Sekanina 2002) to
replace the traditional view that sungrazers split only at perihelion.
Based in part on Sekanina et al.'s (1998) experience with a breakup
sequence of the ill-fated comet Shoemaker-Levy (D/1993~F2), evidence
suggested that the fragmentation rate was tapering off --- the reason
for introducing the ``slowdown'' parameter $\Lambda$ (Sekanina 2002).

Among the additional parameters of the proposed geometric progression
of fragmentation times, the most important is the number of fragment
generations, $n_{\rm gen}$.  Because any fragment is considered to
split into two equal subfragments, there is a relationship between the
number of fragments and the number of their generations.  This rule
also resolves the issue of the fragments' momentum exchange at
breakup.  Potentially confusing is the meaning of the fragmentation
number, $n_{\rm frg}$.  As introduced in Section~4, all fragments of
the $n_{\rm frg}$-th generation return to perihelion at the same time,
regardless of $\Lambda$, as shown in Table~7.  But sungrazers of course
return to perihelion at very different times, when \mbox{$n < n_{\rm
frg}$}.  The fragmentation law is incorporated into the Monte Carlo
simulation routine in Section~6.3, in which the number of fragment
generations, $n_{\rm gen}$, needed to fit the observed arrival rate
of SOHO sungrazers, is introduced.  The arrival rate also depends on
the seed's initial orbital period.

The stochastic nature of the fragmentation process is assumed to be
determined primarily by the complex sequence of fragmentation events,
their timing on the one hand, and by chaotic rotation of fragments,
which results in their unpredictable momentum changes at each event,
on the other hand.  These circumstances have dictated the need for
application of a Monte Carlo simulation approach.

Certain aspects of the fragmentation problem are, for better or worse,
treated deterministically.  For example, all SOHO sungrazers are treated
as fragments of the same generation and equal dimensions.  Similarly,
I assume a constant size (and mass) ratio between the seed and a
fragment.  As already noted, the magnitude of separation velocities
is also kept constant.  The values of the slowdown parameter $\Lambda$
are subject to random variations, but only within a prescribed range.

In conclusion, one of the prime objectives --- to fit the annual arrival
rate of the SOHO sungrazers of Population~I over a period of 10~years ---
has been achieved in Figure~13.  The rates, from 120 per year in 2000
to 200 per year in 2009, have rested on the counts in the C2 coronagraph,
corrected for missed objects due to seasonal effects.  A search by trial
and error suggests that a successful solution requires that the SOHO
sungrazers be 14th generation fragments, which implies a total number of
2$^{14}$ or about 16,400 objects and a 25:1 size ratio between the seed and
an average sungrazer; for example, a seed 400~meters in diameter requires
that an average sungrazer be 16~meters across.  The second prime objective
--- to fit the observed range of 7$^\circ$ in the longitude of the
ascending node of the sungrazers, from 359$^\circ\!$.5 to 6$^\circ\!$.5
--- has been taken care of in Figures~14~and~15.

The stream's activity is predicted to last for 200~years, from about
1950 to about 2150, suggesting that by the end of 2023 approximately
42~percent of the stream's sungrazers had already arrived.  The
stream is predicted to have culminated in the 2010s.  Interestingly,
a swarm of bright SOHO sungrazers of Population~I (not fainter than
magnitude 3 at peak light) appeared in \mbox{2009--2012} (Sekanina
\& Kracht 2013).  These could be fragments of the seed that either
ceased breaking up early or have been fracturing at much slower rates.
On its initial orbit, whose period was 193~years longer than the orbital
period of X/1106~C1's principal fragment, C/1843~D1, the seed would
arrive in 2036.

My description on the proposed Monte Carlo model for the SOHO sungrazer
Population~I stream comes here to an end.  Although not directly related
to the topic, I still should point out that computations carried out in
Section~2.3 suggest that the apparent tidal disruption of the parent
comet X/1106~C1, which appears to have triggered the birth of the stream,
may also account for C/1668~E1 and a possible major fragment that could
arrive in the 2050s or 2060s.\\

It is my sincere hope that this paper will offer a good starting
position for more ambitious investigations in the future.\\

\begin{center}
{\footnotesize REFERENCES}
\end{center}
%
\begin{description}
{\footnotesize
\item[\hspace{-0.3cm}]
Attree, N., Groussin, O., Jorda, L., et al.\ 2018, A\&A, 611, A33
\\[-0.57cm]
\item[\hspace{-0.3cm}]
Battams, K., \& Knight, M.\ M.\ 2017, Phil.\ Trans.\ Roy.\ Soc.\ A,~375,{\linebreak}
 {\hspace*{-0.6cm}}20160257
\\[-0.57cm]
\item[\hspace{-0.3cm}]
Boehnhardt, H.\ 2002, Earth Moon Plan., 89, 91
\\[-0.57cm]
\item[\hspace{-0.3cm}]
Davidsson, B.\ J.\ R.\ 2001, Icarus, 149, 375
\\[-0.57cm]
\item[\hspace{-0.3cm}]
Delsemme, A.\ H., \& Miller, D.\ C.\ 1971, Plan.\ Space Sci., 19, 1229
\\[-0.57cm]
\item[\hspace{-0.3cm}]
Gould, B.\ A.\ 1891, Result. Obs. Nacion. Arg., 13, 600
\\[-0.57cm]
\item[\hspace{-0.3cm}]
Green, D.\ W.\ E.\ 2011, IAU Circ.\ 9246
\\[-0.57cm]
\item[\hspace{-0.3cm}]
Groussin, O., Jorda, L., Auger, A., et al.\ 2015, A\&A, 583, A32
\\[-0.57cm]
\item[\hspace{-03cm}]
Hamid, S.\ E., \& Whipple, F.\ L.\ 1953, AJ, 58, 100
\\[-0.57cm]
\item[\hspace{-0.3cm}]
Henderson, T.\ 1843, Astron.\ Nachr., 20, 333
\\[-0.57cm]
%
%
%
%
%
%
\item[\hspace{-0.3cm}]
Hufnagel, L.\ 1919, Astron.\ Nachr., 209, 17
\\[-0.57cm]
\item[\hspace{-0.3cm}]
Jewitt, D.\ 1997, Earth Moon Plan., 79, 35
\\[-0.57cm]
\item[\hspace{-0.3cm}]
Jewitt, D.\ 2021, AJ, 161, 261
\\[-0.57cm]
\item[\hspace{-0.3cm}]
Jewitt, D.\ 2022, AJ, 164, 158
\\[-0.57cm]
\item[\hspace{-0.3cm}]
Knight, M.\ M., A'Hearn, M.\ F., Biesecker, D.\ A., et al.\ 2010,
 AJ,{\linebreak}
 {\hspace*{-0.6cm}}139, 926
\\[-0.57cm]
\item[\hspace{-0.3cm}]
Kreutz, H.\ 1891, Publ.\ Kiel Sternw., 6
\\[-0.57cm]
\item[\hspace{-0.3cm}]
Kreutz, H.\ 1901, Astron.\ Abhandl., 1, 1
\\[-0.57cm]
\item[\hspace{-0.3cm}]
MacQueen, R.\ M., \& St.\ Cyr, O.\ C.\ 1991, Icarus, 90, 96
\\[-0.57cm]
\item[\hspace{-0.3cm}]
Marsden, B.\ G.\ 1967, AJ, 72, 1170
\\[-0.57cm]
\item[\hspace{-0.3cm}]
Marsden, B.\ G.\ 1989, AJ, 98, 2306
\\[-0.57cm]
%
%
%
\item[\hspace{-0.3cm}]
McCauley, P., Saar, S.\,H., Raymond, J.\,C.,\,et al.\,2013, ApJ,\,768,\,161
\\[-0.57cm]
\item[\hspace{-0.3cm}]
Michels, D.\ J., Sheeley, N.\ R., Jr., Howard, R.\ A., \& Koomen,
M.~J.{\linebreak}
 {\hspace*{-0.6cm}}1982, Science, 215, 1097
\\[-0.57cm]
\item[\hspace{-0.3cm}]
Raymond, J.\,C., Downs, C., Knight, M.\,M., et al.\,2018, ApJ,\,858,\,19
\\[-0.57cm]
\item[\hspace{-0.3cm}]
Seargent, D.\ 2009, The Greatest Comets in History:\ Broom Stars{\linebreak}
 {\hspace*{-0.6cm}}and Celestial Scimitars.  New York:\ Springer
 Science+Business{\linebreak}
 {\hspace*{-0.6cm}}Media, LLC, 260pp
\\[-0.57cm]
\item[\hspace{-0.3cm}]
Sekanina, Z.\ 1982, in Comets, ed.\ L.\ L.\ Wilkening (Tucson:\
 Univ.{\linebreak}
 {\hspace*{-0.6cm}}Arizona), 251
\\[-0.57cm]
\item[\hspace{-0.3cm}]
Sekanina, Z.\ 1984, Icarus, 58, 81
\\[-0.57cm]
\item[\hspace{-0.3cm}]
Sekanina, Z.\ 2000, ApJ, 542, L147
\\[-0.57cm]
\item[\hspace{-0.3cm}]
Sekanina, Z.\ 2002, ApJ, 566, 577
\\[-0.57cm]
\item[\hspace{-0.3cm}]
Sekanina, Z.\ 2003, ApJ, 597, 1237
\\[-0.57cm]
%
%
\item[\hspace{-0.3cm}]
Sekanina, Z.\ 2021a, eprint arXiv:2110.10889
\\[-0.57cm]
\item[\hspace{-0.3cm}]
Sekanina, Z.\ 2021b, eprint arXiv:2109.01297
\\[-0.57cm]
\item[\hspace{-0.3cm}]
Sekanina, Z.\ 2022a, eprint arXiv:2212.11919
\\[-0.57cm]
\item[\hspace{-0.3cm}]
Sekanina, Z.\ 2022b, eprint arXiv:2211.03271
\\[-0.57cm]
\item[\hspace{-0.3cm}]
Sekanina, Z., \& Chodas, P.\ W.\ 2007, ApJ, 663, 657
\\[-0.57cm]
\item[\hspace{-0.3cm}]
Sekanina, Z., \& Chodas, P.\ W.\ 2008, ApJ, 687, 1415
\\[-0.57cm]
\item[\hspace{-0.3cm}]
Sekanina, Z., \& Chodas, P.\ W.\ 2012, ApJ, 757, 127
\\[-0.57cm]
\item[\hspace{-0.3cm}]
Sekanina, Z., \& Kracht, R.\ 2013, ApJ, 778, 24
\\[-0.57cm]
%
%
\item[\hspace{-0.3cm}]
Sekanina, Z., \& Kracht, R.\ 2015a, ApJ, 815, 52
\\[-0.57cm]
\item[\hspace{-0.3cm}]
Sekanina, Z., \& Kracht, R.\ 2015b, ApJ, 801, 135
\\[-0.57cm]
\item[\hspace{-0.3cm}]
Sekanina, Z., \& Kracht, R.\ 2022, eprint arXiv:2206.10827
\\[-0.57cm]
\item[\hspace{-0.3cm}]
Sekanina, Z., Chodas, P.\ W., \& Yeomans, D.\ K.\ 1998, Plan.\ Space{\linebreak}
 {\hspace*{-0.6cm}}Sci., 46, 21
\\[-0.57cm]
\item[\hspace{-0.3cm}]
Sheeley, N.\ R., Jr., Howard, R.\ A., Koomen, M.\ J., \& Michels,
 D.~J.{\linebreak}
 {\hspace*{-0.6cm}}1982, Nature, 300, 239
\\[-0.57cm]
\item[\hspace{-0.3cm}]
Strom, R. 2002, A\&A, 387, L17
\\[-0.57cm]
%
%
\item[\hspace{-0.3cm}]
Vincent,\ J.-B., Hviid,\ S.\ F., Mottola,\ S., et al.\ 2017,
 Mon.\ Not.\ Roy.{\linebreak}
 {\hspace*{-0.6cm}}Astron.\ Soc., 469, S329
\\[-0.57cm]
\item[\hspace{-0.3cm}]
Warner, B.\ 1980, Mon.\ Not.\ Astron.\ Soc.\ South Africa, 39, 69
\\[-0.57cm]
\item[\hspace{-0.3cm}]
Whipple, F.\ L., \& Stefanik, R.\ P.\ 1966, M\'em.\ Soc.\ Roy.\ Sci.\
 Li\`ege{\linebreak}
 {\hspace*{-0.6cm}}(S\'er.\ 5), 12, 33
\\[-0.67cm]
\item[\hspace{-0.3cm}]
Ye, Q., Jewitt, D., Hui, M.-T., et al.\ 2021, AJ, 162, 70}
%
\vspace{-0.39cm}
\end{description}
\end{document}